\documentclass[twocolumn,a4paper,reprint,amssymb,aps,prd,groupedaddress,superscriptaddress,nofootinbib,floatfix]{revtex4-2}
\pdfoutput=1
\usepackage{graphicx}
\usepackage{epsf}
\usepackage{bm}
\usepackage{amsmath}
\usepackage{amsfonts}
\usepackage{amssymb}
\usepackage{epstopdf}
\usepackage{natbib}
\usepackage{color}
\usepackage[dvipsnames]{xcolor}
\usepackage{siunitx}
\usepackage{booktabs}
\usepackage{physics}
\usepackage{bm}
\usepackage[colorlinks = true,
            linkcolor = blue,
            urlcolor = blue,
            citecolor = blue,
            anchorcolor = blue]{hyperref}
\usepackage{lipsum}
\providecommand{\U}[1]{\protect\rule{.1in}{.1in}}

\usepackage[capitalize]{cleveref}
\usepackage{soul}
\usepackage{braket}
\usepackage{enumitem}
\usepackage{subcaption}
\usepackage{hyperref}

\begin{document}

\title{Addressing the DESI DR2 Phantom-Crossing Anomaly and Enhanced $H_0$ Tension with Reconstructed Scalar-Tensor Gravity} 

\author{Dimitrios Efstratiou}
\email{d.efstratiou@uoi.gr}
\affiliation{Department of Physics, University of Ioannina, GR-45110, Ioannina, Greece}

\author{Evangelos Achilleas Paraskevas}
\email{e.paraskevas@uoi.gr}
\affiliation{Department of Physics, University of Ioannina, GR-45110, Ioannina, Greece}
\author{Leandros Perivolaropoulos}
\email{leandros@uoi.gr}
\affiliation{Department of Physics, University of Ioannina, GR-45110, Ioannina, Greece}

\begin{abstract}
Recent cosmological data, including DESI DR2, highlight significant tensions within the $\Lambda$CDM paradigm. When analyzed in the context of General Relativity (GR), the latest DESI data favor a dynamical dark energy (DDE) equation of state, $w(z)$, that crosses the phantom divide line $w=-1$. However, this framework prefers a lower Hubble constant, $H_0$, than Planck 2018, thereby worsening the tension with local measurements. This phantom crossing is a key feature that cannot be achieved by minimally coupled scalar fields (quintessence) within GR. This suggests the need for a new degree of freedom that can simultaneously: (A) increase the best-fit value of $H_0$ in the context of the DESI DR2 data, and (B) allow the crossing of the $w=-1$ line within a new theoretical approach. We argue that both of these goals may be achieved in the context of Modified Gravity (MG), and in particular, Scalar-Tensor (ST) theories, where phantom crossing is a natural and viable feature. We demonstrate these facts by analyzing a joint dataset including DESI DR2, Pantheon+, CMB, and growth-rate (RSD) data in the context of simple parametrizations for the effective gravitational constant, $\mu_G(z) \equiv G_{\rm eff}/G_N$, and the DDE equation of state, $w(z)$. This MG framework significantly alleviates the tension, leading to a higher inferred value of $H_0 = 70.6 \pm 1.4 \, \text{km s}^{-1} \text{Mpc}^{-1}$. We also present a systematic, data-driven reconstruction of the required underlying ST Lagrangian and provide simple, generic analytical expressions for both the non-minimal coupling $F(\Phi) = 1+\xi\Phi^{2}e^{n\Phi}$ and the scalar potential $U(\Phi) = U_{0}+ae^{b\Phi^{2}}$, which well-describe the reconstructed functions.
\end{abstract}

\maketitle

\section{\label{sec:introduction}Introduction}

The \(\Lambda\)CDM model is the standard framework for cosmic evolution, defined by six parameters. It has successfully described cosmic expansion and structure formation for decades. However, new observations show persistent tensions in the \(\Lambda\)CDM model~\cite{2008arXiv0811_4684P,Perivolaropoulos:2021jda,Abdalla:2022yfr,Efstathiou:2024dvn,Peebles:2024txt,Peebles:2022akh,Bull:2015stt,Bullock:2017xww,DiValentino:2017gzb}. These tensions challenge the model's completeness and drive the search for alternatives.
The most prominent cosmological tensions concern the Hubble constant, $H_0$, and the weighted amplitude of matter fluctuations, defined as $S_8 \equiv \sigma_8 \sqrt{\Omega_{\rm m0} / 0.3}$. The so-called $H_0$ tension~\cite{Perivolaropoulos:2021jda,Abdalla:2022yfr,DiValentino:2021izs,Hu:2023jqc,CosmoVerseNetwork:2025alb,Freedman:2017yms,DiValentino:2020zio,Verde:2019ivm,Shah:2021onj,Schoneberg:2021qvd,Verde:2023lmm,Perivolaropoulos:2024yxv,Pantos:2026cxv} remains one of the most significant challenges in modern cosmology. It manifests as a more-than-$5\sigma$ discrepancy between the Planck--$\Lambda$CDM inference, $H_0 = 67.36 \pm 0.54\,{\rm km\,s^{-1}\,Mpc^{-1}}$, based on CMB data~\cite{Planck:2018vyg}, and the local SH0ES determination from Cepheid-calibrated distance ladders, $H_0 = 73.04 \pm 1.04\,{\rm km\,s^{-1}\,Mpc^{-1}}$ (rising to $73.30 \pm 1.04\,{\rm km\,s^{-1}\,Mpc^{-1}}$ when high-redshift Type~Ia supernovae are included)~\cite{Riess:2021jrx}. Latest analysis refines the local measurement to \(H_0 = 73.17 \pm 0.86~\mathrm{km\,s^{-1}\,Mpc^{-1}}\)~\cite{Breuval:2024lsv}. Recent JWST observations provide the most precise validation of Cepheid-based distance measurements to date. Combining JWST Cepheid data for 19 hosts (24 SNe~Ia) with HST measurements for 37 hosts (42 SNe~Ia) yields \(H_0 = 73.49 \pm 0.93~\mathrm{km\,s^{-1}\,Mpc^{-1}}\)\citep{Riess:2025chq}. Including 35 TRGB-based calibrations (from HST and JWST) expands the sample to 55 SNe~Ia, giving \(H_0 = 73.18 \pm 0.88~\mathrm{km\,s^{-1}\,Mpc^{-1}}\), approximately \(6\sigma\) higher than the \(\Lambda\)CDM+CMB prediction \citep{Riess:2025chq}. In contrast, a recent CMB analysis using SPT-3G data finds \(H_0 = 66.66 \pm 0.60~\mathrm{km\,s^{-1}\,Mpc^{-1}}\) within \(\Lambda\)CDM, lying \(6.2\sigma\) below the SH0ES result and reinforcing the early–late Universe tension~\cite{SPT-3G:2025bzu}. Combining SPT-3G with ACT and Planck further tightens the constraint to \(H_0 = 67.24 \pm 0.35~\mathrm{km\,s^{-1}\,Mpc^{-1}}\), consistent with Planck and confirming the persistence of the Hubble tension. The extent to which $H_0$ and $H(z)$ can deviate from $\Lambda$CDM while remaining consistent with CMB observations has been quantified in previous works~\cite{Das:2023rvg,Das:2013sca}, providing critical boundaries for proposed solutions to these cosmological tensions. Recent binned analyses of Type Ia supernova and BAO datasets, using samples such as Pantheon, Pantheon+, Joint Light-curve Analysis (JLA) \cite{SDSS:2014iwm}, Dark Energy Survey (DES) \cite{DES:2021jns,DES:2021wwk,DES:2025llp}, and their combinations, have revealed a slowly decreasing trend for the Hubble constant with redshift. This evolutionary behavior provides valuable insight into the $H_0$ tension and has been reported by multiple studies~\cite{Dainotti:2021pqg,Dainotti:2022bzg,Schiavone:2022wvq,Dainotti:2025qxz,DeSimone:2024lvy}.

The \(S_8\) tension arises from a mismatch between the inferred value of \(S_8\) when a cosmological model—such as the \(\Lambda\)CDM model—is constrained using CMB data, which primarily probe the early universe, and when it is constrained using large-scale structure (LSS) probes such as weak lensing, cluster counts, and redshift-space distortions, which are sensitive to the late universe (see also in \cite{CosmoVerseNetwork:2025alb,Pantos:2026koc}). Planck-\(\Lambda\)CDM predicts a higher weighted amplitude of matter fluctuations, specifically \(S_8 = 0.832 \pm 0.013\)~\cite{Planck:2018vyg}. Cross-correlation of DESI Legacy Imaging Survey LRGs with CMB lensing maps yields \(S_8 = 0.765 \pm 0.023\) from Planck and \(S_8 = 0.790^{+0.024}_{-0.027}\) from ACT~DR6~\cite{Sailer:2024jrx}.
 Cosmic shear measures the weak lensing signal from LSS imprinted on galaxy shapes, tracing the projected matter distribution and constraining cosmological parameters.  Cosmic shear is most sensitive to \(S_8\), with current surveys finding values typically \(1\text{--}3\sigma\) lower than CMB predictions. Weak-lensing surveys such as KiDS-1000 reported \(S_8 = 0.759^{+0.024}_{-0.021}\)~\cite{KiDS:2020suj}, showing a \(\sim3\sigma\) tension with the Planck-\(\Lambda\)CDM prediction. A reanalysis using a more robust shape-measurement pipeline yields \(S_8 = 0.789^{+0.020}_{-0.024}\), consistent with the previous KiDS-1000 lensfit result of \(S_8 = 0.776^{+0.029+0.002}_{-0.027-0.003}\) \citep{Yoon:2025eem}.  The tension with Planck remains at the \(1.8\sigma\) level, suggesting it is not driven by shear calibration. This discrepancy appears reduced in the recent KiDS-Legacy release, which reports \(S_8 = 0.815^{+0.016}_{-0.021}\)~\cite{Wright:2025xka}.  Similarly, DES~Y3 finds \(S_8 = 0.780 \pm 0.015\) within the \(\Lambda\)CDM model~\cite{DES:2025llp}. Recent results from the DES Y6 analysis \cite{DES:2026zjp} report $S_8 = 0.789^{+0.012}_{-0.012}$ for $\Lambda$CDM from the combined large-scale structure probes, while the DES Y6 $3\times2$pt analysis in $w$CDM yields $S_8 = 0.782^{+0.021}_{-0.020}$. 
 The status of the \(S_8\) tension thus remains uncertain and is still a subject of active debate within the cosmology community\cite{Pantos:2026koc}.

These  tensions, within the $\Lambda$CDM model, are  often considered signatures of new physics. Various models have been proposed to address them, categorized as follows:
\begin{itemize}[nosep]
    \item{\textit{Early time models} ($z \gtrsim 1100$):} These introduce new physics before recombination, aiming to reduce the sound horizon and increase $H_0$. Examples include: Early Dark Energy (EDE)~\cite{Karwal:2016vyq,Poulin:2018cxd,Poulin:2018dzj,Agrawal:2019lmo,Kamionkowski:2022pkx,Odintsov:2023cli}, New EDE~\cite{Niedermann:2019olb,Cruz:2023lmn,Niedermann:2023ssr}, AdS--EDE~\cite{Ye:2020btb,Ye:2020oix,Ye:2021iwa}, and modified gravity models~\cite{Rossi:2019lgt,Braglia:2020iik,Adi:2020qqf,Braglia:2020auw,Ballardini:2020iws,FrancoAbellan:2023gec,Petronikolou:2023cwu}. These models often struggle to simultaneously resolve both $H_0$ and $S_8$ tensions.
    \item{\textit{Intermediate/Late time models} ($0.1 \lesssim z \lesssim 3.0$):} These modify cosmic evolution at intermediate to late times, adjusting the expansion rate history $H(z)$. Examples include: $\Lambda_{\rm s}$CDM~\cite{Akarsu:2019hmw,Akarsu:2021fol,Akarsu:2022typ,Akarsu:2023mfb,Yadav:2024duq,Paraskevas:2024ytz, universe11010002, Akarsu:2024nas, Akarsu:2024qsi, Akarsu:2024eoo,Bousis:2024rnb,Akarsu:2025dmj,Akarsu:2025gwi,Akarsu:2025ijk,Hogas:2025ahb}, Phantom Crossing Dark Energy~\cite{DiValentino:2020naf,Alestas:2020mvb,Alestas:2020zol,Gangopadhyay:2022bsh,Basilakos:2023kvk,Adil:2023exv,Specogna:2025guo,Gangopadhyay:2023nli}. These models often show promise in addressing  $H_0$, even though they appear to have a problem in simultaneously fitting BAO and SNe Ia data~\cite{Bousis:2024rnb}.
    \item{\textit{Ultra late time models} ($z \lesssim 0.01$):} These models propose changes to fundamental or stellar physics in the recent universe, aiming to alleviate current tensions by revising our understanding of local astrophysics~\cite{Marra:2021fvf,Alestas:2020zol,Alestas:2021nmi,Alestas:2021luu,Perivolaropoulos:2021bds,Ruchika:2024ymt,Efstratiou:2025xou,Perivolaropoulos:2025gzo}.
\end{itemize}

Beyond the above  classification\footnote{Additional mechanisms, such as gravitational vacuum polarization and particle production, have also been proposed, which do not fit neatly into this classification and may provide a viable solution to the Hubble tension in both $\Lambda$CDM and phantom-crossing models~\cite{Erdem:2024vsr,Erdem:2025xtr}.}, recent work has reignited the debate on whether either early- or late-time modifications alone can meaningfully resolve the Hubble tension.  \citet{Vagnozzi:2023nrq} argues that early-time new physics, despite its ability to shrink the sound horizon, cannot by itself provide a complete solution. His analysis highlights seven independent clues—ranging from age estimates of old astrophysical objects to consistency relations involving the early ISW effect and equality-scale measurements—which jointly point to the need for a more intricate interplay between early, intermediate/high-redshift, and even local new physics. Meanwhile, \citet{Pedrotti:2025ccw} test whether systematic biases in BAO reconstruction could permit post-recombination solutions. By testing a range of dark energy scenarios under deliberately exaggerated BAO miscalibrations, and by leveraging the constraining power of unanchored SNe~Ia, it is shown that late-time modifications alone remain unable to accommodate the required increase in $H_0$. Together, these analyses underscore a growing perspective in the literature: a successful resolution of the Hubble tension may require a synthesis of early- and late-time mechanisms rather than relying on either class of models in isolation.

The recent Data Release 2 (DR2) from the Dark Energy Spectroscopic Instrument (DESI) \citep{DESI:2025zgx,DESI:2025fii} provides percent-level baryon acoustic oscillation (BAO) measurements up to $z\simeq1.1$, delivering unprecedented precision for probing dark energy dynamics. These data, from a survey aiming to collect spectra of over 35 million galaxies and quasars \cite{DESI:2016fyo,DESI:2016igz,DESI:2018kpn,DESI:2022xcl}, hint at mild deviations from a cosmological constant and suggest the dark energy equation of state $w(z)$ may have evolved across the phantom divide $w = -1$.
This possibility of phantom crossing further challenges the standard $\Lambda$CDM paradigm. Observational evidence from DESI \cite{DESI:2024aqx,DESI:2024hhd,DESI:2025wyn,DESI:2025zgx,DESI:2025fii,Nesseris:2025lke,Wang:2025bkk,Cortes:2024lgw,RoyChoudhury:2024wri,RoyChoudhury:2025iis}, combined with other probes including SNe Ia, CMB, and large-scale structure, increasingly favors such evolving dark energy behavior \cite{Mortonson:2013zfa,Scolnic:2021amr,Escamilla:2021uoj,Escamilla:2023shf,Shajib:2025tpd,Specogna:2025guo,RoyChoudhury:2025dhe,Lu:2025gki,Li:2024qso,Li:2025owk,Cai:2025mas,Alestas:2025syk}. Furthermore, when these new data are analyzed within a standard DDE-GR framework, the resulting expansion history favors a lower $H_0$, exacerbating the tension with local measurements \cite{Pang:2025lvh,Mirpoorian:2025rfp}. 

Recent model-independent studies~\cite{Wang:2025xvi} present clear evidence for dynamical dark energy with an evolving Hubble parameter, reinforcing the case for modified-gravity solutions to the Hubble tension and the phantom-crossing anomaly. Moreover, \cite{Scherer:2025esj} report $>5\sigma$ evidence and a large Bayes factor for a new dynamical dark-energy parametrization. Independent analyses~\cite{Sabogal:2025jbo,Silva:2025twg} also confirm the phantom-crossing anomaly, even without primary CMB anisotropy data.

A common approach to modeling dark energy uses a homogeneous, minimally coupled scalar field $\phi$ with a potential $V(\phi)$, described by the Lagrangian
\begin{equation}
\mathcal{L} = \pm \frac{1}{2} \dot{\phi}^2 - V(\phi),
\label{eq:minimal_lagrangian}
\end{equation}
where the plus sign denotes quintessence, while the minus sign denotes a phantom field. In a flat FLRW background, the energy density and pressure of a scalar field are $\rho = \pm\frac{1}{2}\dot{\phi}^2 + V(\phi)$ and $p = \pm\frac{1}{2}\dot{\phi}^2 - V(\phi)$, where the upper ($+$) sign corresponds to standard quintessence and the lower ($-$) sign to a phantom field. This leads to an equation of state\begin{equation}
    w = \frac{p}{\rho} = \frac{\pm\frac{1}{2}\dot{\phi}^2 - V(\phi)}{\pm\frac{1}{2}\dot{\phi}^2 + V(\phi)}.
\end{equation}  
Assuming $\rho > 0$ and requiring $w < -1/3$ to drive late-time acceleration (which implies $p < 0$), the condition for quintessence becomes $\frac{\frac{1}{2}\dot{\phi}^2 - V(\phi)}{\frac{1}{2}\dot{\phi}^2 + V(\phi)} < -1/3$, yielding $\dot{\phi}^2 < V(\phi)$. Note that $w < -1/3$ is required by the Friedmann equations for acceleration, and that in a kinetic-dominated phase ($\dot{\phi}^2 \gg V(\phi)$), the quintessence field may have $w \simeq 1$. In the phantom case, imposing $\rho > 0$ requires $V(\phi) > \frac{1}{2}\dot{\phi}^2$. Substituting the lower ($-$) signs into the equation of state yields $w = -\frac{V(\phi) + \frac{1}{2}\dot{\phi}^2}{V(\phi) - \frac{1}{2}\dot{\phi}^2}$, which guarantees $w < -1$. Despite their simplicity, these models have a critical limitation: a single field of this type cannot continuously cross the phantom divide $w = -1$ \cite{Perivolaropoulos:2005yv,Kunz:2006wc,Chimento:2008ws,Cai:2009zp,Saitou:2012xw,Wolf:2024eph}.
This obstruction occurs because a smooth crossing requires the kinetic energy term to vanish and change sign, which is forbidden by the fixed sign in the Lagrangian. This has motivated more complex frameworks. However, even generalized k-essence models with a Lagrangian of the form $\mathcal{L} = \frac{1}{2} f(\phi) \dot{\phi}^2 - V(\phi),$
are generally unable to achieve a stable phantom crossing without introducing instabilities or multiple fields with fine-tuning \cite{Sen:2005ra, Sur:2008tc,Saitou:2012xw}.
This obstruction occurs because a smooth crossing requires the kinetic energy term to vanish and change sign, which is forbidden by the fixed sign in the Lagrangian. This has motivated more complex frameworks. However, even generalized k-essence models with a Lagrangian of the form $\mathcal{L} = \frac{1}{2} f(\phi) \dot{\phi}^2 - V(\phi),$
are generally unable to achieve a stable phantom crossing without introducing instabilities or multiple fields with fine-tuning \cite{Sen:2005ra, Sur:2008tc,Saitou:2012xw}.

 In other scenarios, the dark energy may consist of multiple coexisting components, including at least one non-canonical \emph{phantom} fluid \cite{Nesseris:2006er, Hu:2004kh,Chaves:2008gd}. The most prominent paradigm for this mechanism is the ``Quintom'' model \cite{Guo:2004fq, Feng:2004ad, Cai:2025mas}, which serves as the simplest two-field realization of this scenario. It is established by a well-known no-go theorem that a single minimally coupled scalar field cannot cross the $w = -1$ boundary without suffering from fatal physical pathologies, such as a diverging sound speed or severe classical instabilities. To achieve a smooth and stable transition between the quintessence ($w > -1$) and phantom ($w < -1$) regimes, the Quintom mechanism employs a hybrid dark sector comprising two distinct degrees of freedom: a canonical scalar field equipped with a positive kinetic energy term, and a phantom scalar field driven by a negative kinetic energy term.

Alternatively, an apparent phantom crossing naturally emerges if dark energy interacts directly with dark matter \cite{Huey:2004qv, Das:2005yj}. This broad class of scenarios also encompasses dynamical vacuum models, where a time-varying dark energy component exchanges energy with the matter sector \cite{Sola:2005et,Sola:2016jky,Sola:2017znb,SolaPeracaula:2021yid}. In such interacting scenarios, the energy exchange inherently alters the standard redshift evolution of the matter density. Specifically, energy transfer from the dark energy component causes the dark matter density to dilute more slowly in the past than the standard $(1+z)^3$ scaling dictates. Consequently, an observer fitting cosmological data under the standard assumption of a separately conserved dark matter fluid will perceive a relative deficit in the past matter density. To make the expansion history fit the observations, this missing energy is implicitly misattributed to the dark energy component, causing its effective energy density to artificially appear as if it is growing over time. This mathematical compensation forces the inferred effective equation of state into the phantom regime ($w < -1$), even though the true underlying dark energy field remains canonical, dynamically stable, and strictly avoids Big Rip singularities \cite{Huey:2004qv, Das:2005yj}. Furthermore, interacting dark sector frameworks \cite{Khoury:2025txd,Zhang:2025dwu} have demonstrated that this mechanism can resolve the DESI-preferred phantom crossing as well as the $H_0$ and $S_8$ tensions.

If the dark–energy equation of state parameter $w(z)$ crosses the phantom divide line $w=-1$ there are three  main classes of cosmological scenarios:
\begin{enumerate}
  \item The dark energy consists of multiple components, with at least one non-canonical \emph{phantom} component.
  \item An interaction between dark energy and dark matter produces an effective phantom crossing while the underlying scalar field remains stable.
  \item General relativity must be extended to a more general theory on cosmological scales.
\end{enumerate}
One of the most promising theoretical frameworks to describe dynamical dark energy phenomena is scalar-tensor theories of gravity \cite{Brans:1961sx,Singh:1987is,Faraoni:2004pi,Quiros:2019ktw}, as well as by Horndeski theories and their generalizations~\cite{Horndeski1974,Langlois:2018dxi}. These models extend general relativity by introducing a scalar degree of freedom that couples non-minimally to the metric or to the Ricci scalar via a coupling function. Such theories include Brans-Dicke theory and its generalizations \cite{Brans:1961sx,Damour:1992we,Barrow:1990nv}, $f(R)$ gravity in its scalar-tensor representation \cite{Sotiriou:2006hs,Capozziello:2011wg,Goncalves:2021vci}. A key advantage of scalar-tensor models is their inherent flexibility in constructing $w(z)$ behaviors that allow for transitions across the phantom divide without invoking ghosts or instabilities, at least in certain subclasses \cite{Nojiri:2003ft,Perivolaropoulos:2005yv,Gannouji:2006jm,Wolf:2024stt,Wolf:2025jed,Wolf:2025jlc}. A general model for dark energy crossing the phantom divide, which independently controls the cosmological background and perturbations and avoids ghost and gradient instabilities in linear perturbations, was recently presented in \cite{Yao:2025wlx}. Specifically,   scalar-tensor  may violate a constraint which minimally coupled quintessence must satisfy. Defining the dimensionless Hubble parameter as $q(z) \equiv H(z)^2/H_0^2$, it has been shown that for any minimally coupled model, the following inequality must hold \cite{Papantonopoulos:2007zz}:
\begin{equation}
\frac{d q(z)}{dz} \geq 3 \Omega_{0m} (1+z)^2.
\label{eq:ineq_minimal}
\end{equation}
This can be equivalently expressed in terms of the dark energy equation of state as:
\begin{equation}
w(z) = \frac{2(1+z)\frac{d\ln H}{dz} - 1}{1 - \Omega_{0m}(1+z)^3 \left( \frac{H_0}{H(z)} \right)^2 } \geq -1.
\label{eq:w_constraint}
\end{equation}
Violation of the inequality above implies a crossing of the phantom divide, and therefore cannot be realized within any minimally coupled quintessence model\cite{Park:2025fbl,Gialamas:2025pwv}. In contrast, scalar-tensor theories allow for such behavior and remain consistent with both theoretical and observational constraints\cite{Perivolaropoulos:2005yv,SanchezLopez:2025uzw}. Recently, Wolf et al.~[162] demonstrated that a non-minimally coupled scalar field induces time variations in the gravitational constant on cosmological scales and fifth forces on small scales, which are tightly constrained by Solar System tests and thus require a screening mechanism. This tension motivates the study of modified-gravity scenarios that evade local constraints by construction.

The Lagrangian density considered in this work is given by:
\begin{equation}\label{eq:lagrangian}
\mathcal{L}
= 
\frac{F(\Phi)}{2}\,R
-\frac{Z(\Phi)}{2}\,g^{\mu\nu}\,\partial_{\mu}\Phi\,\partial_{\nu}\Phi
- U(\Phi)
+ \mathcal{L}_{m}[\psi_{m};\,g_{\mu\nu}],
\end{equation}
where we have set $8\pi G_N = 1$ (equivalent to $F_0 = 1$). 
We work in the Jordan frame, where physical observables correspond directly to measurements, using the Lagrangian density of Eq.\eqref{eq:lagrangian}  and matter fields independent of \(\Phi\) to preserve the weak equivalence principle\footnote{The weak equivalence principle  establishes the existence of a Jordan-frame metric with universal matter coupling, guaranteeing that test bodies follow geodesics independent of their composition. The strong equivalence principle  extends this universality to include gravitational binding energy, requiring compact objects like black holes to also follow geodesics. This distinction provides a classification criterion: theories satisfying the SEP are categorized as dark energy, while those violating it constitute modified gravity \citep{Joyce:2016vqv}.}. We adopt  \( Z(\Phi) = 1 \) for the kinetic term and  \( F(\Phi) > 0 \) for the non-minimal coupling.
 Although a transformation to the Einstein frame diagonalizes the kinetic terms, it introduces an explicit scalar-matter coupling, making the Jordan frame preferable for direct experimental comparison. The scalar potential \(U(\Phi)\) provides the effective dark energy density, with the non-minimal coupling permitting a broader class of viable potentials than quintessence models \cite{Boisseau:2000pr, Esposito-Farese:2000pbo, Perivolaropoulos:2005yv, Poisson:1995ef, Scharre:2001hn}.

The availability of DESI DR2 data provides a robust foundation to test this hypothesis. This paper has three primary objectives. \textbf{First}, we aim to demonstrate that a Modified Gravity (MG) framework, specifically ST theory, can resolve the \textit{worsened} $H_0$ tension by yielding a higher best-fit value for $H_0$, consistent with local measurements. \textbf{Second}, we intend to show that this \textit{same} framework naturally and viably accommodates the phantom-crossing DDE behavior ($w < -1$) hinted at by the DESI data, a feat forbidden in standard GR/quintessence. \textbf{Finally}, moving beyond simple parametrizations, our third goal is to use the combined power of recent high-precision data—including DESI DR2, Pantheon+, CMB, and growth data—to perform a systematic, data-driven \textbf{reconstruction} of the underlying ST Lagrangian \cite{Boisseau:2000pr,Esposito-Farese:2000pbo,Perivolaropoulos:2005yv,Nesseris:2017vor,Capozziello:2007iu,Kazantzidis:2019nuh}, deriving the fundamental coupling $F(\Phi)$ and potential $U(\Phi)$ functions that this new cosmological picture requires.

\vspace{0.3cm}

\noindent
The paper is organized as follows. Section~\ref{sec:datanalysisprocedure} describes the data analysis methodology employed to determine the best-fit parameters and discusses the corresponding results. In particular, we investigate the effects of allowing the ratio $G_{\rm eff}/G_{N}$ to vary freely and reconstruct its evolution using the model-independent parametrization of Eq.(\ref{pantazisparam}), examining how this variation affects the reconstructed Hubble evolution under different $w(z)$ parametrizations. In Section~\ref{sec:tensions}, we analyze the physical implications of these parametrizations for the Hubble and $S_8$ tensions. Section~\ref{sec:theory} introduces the theoretical framework of scalar–tensor theories and outlines a systematic reconstruction procedure—given $\mu_G(z)$ and $H(z)$—to derive $F(z)$, $U(z)$, $\Phi(z)$, and their functional relations $F(\Phi)$ and $U(\Phi)$, followed by a discussion of their viability. Finally, Section~\ref{sec:conclusion} summarizes our main findings and suggests directions for future research.

\section{Reconstructing Cosmic Evolution: Dark Energy and the Effective Gravitational Constant from Observational Data}\label{sec:datanalysisprocedure}

\subsection{Data analysis}\label{data_analysis}
 We present the framework for constraining cosmological parameters using multiple observational probes. The methodology combines up-to-date measurements from baryon acoustic oscillations (BAO), cosmic microwave background (CMB), Type Ia supernovae (SNe Ia), and redshift-space distortions (RSD).

  We construct the function \( H(z) \) (and further $q(z)$) by adopting a parametric form for the dark energy equation of state \( w_{\mathrm{DE}}(z) \), selecting from the following representations in each case:
\begin{itemize}

\item \textbf{CPL (Chevallier–Polarski–Linder)}\cite{Chevallier:2000qy,Linder:2002et}:
\begin{equation}\label{cpl}
w(z) = w_0 + w_a \frac{z}{1+z}
\end{equation}
The CPL parametrization is one of the most widely used due to its simplicity and physical interpretability. It captures the late-time evolution of dark energy by linearly interpolating between $w_0$ (today's value) and $w_0 + w_a$ (early-time value) as $z \to \infty$. It avoids divergence at high redshift, making it suitable for a broad range of cosmological analyses including CMB, BAO, and supernova data.

\item \textbf{BA (Barboza–Alcaniz)}\cite{Barboza:2008rh,Barboza:2011gd}:
\begin{equation}\label{ba}
w(z) = w_0 + w_a \frac{z(1+z)}{1+z^2}
\end{equation}
The BA parametrization is designed to remain well-behaved at both low and high redshifts. It provides a symmetric evolution of $w(z)$ around $z = 1$, making it useful for probing transitions in the dark energy equation-of-state near the onset of cosmic acceleration. Compared to CPL, BA allows for richer late-time dynamics and can accommodate observational hints of phantom crossing more naturally.

\item \textbf{Logarithmic} \cite{Feng:2011zzo}:
\begin{equation}\label{log}
w(z) = w_0 + w_a \ln(1 + z)
\end{equation}
The logarithmic parametrization introduces a slow, monotonic evolution of the equation-of-state parameter with redshift. It is particularly useful for probing deviations from $\Lambda$CDM at low to intermediate redshifts while avoiding divergence issues at high $z$. This form allows for enhanced sensitivity to late-time dynamics, often leading to better fits in models that explore the smooth onset of dark energy dominance.

\item \textbf{JBP (Jassal–Bagla–Padmanabhan)}\cite{Jassal:2004ej}:
\begin{equation}\label{jbp}
w(z) = w_0 + w_a \frac{z}{(1+z)^2}
\end{equation}
The JBP parametrization features a maximum deviation from $w_0$ around $z \approx 1$ and asymptotes back to $w_0$ both at $z = 0$ and as $z \to \infty$. This makes it highly suitable for isolating the redshift range around the transition to acceleration, offering sensitivity to possible intermediate-time deviations in the expansion history.
\end{itemize}

Let $\theta$ denote the vector of unknown cosmological parameters (e.g., $\theta = \{w_0, w_a, \Omega_{m0}, \dots\}$) that our model aims to constrain. We define a vector of theoretical predictions, $\textbf{x}_{\text{th}}(\theta)$, which depends on this parameter vector, and a corresponding vector of observed data points, $\textbf{x}_{\text{obs}}$. The goodness-of-fit is determined by minimizing the $\chi^2$ statistic, which is constructed from the difference vector (or residual vector), $\textbf{x}(\theta) \equiv \textbf{x}_{\text{th}}(\theta) - \textbf{x}_{\text{obs}}$.

The $\chi^2$ distribution is defined as\footnote{We refer readers to Refs.~\cite{Verde:2009tu,Theodoropoulos:2021hkk} for detailed discussion.}:
\begin{equation}
\begin{aligned}
 \chi^{2}(\theta) &= \textbf{x}(\theta)^{T} \mathbf{C}^{-1} \textbf{x}(\theta) \\
 &= \sum_{i,j} \left[x_{\text{th},i}(\theta) - x_{\text{obs},i}\right] (\mathbf{C}^{-1})_{ij} \left[x_{\text{th},j}(\theta) - x_{\text{obs},j}\right]\,,
\end{aligned}
\end{equation}
where $\mathbf{C}$ is the data covariance matrix, which accounts for the statistical and systematic uncertainties and their correlations. Its components are $C_{ij} \equiv C(x_i, x_j)$, and $(\mathbf{C}^{-1})_{ij}$ represents the components of the inverse covariance matrix.

\subsection*{Growth rate data}
In observational cosmology, the parameter $f\sigma_8$ serves as an important observable that encapsulates information about the growth rate of cosmic structures\footnote{ Cosmic structures are directly driven by peculiar velocity fields and the gravitational clustering of matter. For a comprehensive review of large-scale peculiar velocities and their role in structure formation, see~\cite{Tsagas:2025pxi}.}. The combination $f\sigma_8$ is defined as\footnote{Galaxy redshift surveys constrain the combinations $b_{1}\,\sigma_{8}(\bar z)$ and $f\,\sigma_{8}(\bar z)$ at the sample’s effective redshift $\bar z$, where $b_{1}$ is the linear galaxy bias parameter that relates galaxy clustering to the underlying matter distribution. In practice, $b_{1}\,\sigma_{8}$ is treated as a nuisance parameter, while $f\,\sigma_{8}(z)$ provides a bias-independent measure of the linear growth rate of matter perturbations~\citep{Dodelson:2020bqr}.}
\begin{equation}
  f\sigma_8(a) \equiv \,\frac{a\,\sigma_8}{\delta_m(a=1)}\frac{d\delta_m(a)}{da}\,,
\end{equation}
where $a$ is the scale factor and $\delta_m(a)$ is the linear matter overdensity. Here, $\sigma_8$ represents the root-mean-square fluctuation of the linear matter density field within a comoving sphere of radius $R$, defined as $\sigma_R^2 = \int \frac{d^3k}{(2\pi)^3}\, P_L(k)\, |W_R(k)|^2$, and evaluated on a scale of $8\,h^{-1}\,\mathrm{Mpc}$. The quantity $P_L(k)$ denotes the linear matter power spectrum, while $W_R(k)$ is the Fourier transform of the real-space tophat window function, given by $W_R(k) = \frac{3}{(kR)^3}\left[\sin(kR) - kR\cos(kR)\right]$, which in configuration space corresponds to $W_R(x) = \frac{3}{4\pi R^3}$ for $x < R$ and $0$ otherwise (see, e.g., \citep{Dodelson:2020bqr}).

We constrain the evolution of $f\sigma_8(a)$ using measurements from redshift-space distortion (RSD) data compiled in Ref.\cite{Euclid:2025bxg}, summarized in Table~\ref{growth_rate_RSD}.
 Note that the equation for the matter density contrast $\delta_m$ can be written as
\begin{equation}\label{delta}
  \ddot{\delta}_m + 2 H\,\dot{\delta}_m - 4\pi\,G_{\rm eff}\,\rho_m\,\delta_m \;=\; 0\,,
\end{equation}
where Eq.(\ref{delta}) implies 
\begin{widetext}
   \begin{equation}\label{density}
\frac{\mathrm{d}^2\delta_m(a)}{\mathrm{d}a^2}
\;+\;\Biggl[\frac{3}{a} + \frac{1}{H(a)}\,\frac{\mathrm{d}H(a)}{\mathrm{d}a}\Biggr]\,
\frac{\mathrm{d}\delta_m(a)}{\mathrm{d}a}
\;-\;\frac{3}{2}\,\Omega_{0m}\,
\mu_{G}(a)\;\frac{H_{0}^2}{a^5 H(a)^2}\;\delta_m(a)
\;=\;0,
\end{equation} 
\end{widetext}
 We  have introduced the parameter $\mu_G(a)$, which quantifies the deviation from standard gravity, defined as the ratio
\begin{equation}\label{eq:geff_param}
  \mu_G(a) \equiv\frac{G_{\mathrm{eff}}(a)}{G_N}.
\end{equation}
\begin{figure}[htbp]
    \centering
    \includegraphics[width=1\linewidth]{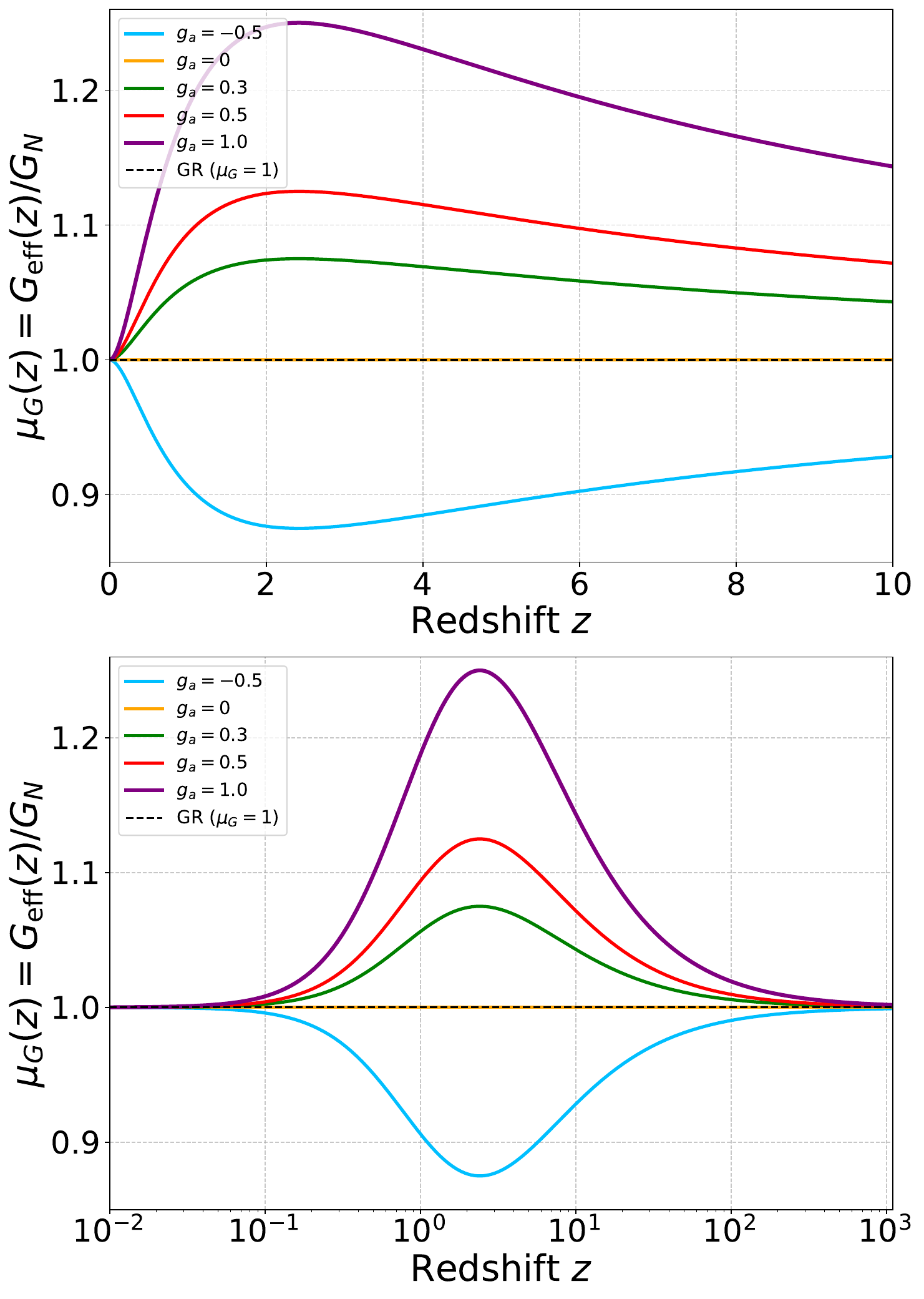}
  \caption{Evolution of the effective gravitational constant $\mu_G(z) = G_{\rm eff}(z)/G_N$ for different values of the parameter $g_a$ with $n = 2$ [Eq.\eqref{pantazisparam}]. The top panel illustrates the late-time and intermediate-redshift behavior on a linear scale up to $z = 10$, highlighting a transient modification to gravity that peaks at $z \sim 2$. The bottom panel extends the evolution up to $z = 1100$ on a logarithmic scale, clearly demonstrating that the modification vanishes at early times. This transient behavior ensures consistency with Big Bang Nucleosynthesis (BBN) at high redshifts, just as it vanishes today ($z=0$) to satisfy local gravity constraints. The horizontal dashed line indicates General Relativity ($\mu_G = 1$).}
    \label{muG_evolution}
\end{figure}

 Accordingly, we employ the measurements listed in Table \ref{growth_rate_RSD}  to reconstruct the  function \(\mu_G\). 
The reconstruction is performed using the phenomenological parametrization introduced in \cite{Nesseris:2017vor}:
\begin{align}
\mu_G &= \mu_{\rm G,0} + g_a (1 - a)^n - g_a (1 - a)^{2n} \nonumber\\
&= \mu_{\rm G,0} + g_a \left( \frac{z}{1 + z} \right)^n - g_a \left( \frac{z}{1 + z} \right)^{2n},\label{pantazisparam}
\end{align}
with $\mu_{\rm G}(z=0) \equiv \mu_{\rm G,0}$ and $n = 2$ throughout our analysis, as this parametrization is only viable for $n \ge 2$ due to Solar System tests requiring the first time derivative of $G_{\rm eff}$ to vanish. In this case, $\lim_{z \to \infty} \mu_G = \mu_{\rm G,0}$ (equivalently $a \to 0$), and the derivative is $d\mu_G/dz = 2 g_a z (1+z)^{-5} [1-(z-2)z]$, which satisfies $\lim_{z \to 0} d\mu_G/dz = \lim_{z \to \infty} d\mu_G/dz = 0$.

As shown in the top panel of Figure~\ref{muG_evolution}, the parameter $g_a$ sets the amplitude of deviations from $\mu_{\rm G,0}$, producing a transient modification that peaks at $z \approx 2$. For $\mu_{\rm G,0}=1$, positive (negative) values correspond to an enhancement (suppression) of gravity relative to $G_{\rm N}$ in the redshift range where large-scale structure surveys are most sensitive \cite{DESI:2025zgx,Alam:2016hwk,Blake:2011rj,delaTorre:2013rpa}. This parametrization satisfies key constraints without invoking screening mechanisms \cite{Khoury:2003aq,Mota:2006tg}: the effective gravitational coupling $G_{\rm eff}$ remains consistent with Big Bang Nucleosynthesis ($G_{\rm eff}/G_{\rm N} = 1.09 \pm 0.2$)—as explicitly demonstrated by the high-redshift asymptotic behavior in the bottom panel—and is normalized to match Newton's constant today ($G_{\rm eff}(a=1)/G_{\rm N} = 1$).

\begin{table}[tbp]  
    \centering
  \caption{Summary of the growth rate RSD data from various astronomical surveys which can be found in \cite{Euclid:2025bxg}.}
   \label{growth_rate_RSD}
    \begin{ruledtabular}
    \begin{tabular}{cllll}
    ID & $z_{\rm eff}$ & $f\sigma_8(z)$ & Survey & Reference \\
    \hline
    1 & 0.17 & $0.510 \pm 0.060$ & 2dFGRS & \cite{Song:2008qt} \\
    2 & 0.02 & $0.314 \pm 0.048$ & 2MRS & \cite{Davis:2010sw}\\
    3 & 0.02 & $0.398 \pm 0.065$ & PSCz & \cite{10.1111/j.1365-2966.2011.20050.x} \\
    4 & 0.44 & $0.413 \pm 0.080$ & WiggleZ &\cite{Blake:2012pj} \\
    5 & 0.60 & $0.390 \pm 0.063$ & WiggleZ & \cite{Blake:2012pj} \\
    6 & 0.73 & $0.437 \pm 0.072$ & WiggleZ & \cite{Blake:2012pj} \\
    7 & 0.18 & $0.36 \pm 0.09$ & GAMA & \cite{Blake:2013nif} \\
    8 & 0.38 & $0.44 \pm 0.06$ & GAMA & \cite{Blake:2013nif} \\
    9 & 1.4 & $0.482 \pm 0.116$ & FastSound & \cite{Okumura:2015lvp} \\
    10 & 0.02 & $0.428 \pm 0.048$ & 6dFGS+SN Ia & \cite{Huterer:2016uyq} \\
    11 & 0.6 & $0.55 \pm 0.12$ & VIPERS & \cite{Mohammad:2017lzz} \\
    12 & 0.86 & $0.40 \pm 0.11$ & VIPERS & \cite{Mohammad:2017lzz}\\
    13 & 0.03 & $0.404 \pm 0.082$ & 2MTF/6dFGSv & \cite{Qin:2019axr} \\
    14 & 0.013 & $0.46 \pm 0.06$ &  ALFALFA  & \cite{Avila:2021dqv}\\
    15 & 0.15 & $0.53 \pm 0.16$ & SDSS-IV eBOSS & \cite{eBOSS:2020yzd} \\
    16 & 0.38 & $0.500 \pm 0.047$ & SDSS-IV eBOSS& \cite{eBOSS:2020yzd}  \\
    17 & 0.51 & $0.455 \pm 0.039$ & SDSS-IV eBOSS & \cite{eBOSS:2020yzd}  \\
    18 & 0.70 & $0.448 \pm 0.043$ & SDSS-IV eBOSS & \cite{eBOSS:2020yzd}  \\
    19 & 0.85 & $0.315 \pm 0.095$ & SDSS-IV eBOSS & \cite{eBOSS:2020yzd}  \\
    20 & 1.48 & $0.462 \pm 0.045$ & SDSS-IV eBOSS & \cite{eBOSS:2020yzd}  \\
    \end{tabular}
    \end{ruledtabular}
\end{table}

The total $\chi^2$ for the growth data is computed as: \begin{equation} \label{chi2_total} \chi^2_{\rm Growth Data}=\chi^2_{\rm WiggleZ}+\chi^2_{\rm diag}\,, \end{equation} where $\chi^2_{\rm diag}$ is the sum over all uncorrelated data points (all points in Table \ref{growth_rate_RSD} except ID 4, 5, and 6): \begin{equation} \chi^{2}_{\rm diag} = \sum_{i \in \rm diag} \frac{\left[f\sigma_8(z_i, \theta) - (f\sigma_8)^{\rm obs}_{i}\right]^2}{\sigma_i^2}\,, \end{equation} and $\chi^2_{\rm WiggleZ}$ is calculated for the correlated WiggleZ data points (ID 4, 5, 6) using their published covariance matrix \cite{2012MNRAS.425..405B, Kazantzidis:2018rnb}: \begin{equation}
\begin{split}
    \chi^2_{\rm WiggleZ} = \sum_{i,j \in \rm Wig.} & \left[f\sigma_8(z_i, \theta) - (f\sigma_8)^{\rm obs}_{i}\right] (C_{\rm WiggleZ}^{-1})_{ij} \\
    & \times \left[f\sigma_8(z_j, \theta) - (f\sigma_8)^{\rm obs}_{j}\right]\,.
\end{split}
\end{equation} The parameter vector is $\theta = \{\Omega_{\rm m0}, w_0, w_a, g_a, \sigma_8\}$, and the $N \times N$ WiggleZ covariance matrix $\mathbf{C}_{\rm WiggleZ}$ (for $N=3$) is: \begin{equation} \label{wigglecov} \left[C^{\rm WiggleZ}_{ij}\right] = \begin{pmatrix} 0.00640 & 0.002570 & 0.000000 \\ 0.00257 & 0.003969 & 0.002540 \\ 0.00000 & 0.002540 & 0.005184 \end{pmatrix}\,. \end{equation} The parameters that minimize the total $\chi^2_{\rm Growth Data}$ are the most probable values, referred to as the best fit parameters.

\subsection*{Type Ia supernovae}

Type Ia supernovae, characterized by the absence of a spectral line of hydrogen and the presence of an absorption line attributed to singly ionized silicon, result from the explosion of a white dwarf in a binary system that surpasses the Chandrasekhar limit due to gas accretion from a companion star.

Importantly, type Ia supernovae exhibit a nearly constant absolute luminosity at the peak of their brightness, denoted by an established absolute magnitude of approximately $M \approx -19$. As a result, the distance to a type Ia supernova can be deduced through the observation of its apparent luminosity. By concurrently measuring the apparent magnitude and the light curve, it becomes feasible to predict the corresponding absolute \mbox{magnitude \cite{Amendola:2015ksp}}.

Brighter supernovae exhibit broader light curves (flux or luminosity of the supernova as a function of time). It is important to note that when referring to the universal absolute magnitude of type Ia supernovae hereafter, it is implied that the magnitude has been appropriately adjusted to account for the light curve width.

When considering the luminosity distance $d_{L}$ measured in megaparsecs (Mpc), the concepts of absolute magnitude, apparent magnitude and luminosity distance can be formally related as follows:

\begin{equation}\label{Magn}
\mu \equiv m - M = 5 \log_{10}\left(\frac{d_{L}}{\text{Mpc}}\right) + 25.
\end{equation}

Here, $\mu$ represents the distance modulus, which quantifies the difference between the apparent magnitude $m$ and the absolute magnitude $M$ of an object. The luminosity distance $d_L$ can be expressed as:
\begin{equation}\label{luminosity}
d_L(z) = (1 + z) \int_{0}^{z} \frac{c}{H(z')}  dz'
\end{equation}
where $c$ is the speed of light and $H(z)$ is the Hubble parameter as a function of the redshift $z$.

The Pantheon+ dataset comprises a collection of $1550$ type Ia supernovae and $1701$ corresponding light curves, spanning a redshift range of $0.001<z<2.26$ \cite{Brout:2022vxf}. To analyze the Pantheon+ data, we adopt the methodology outlined in the study of \mbox{Brout et al. (2022) \cite{Brout:2022vxf}.}

  Denoting the $1701 \times 1701$ covariance matrix, including both statistical and systematic uncertainties, as $[\mathbf{C_{stat+syst}}]$, the standard $\chi^2$ is given by
\begin{equation}
    \chi^{2} = \mathbf{Q}^{T} [\mathbf{C_{stat+syst}}]^{-1} \mathbf{Q},
\end{equation}
where the $1701$-component vector $\mathbf{Q}$ is defined as
\begin{equation}
    Q_{i} = m_{B_{i}} - M - \mu_{\rm model}(z_{i}; \Omega_{\rm m0}, w_0, w_a, \Omega_b h^2, h),
\end{equation}
with the model distance modulus
\begin{equation}
    \mu_{\rm model}(z_{i};\Omega_{\rm m0}, w_0, w_a, \Omega_b h^2, h) = 5 \log_{10} \left( \frac{d_{L}(z_i)}{\rm Mpc} \right) + 25.
\end{equation}
Here $d_L$ is the luminosity distance given in Eq.(\ref{luminosity}).  

Due to the degeneracy between $H_0$ and the absolute magnitude $M$ of SNe Ia, it is not possible to estimate $H_0$ from the Pantheon+ data alone \cite{Pan-STARRS1:2017jku}. This degeneracy is broken by incorporating the distance moduli of SNe Ia in Cepheid hosts, $\mu^{\rm Ceph}$, which constrain $M$ independently \cite{Riess:2021jrx}. The modified vector $\mathbf{Q}'$ is then
\begin{equation} \label{pantheonvec}
 Q_{i}' =
  \begin{cases}
    m_{B_{i}} - M - \mu^{\rm Ceph}(z_i), & i \in \text{Cepheids} \\
    m_{B_{i}} - M - \mu_{\rm model}(z_i; \Omega_{\rm m0}, h, \alpha, z_s), & \text{otherwise}
  \end{cases}
\end{equation}
where $\mu^{\rm Ceph}(z_i)$ is the corrected distance modulus of the Cepheid host of the $i^{\rm th}$ SNe Ia \cite{Riess:2021jrx}. Finally, the Pantheon+ $\chi^2$ is
\begin{equation}
    \chi^2_{\rm Pant} = \mathbf{Q}'^{T} [\mathbf{C_{stat+syst}}]^{-1} \mathbf{Q}'.
\end{equation}

\subsection*{BAO}

The baryon acoustic oscillation (BAO) measurements, which detect the presence of a characteristic scale in the matter distribution, offer a standard ruler that is valuable for  estimating cosmological parameters. 
In the early universe prior to recombination, initial perturbations evolved into overdensities through gravitational interactions with dark matter. Baryonic matter was embedded within these dark matter overdensities and the collapse of these overdensities was followed by radiation-induced overpressure. This overpressure, in turn, generated an expanding sound wave that propagated through plasma at a velocity of \citep{Eisenstein:1997ik}

\begin{equation}
c_{s}=\frac{c}{\sqrt{3(1+R_{s})}}.
\end{equation}

Here, $R_{s}\equiv\frac{3 \rho_{b}}{4 \rho_{\gamma}}$, where $\rho_{b}$ represents the baryon density and $\rho_{\gamma}$ represents the photon density. The fluid undergoes damped oscillations in both space and time, wherein the oscillation period depends on the sound speed.  The sound speed $c_{s}$  depends on the density of baryonic matter. When the density of baryons is considerably lower than that of radiation, the sound speed assumes the typical value for a relativistic fluid, i.e., $c_s = c/\sqrt{3}$. However, the introduction of baryonic matter increases the mass of the fluid, leading to a decrease in the sound speed. 

 The redshift $z_{\text{drag}}$ denotes the period of the drag epoch, i.e., the epoch when the baryons were released from the Compton drag of photons, which occurred slightly after recombination in the early universe. For the parameter ranges $\Omega_m h^2 \in [0.13, 0.15]$ and $\Omega_b h^2 \in [0.0214, 0.0234]$, the redshift at the drag epoch, $z_d$, is computed using the following fitting formula \citep{Aizpuru:2021vhd}:
\begin{equation}
z_{\rm drag} = \frac{1 + 428.169\,\omega_b^{0.256459}\,\omega_m^{0.616388} + 925.56\,\omega_m^{0.751615}}{\omega_m^{0.714129}},
\label{eq:zd_approximation}
\end{equation}
where $\omega_m \equiv \Omega_m h^2$ and $\omega_b \equiv \Omega_b h^2$.  This approximation achieves an accuracy of approximately $0.001\%$. 
The photon-decoupling surface, $z=z_*$, over the parameter ranges $\Omega_m h^2 \in [0.13, 0.15]$ and $\Omega_b h^2 \in [0.0214, 0.0234]$ has the fitting formula \citep{Aizpuru:2021vhd}:
\begin{equation}
z_* = \frac{ 391.672\,\omega_m^{-0.372296} + 937.422\,\omega_b^{-0.97966} }{ \omega_m^{-0.0192951}\,\omega_b^{-0.93681}} + \omega_m^{-0.731631} .
\label{eq:zstar_approximation}
\end{equation}
This approximation is accurate to within approximately $0.0005\%$. 

If we denote as $r_{s}(z)$ the sound horizon, i.e., the comoving distance traveled by a sound wave from the Big Bang until a corresponding redshift $z$, then \cite{Amendola:2015ksp}
\begin{equation}\label{sound}
r_{s}(z)=\frac{1}{H_{0}}\int_{z}^{\infty} dz' \,\frac{c_{s}(z')}{H(z')/H_0}.\end{equation} 
The sound horizon at the drag epoch,
 within the $\Lambda$CDM framework, is well-approximated by the relation \citep{Aizpuru:2021vhd}

\begin{equation}
r_s(z_d) = \frac{1}{a_1 \omega_b^{a_2} \omega_m^{a_3} \left[(\alpha/\alpha_0)^{a_4} + \omega_b^{a_5} \omega_m^{a_6} \right] + a_7 \omega_m^{a_8}} \ \text{Mpc},
\end{equation}

where the coefficients take the following values
\begin{align*}
a_1 &= 0.00730258, \quad a_2 = 0.088182, \quad a_3 = 0.099958, \\
a_4 &= 1.97913, \quad a_5 = 0.346626, \quad a_6 = 0.0092295, \\
a_7 &= 0.0074056, \quad a_8 = 0.8659935,
\end{align*}
which is accurate to within $0.0077\%$.

\begin{table}[h!]
\centering
\caption{DESI BAO measurements at different redshifts (see Table IV - \citep{DESI:2025zgx})}
\begin{tabular}{cccc}
\toprule
Tracer & Redshift ($z$) & Observable & Value \\
\midrule
BGS & 0.295 & $d^{\rm obs}_1\equiv d_V/r_d$ & 7.942 \\
LRG1 & 0.510 & $d^{\rm obs}_2\equiv d_V/r_d$ & 12.720 \\
LRG1 & 0.510 & $d^{\rm obs}_3\equiv d_M/d_H$ & 0.622 \\
LRG2 & 0.706 & $d^{\rm obs}_4\equiv d_V/r_d$ & 16.048 \\
LRG2 & 0.706 & $d^{\rm obs}_5\equiv d_M/d_H$ & 0.892 \\
LRG3+ELG1 & 0.934 & $d^{\rm obs}_6\equiv d_V/r_d$ & 19.721 \\
LRG3+ELG1 & 0.934 & $d^{\rm obs}_7\equiv d_M/d_H$ & 1.223 \\
ELG2 & 1.321 & $d^{\rm obs}_8\equiv d_V/r_d$ & 24.256 \\
ELG2 & 1.321 & $d^{\rm obs}_9\equiv d_M/d_H$ & 1.948 \\
QSO & 1.484 & $d^{\rm obs}_{10}\equiv d_V/r_d$ & 26.059 \\
QSO & 1.484 & $d^{\rm obs}_{11}\equiv d_M/d_H$ & 2.386 \\
Lya & 2.330 & $d^{\rm obs}_{12}\equiv d_V/r_d$ & 31.267 \\
Lya & 2.330 & $d^{\rm obs}_{13}\equiv d_M/d_H$ & 4.518 \\
\bottomrule
\end{tabular}\label{Tabledesibaodat}
\end{table}

The Hubble distance, denoted by $d_H$, is a characteristic length scale of the universe and is defined as
\begin{equation}
d_{H}(z)=cH^{-1}(z).
\end{equation}
We also define the angular diameter distance, $d_A$, and the proper motion distance, $d_M$, which are related by
\begin{equation}\label{dm}
d_{M}(z)=(1+z)d_{A}(z)=\frac{d_{L}(z)}{1+z}.
\end{equation}
The related effective distance, $d_{V}(z)$, is given by the equation:
\begin{equation}
d_{V}(z)=\bigg{[}c z\frac{ d^{2}_{M}(z)}{H(z)}\bigg{]}^{\frac{1}{3}}.
\end{equation}

We construct the $\chi^2$ distribution for the DESI BAO dataset which can be seen in Table \ref{Tabledesibaodat} as
\begin{equation}
      \chi^{2}_{\text{DESI BAO}}\left(\Omega_bh^2,\Omega_{\rm m0},w_0,w_a,h\right)=\textbf{v}^{T}[\mathbf{C_{ \text{DESI BAO}}}]^{-1}\textbf{v},
      \end{equation}
where we define $\mathbf{v} \equiv \mathbf{d}^{\rm model} - \mathbf{d}^{\rm obs}$. The covariance matrix takes a block-diagonal form, with a single scalar block and six $2\times2$ submatrices capturing correlations among observables:
\[
\scriptsize
[\mathbf{C}_{\text{DESI BAO}}] =
\begin{pmatrix}
A_1 & \mathbf{0}_{1\times 2} & \mathbf{0}_{1\times 2} & \mathbf{0}_{1\times 2} & \mathbf{0}_{1\times 2} & \mathbf{0}_{1\times 2} & \mathbf{0}_{1\times 2} \\[4pt]
\mathbf{0}_{2\times 1} & A_2 & \mathbf{0}_{2\times 2} & \mathbf{0}_{2\times 2} & \mathbf{0}_{2\times 2} & \mathbf{0}_{2\times 2} & \mathbf{0}_{2\times 2} \\[4pt]
\mathbf{0}_{2\times 1} & \mathbf{0}_{2\times 2} & A_3 & \mathbf{0}_{2\times 2} & \mathbf{0}_{2\times 2} & \mathbf{0}_{2\times 2} & \mathbf{0}_{2\times 2} \\[4pt]
\mathbf{0}_{2\times 1} & \mathbf{0}_{2\times 2} & \mathbf{0}_{2\times 2} & A_4 & \mathbf{0}_{2\times 2} & \mathbf{0}_{2\times 2} & \mathbf{0}_{2\times 2} \\[4pt]
\mathbf{0}_{2\times 1} & \mathbf{0}_{2\times 2} & \mathbf{0}_{2\times 2} & \mathbf{0}_{2\times 2} & A_5 & \mathbf{0}_{2\times 2} & \mathbf{0}_{2\times 2} \\[4pt]
\mathbf{0}_{2\times 1} & \mathbf{0}_{2\times 2} & \mathbf{0}_{2\times 2} & \mathbf{0}_{2\times 2} & \mathbf{0}_{2\times 2} & A_6 & \mathbf{0}_{2\times 2} \\[4pt]
\mathbf{0}_{2\times 1} & \mathbf{0}_{2\times 2} & \mathbf{0}_{2\times 2} & \mathbf{0}_{2\times 2} & \mathbf{0}_{2\times 2} & \mathbf{0}_{2\times 2} & A_7
\end{pmatrix}
\]
where $\mathbf{0}_{i\times j}$ denote the zero $i\times j$ matrix  and 
\begin{align*}
A_1 &= 0.005625, \\
A_2 &= \begin{pmatrix}0.009801& 0.00008415\\0.00008415 & 0.000289\end{pmatrix}, \\
A_3 &= \begin{pmatrix}0.0121 & -0.00004158\\-0.00004158 & 0.000441\end{pmatrix}, \\
A_4 &= \begin{pmatrix}0.008281 & 0.000096824\\0.000096824 & 0.000361\end{pmatrix}, \\
A_5 &= \begin{pmatrix}0.030276 & 0.00158166\\0.00158166 & 0.002025\end{pmatrix}, \\
A_6 &= \begin{pmatrix}0.158404 & 0.00238163\\0.00238163 & 0.018496\end{pmatrix}, \\
A_7 &= \begin{pmatrix}0.065536 & 0.0142536\\0.0142536 & 0.009409\end{pmatrix}.
\end{align*}

\subsection*{CMB}
The BAO leaves a characteristic imprint on the power spectrum of the cosmic microwave background (CMB) anisotropies that is observed as a series of peaks and troughs. The characteristic angle, $\theta_{A}$, that defines the location of the peaks can be calculated by the following equation \cite{Amendola:2015ksp}:
\begin{equation}\label{thetaA}
\theta_{A}=\frac{r_{s}(z_{*})}{d_{M}(z_{*})}.
\end{equation}
  The angular power spectrum of the CMB is decomposed into its multipole moments, where the low multipole moments correspond to the large angular scales and the high multipole moments correspond to the small angular scales. Each multipole $l$ that corresponds to the characteristic angle $\theta_{A}$ can be determined by the following equation \cite{Amendola:2015ksp}:

\begin{equation}\label{lACMBcomp}
l_{A}=\frac{\pi}{\theta_{A}}=\pi \frac{d_{M}(z_{*})}{r_{s}(z_{*})}.
\end{equation}
By adopting the luminosity distance Eq.(\ref{luminosity}) and  the proper motion distance, as defined in Eq.(\ref{dm}), then for a theoretical model:
\begin{equation}
l_{A}=\pi\frac{d_{M}\left[z_{*}(\Omega_bh^2,\Omega_{\rm m0},h);\Omega_{\rm m0},w_0,w_a,h\right]}{r_{s}\left[z_{*}(\Omega_bh^2,\Omega_{\rm m0},h);\Omega_{\rm m0},w_0,w_a,h\right]}.
\end{equation}

\begin{table*}[!ht]
    \centering
    \caption{\label{tab:fs8_constraint}We present observational constraints  (Pantheon+, CMB, DESI DR2, RSD, BBN constraint)  on the parameters of modified gravity models featuring a phenomenological $\mu_{\mathrm{G}}$ parametrization and a Hubble function described by the CPL, BA, JBP, and Logarithmic models. The minimum $\chi^2$ value ($\chi^2_{\mathrm{min}}$) for each model are listed in the bottom rows (note that $(\chi^2_{\rm min})^{\Lambda\rm CDM}=1588.15$).  }
    \begin{ruledtabular}
\begin{tabular}{ccccc}
    & CPL & BA & Log & JBP \\
    \hline
    $M_{\rm B0}$ & $-19.337 \pm 0.033$ & $-19.337 \pm 0.038$ & $-19.338 \pm 0.032$ & $-19.333 \pm 0.024$ \\
    $\Omega_{\rm m0}$ & $0.286 \pm 0.013$ & $0.286 \pm 0.014$ & $0.286 \pm 0.012$ & $0.284 \pm 0.008$ \\
    $\Omega_bh^2$ & $0.02245 \pm 0.00014$ & $0.02246 \pm 0.00014$ & $0.02245 \pm 0.00014$ & $0.02248 \pm 0.00013$ \\
    $w_0$ & $-1.021 \pm 0.134$ & $-1.033\pm 0.139$ & $-1.023 \pm 0.115$ & $-1.031 \pm 0.094$ \\
    $w_a$ & $-0.271 \pm 0.413$ & $-0.128 \pm 0.235$ & $-0.205 \pm 0.272$ & $-0.377 \pm 0.491$ \\
    $g_a$ & $0.318 \pm 0.181$ & $0.318\pm 0.209$ & $0.309 \pm 0.173$ & $0.349 \pm 0.123$ \\
    $\sigma_8$ & $0.783 \pm 0.026$ & $0.783 \pm 0.027$ & $0.783 \pm 0.026$ & $0.781 \pm 0.026$ \\
    $h$ & $0.706 \pm 0.014$ & $0.706 \pm 0.016$ & $0.706 \pm 0.014$ & $0.708 \pm 0.010$ \\
    \hline
    $\chi^2_{\rm min}$ & $1569.88$ & $1570.01$ & $1569.53$ & $1570.67$\\
    \hline
    $\Delta\chi^2_{\rm min}\equiv(\chi^2_{\rm min})^{\rm model}-(\chi^2_{\rm min})^{\rm \Lambda CDM}$ & $-18.27$ & $-18.14$ & $-18.62$ & $-17.48$
\end{tabular}
    \end{ruledtabular}
\end{table*}

\begin{table*}[!ht]
    \centering
    \caption{\label{tab:constraint}We present observational constraints  (Pantheon+, CMB, DESI DR2, RSD, BBN constraint)  on the parameters of the CPL model with $g_a = 0$ 
(note that in Appendix~\ref{AppA}, we also provide observational constraints for the dynamical dark energy case, 
where GR is assumed, and for the Hubble function described by the CPL, BA, JBP, and Logarithmic models). 
For comparison, results for the $\Lambda$CDM model are also shown. 
The minimum $\chi^2$ value ($\chi^2_{\mathrm{min}}$) for each model is listed in the bottom row. }
       \begin{ruledtabular}
   \begin{tabular}{ccccc}
    & CPL & $\Lambda$CDM  \\
    \hline
    $M_{\rm B}$ & $-19.379 \pm 0.014$ & $-19.399 \pm 0.009$  \\
    $\Omega_{\rm m0}$ & $0.298 \pm 0.006$ & $0.297 \pm 0.004$ \\
    $\Omega_bh^2$ & $0.02247 \pm 0.00014$ & $0.0226\pm 0.00012$  \\
    $w_0$ & $-0.892 \pm 0.060$ & -1  \\
    $w_a$ & $-0.578 \pm 0.26$ & 0\\
    $\sigma_8$ & $0.780 \pm 0.026$ & $0.802\pm 0.026$ \\
    $h$ & $0.691 \pm 0.006$ & $0.688\pm 0.003$  \\
    \hline
    $\chi^2_{\rm min}$ & 1578.44 & 1588.15  \\
\end{tabular}
    \end{ruledtabular}
\end{table*}

The shift parameter, denoted as $R$, is a dimensionless parameter that  encompasses information related to the comparison of predicted and observed positions of the acoustic peaks in the cosmic microwave background (CMB). Its definition is as follows \cite{Amendola:2015ksp}:
\begin{equation}
R\equiv\frac{\sqrt{\Omega_{\rm m0}H_{0}^{2}}}{c}d_{M}\left[z_{*}(\Omega_bh^2,\Omega_{\rm m0},h);\Omega_{\rm m0},w_0,w_a,h\right].
\end{equation}

 The compressed Planck distance priors provide a compact and numerically efficient summary of the full CMB likelihood, typically expressed in terms of the shift parameters $R$ and $l_A$ \citep{Zhai:2018vmm,Zhai:2019nad}. Within $\Lambda$CDM the CMB anisotropies can be characterized by three angular scales $\ell_X \equiv \pi d_A(z_*)/r_X(z_*)$\cite{Poulin:2018cxd}, where $r_X(z_*)$ is the relevant physical scale at recombination and $X=\mathrm{eq},s,D$ correspond to the projected Hubble distance at matter--radiation equality, the projected sound horizon, and the projected Silk damping scale\cite{Dodelson:2020bqr}. In our analysis $l_A$ (Eq.(\ref{lACMBcomp})) relates to the sound-horizon scale $\ell_s$ (see Eq.(\ref{dm})), while $\ell_{\mathrm{eq}}$ is implicitly determined by the fitted present-day matter density $\Omega_{\rm m0}$ and the assumed expansion history.
The damping scale $\ell_D$ is not constrained by the compressed CMB spectrum. This is justified since our model is assumed to be effectively identical to $\Lambda$CDM at recombination and to share the same pre-recombination physics, differing only through late-time modifications to the expansion history. In particular, by construction $\mu_G(z_*)\simeq1$ (see Eq.~(\ref{pantazisparam})) and $\Omega_m(z_*)+\Omega_r(z_*)\simeq1$, implying that both the background evolution and the photon--baryon perturbations at recombination remain essentially unchanged relative to $\Lambda$CDM. Consequently, the damping scale is expected to remain unaltered and therefore does not provide an additional independent constraint for our model. By contrast, early-time solutions to the $H_0$ tension (e.g.\ Early Dark Energy) modify the expansion near recombination and affect both the sound horizon and the damping scale, requiring damping-tail information from the full CMB power spectrum \citep{Poulin:2018cxd,Lin:2019qug}.

Using the compressed CMB data from the Planck satellite \citep{Zhai:2018vmm} for a flat universe, we consider three key observables: the CMB shift parameter ($R$), the acoustic scale ($l_A$), and the physical baryon density ($\Omega_b h^2$). 
The central values of these CMB observables are summarized in the following data vector:
\begin{equation}
\mathbf{v}_{\text{CMB}} = \begin{pmatrix}
    R_{\text{Planck}} \\
    l_{A,\text{Planck}} \\
    \Omega_b h^2_{\text{Planck}}
\end{pmatrix} = \begin{pmatrix}
    1.74963  \\
    301.80845 \\
    0.02237
\end{pmatrix}.
\end{equation}

\begin{figure}
\begin{subfigure}{\linewidth}
        \centering
    \includegraphics[width=.7\linewidth]{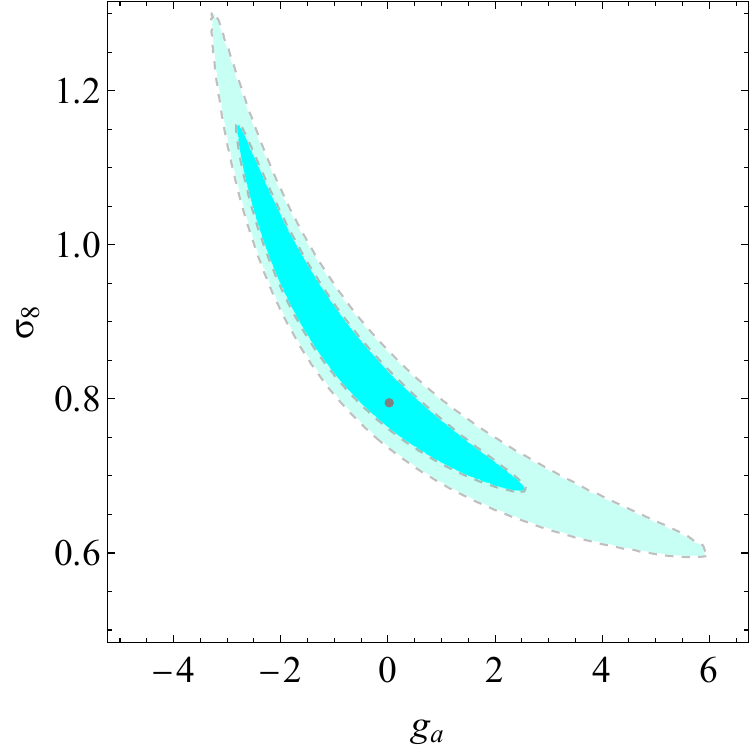}
        \caption{The quasi-degeneracy between $g_a$ and $\sigma_8$ is illustrated in the $2\sigma$ contour plot for the CPL parametrization. This analysis precedes our reanalysis of the Pantheon+ dataset using Eqs.(\ref{eq:luminosity_scaling})--(\ref{absolutmagn}) in Eq.(\ref{pantheonvec}).}
        \label{contour5a}
      \end{subfigure}
    \begin{subfigure}{\linewidth}
        \centering
    \includegraphics[width=.7\linewidth]{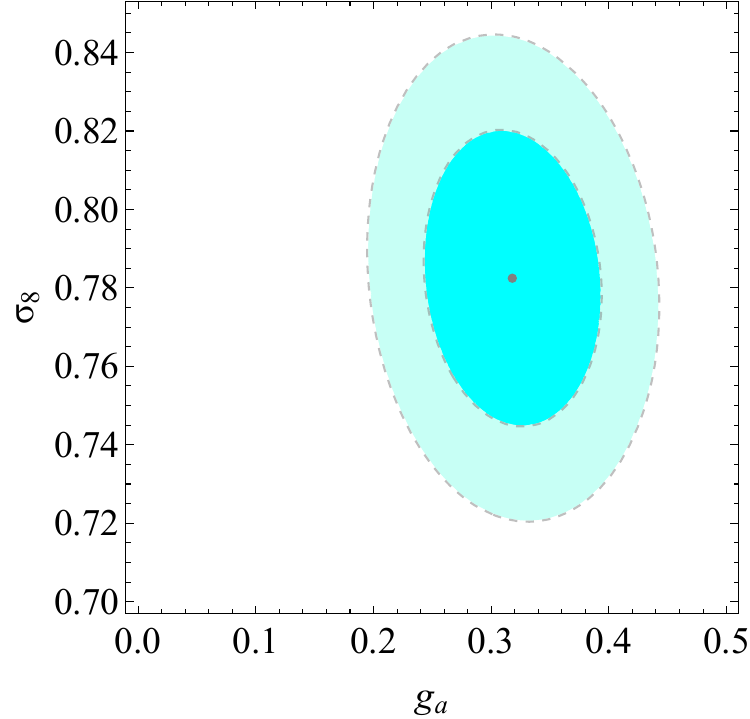}
        \caption{The reanalysis of the Pantheon+ dataset using Eqs.(\ref{eq:luminosity_scaling})--(\ref{absolutmagn}) in Eq.(\ref{pantheonvec}) break the quasi-degeneracy between $g_a$ and $\sigma_8$, as illustrated in the $2\sigma$ contour plot for the CPL parametrization. }
        \label{contour5b}
        \end{subfigure}
        
    \caption{Contour plots showing the 1$\sigma$ and 2$\sigma$ confidence regions in the $(g_a, \sigma_8)$ parameter space, obtained from the profile likelihood with all other parameters fixed at their best-fit values. The dataset combination used here is: Pantheon+, CMB, DESI DR2, RSD, and BBN constraint.}
    \label{contour5}
\end{figure}

The uncertainties and correlations between these observables are encoded in the covariance matrix $C_{\text{CMB}}$. We use the  covariance matrix provided by \citep{Zhai:2018vmm}:
\begin{equation*}
[\mathbf{C_{\text{CMB}}}] = 10^{-8} \times 
\end{equation*}
\begin{equation}
   \times \begin{pmatrix}
    1598.9554 & 17112.007 & -36.311179 \\
    17112.007 & 811208.45 & -494.79813 \\
    -36.311179 & -494.79813 & 2.1242182
\end{pmatrix}.
\end{equation}
The goodness-of-fit between model predictions and observational data is quantified using the $\chi^2$ statistic.
The $\chi^{2}_{\text{CMB}}$ distribution is \cite{Zhai:2018vmm}:

  \begin{equation}
      \chi^{2}_{\text{CMB}}=\textbf{v}^{T}[\mathbf{C_{ CMB}}]^{-1}\textbf{v},
      \end{equation}

 where in a flat universe the vector is written as \cite{Zhai:2018vmm} 
\begin{equation}
\textbf{v} = 
 \begin{pmatrix}
  R-1.74963  \\
  l_{A}-301.80845\\
  \Omega_{b}h^{2}-0.02237
 \end{pmatrix}.\end{equation}

\subsection{Results}\label{results}
The joint $\chi^2$ function incorporates all datasets:
\begin{align}\label{totalchi2}
\chi^2_{\text{total}} = & \chi^2_{\text{CMB}} + \chi^2_{\text{DESI BAO}} + \chi^2_{\text{Pantheon+}}\nonumber \\
& + \chi^2_{\text{GrowthData}} +\chi^2_{\text{BBN}},
\end{align}
Note that we have included the Big Bang Nucleosynthesis constraint, which provides a Gaussian prior on the baryon density~\cite{Schoneberg:2024ifp,Akrami:2025zlb,Nesseris:2025lke,Alestas:2025syk}:
\begin{equation}
\Omega_b h^2 = 0.02218 \pm 0.00055.
\end{equation}
The corresponding \(\chi^2\) function is:
\begin{equation}
\chi_{\text{BBN}}^2(\Omega_b h^2) = \left( \frac{\Omega_b h^2 - 0.02218}{0.00055} \right)^2.
\end{equation}
The  parameters being constrained are:
\begin{itemize}
\item $\Omega_{\rm m0}$: Matter density parameter
\item $\Omega_b h^2$: Baryon density
\item $h$: Hubble parameter
\item $w_0, w_a$: Dark energy equation of state parameters
\item $\sigma_8$: Amplitude of matter fluctuations
\item $g_a$: Modified gravity parameter
\item $M$: Absolute magnitude of SNe Ia
\end{itemize}
\begin{figure*}
    \centering
\includegraphics[width=1\linewidth]{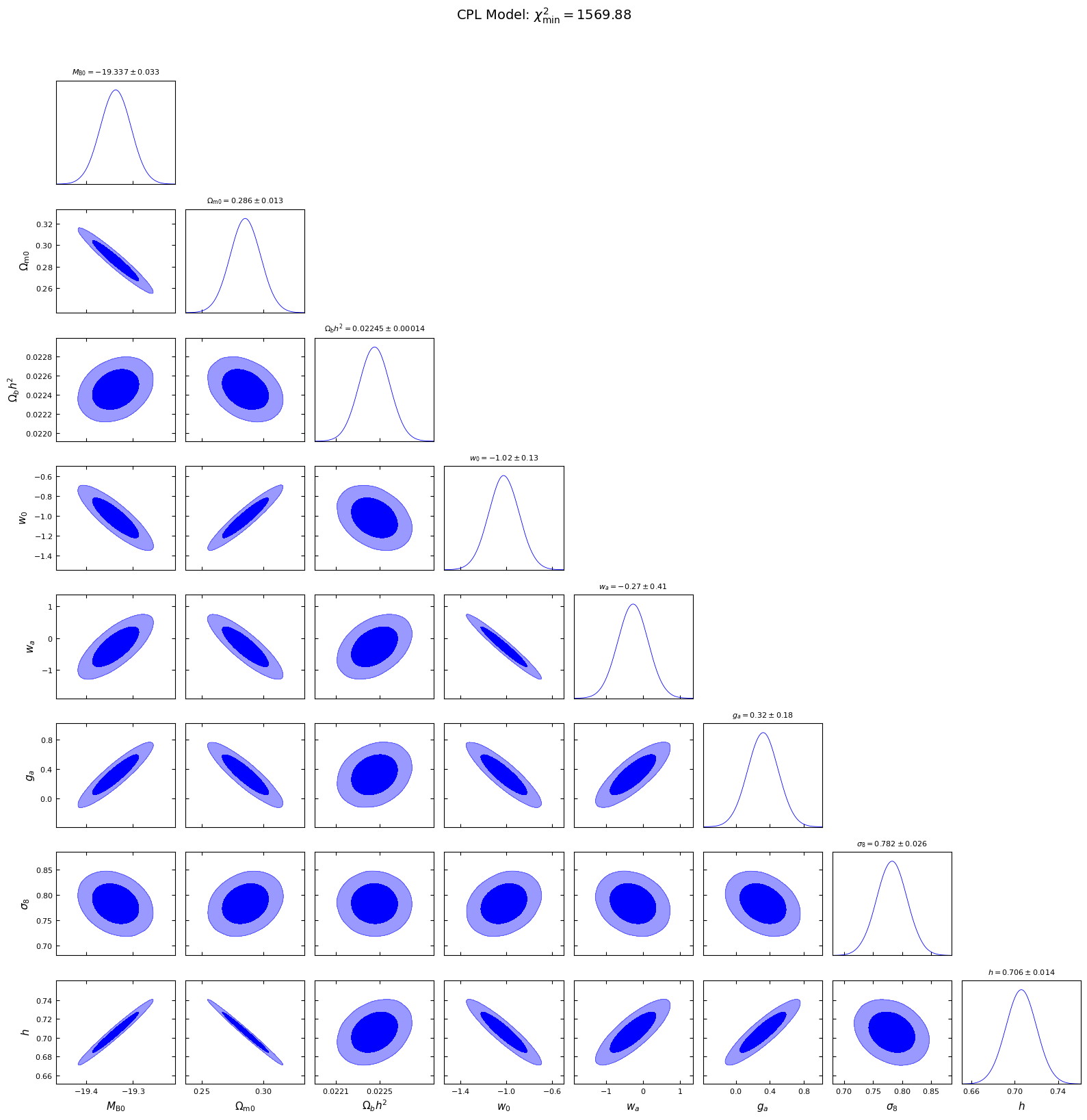}
    \caption{Triangle plot showing the posterior distributions for the CPL cosmological model parameters, assuming a multivariate Gaussian likelihood $P(\theta|{\rm data}) \propto \exp\left(-\frac{1}{2}(\theta - \mu)^T \Sigma^{-1} (\theta - \mu)\right)$ with uniform priors (see Appendix \ref{appB}).  Unlike direct $\chi^2$ mapping, this approach marginalizes over all parameters simultaneously but assumes Gaussianity. The minimum $\chi^2 = 1569.88$ for the best-fit model is indicated. The dataset combination used here is: Pantheon+, CMB, DESI DR2, RSD, and BBN constraints.}
\label{fig:corner_plots_cpl}
\end{figure*}
Notably, when the current data are analyzed using a proper joint $\chi^2$ (see Eq.\ref{totalchi2}), as illustrated in Fig.\ref{contour5a}, we find that the $\chi^2$ minimum is highly non-unique. There exist numerous statistically equivalent minima ($\Delta \chi_{\rm min}^2 \lesssim 1$) spanning a broad region of parameter space ($-2.5 \lesssim g_a \lesssim 2$, $0.7 \lesssim \sigma_8 \lesssim 1.1$), revealing a \textit{quasi-degeneracy} between $\sigma_8$ and $g_a$.

\begin{table*}[!ht]
    \centering
    \caption{\label{tab:aic_bic_results} Summary of the statistical analysis using the Akaike Information Criterion (AIC) and the Bayesian Information Criterion (BIC). We report the minimum $\chi^2$ value ($\chi^2_{\rm min}$) and the number of free parameters ($k$) for each model. The difference $\Delta{\rm AIC} = {\rm AIC}_i - {\rm AIC}_{\Lambda{\rm CDM}}$ is calculated relative to the base $\Lambda$CDM model, while $\Delta{\rm BIC} = {\rm BIC}_i - {\rm BIC}_{\Lambda{\rm CDM}}$ is calculated relative to the  preferred model. For the BIC, the number of data points is $N=1738$. The Log model ($g_a \neq 0$) minimizes the AIC (${\rm AIC}_{\rm min} = 1585.53$), while the base $\Lambda$CDM model minimizes the BIC (${\rm BIC}_{\rm min} = 1625.45$).}
    \begin{ruledtabular}
\begin{tabular}{lcccccc}
    Model & $\chi^2_{\rm min}$ & $k$ & AIC & $\Delta$AIC & BIC & $\Delta$BIC \\
    \hline
    $\Lambda$CDM & 1588.15 & 5 & 1598.15 & 0.00 & 1625.45 & 0.00 \\
    CPL ($g_a=0$) & 1578.44 & 7 & 1592.44 & -5.71 & 1630.66 & 5.21 \\
    CPL ($g_a\neq0$) & 1569.88 & 8 & 1585.88 & -12.27 & 1629.56 & 4.11 \\
    BA ($g_a\neq0$) & 1570.01 & 8 & 1586.01 & -12.14 & 1629.69 & 4.24 \\
    Log ($g_a\neq0$) & 1569.53 & 8 & 1585.53 & -12.62 & 1629.21 & 3.76 \\
    JBP ($g_a\neq0$) & 1570.67 & 8 & 1586.67 & -11.48 & 1630.35 & 4.90 \\
\end{tabular}
    \end{ruledtabular}
\end{table*}

In order to be conclusive we need to break this quasi-degeneracy.
We note that  the effective gravitational constant $G_{\rm{eff}}$ influences supernova luminosity $L$, when there is no screening mechanism. As shown by \citet{Wright:2017rsu}, when accounting for variations in the mass of $^{56}\rm Ni$ (the primary isotope responsible for SN luminosity) and including standardization effects from the light curve width, the relation becomes approximately \citep{Wright:2017rsu,Desmond:2019ygn}:
\begin{equation}\label{eq:luminosity_scaling}
    \frac{L}{L_0} \simeq \left(\frac{G_{\mathrm{eff}}}{G_N}\right)^{1.46}
\end{equation}
  The absolute magnitude $M_B$ relates to luminosity as \citep{Riazuelo:2001mg,Gannouji:2018ncm,Kazantzidis:2020tko,Marra:2021fvf}
\begin{equation}\label{absolutmagn}
    M_B  =M_{0B}- \frac{5}{2} \log_{10} \frac{L}{L_0}\simeq M_{0B}- \frac{5}{2} \log_{10} \left[\mu_G(z)\right]^{1.46},
\end{equation}
so an increase in $G_{\text{eff}}$ increases $L$ and decreases $M_B$.  We reanalyze Pantheon+ dataset by using Eqs.(\ref{eq:luminosity_scaling}-\ref{absolutmagn}) in Eq.(\ref{pantheonvec}). The reconstruction method, which relies on the assumption in Eq.(\ref{eq:luminosity_scaling}), provides an independent constraint on $\mu_G$ that complements the $f\sigma_8$ data. This additional assumption breaks the quasi-degeneracy between $\sigma_8$ and $g_a$, leading to the inference that $g_a > 0$, as illustrated in Fig.\ref{contour5b}. Note, however, that if we instead adopt the simpler scaling relation $L \propto M_{\mathrm{Ch}} \propto G^{-3/2}$~\cite{Gaztanaga:2001fh,Riazuelo:2001mg}, we would conclude that the best fit $g_a < 0$.

Constraints on the eight cosmological parameters for the CPL (MG) parametrization are presented in Fig.\ref{fig:corner_plots_cpl}, showing the approximate $2\sigma$ confidence ellipses. 
Further details, including the corresponding covariance matrix, are provided in Appendix~\ref{appB}.

\begin{figure}
    \begin{subfigure}{\linewidth}
        \centering
        \includegraphics[width=1\linewidth]{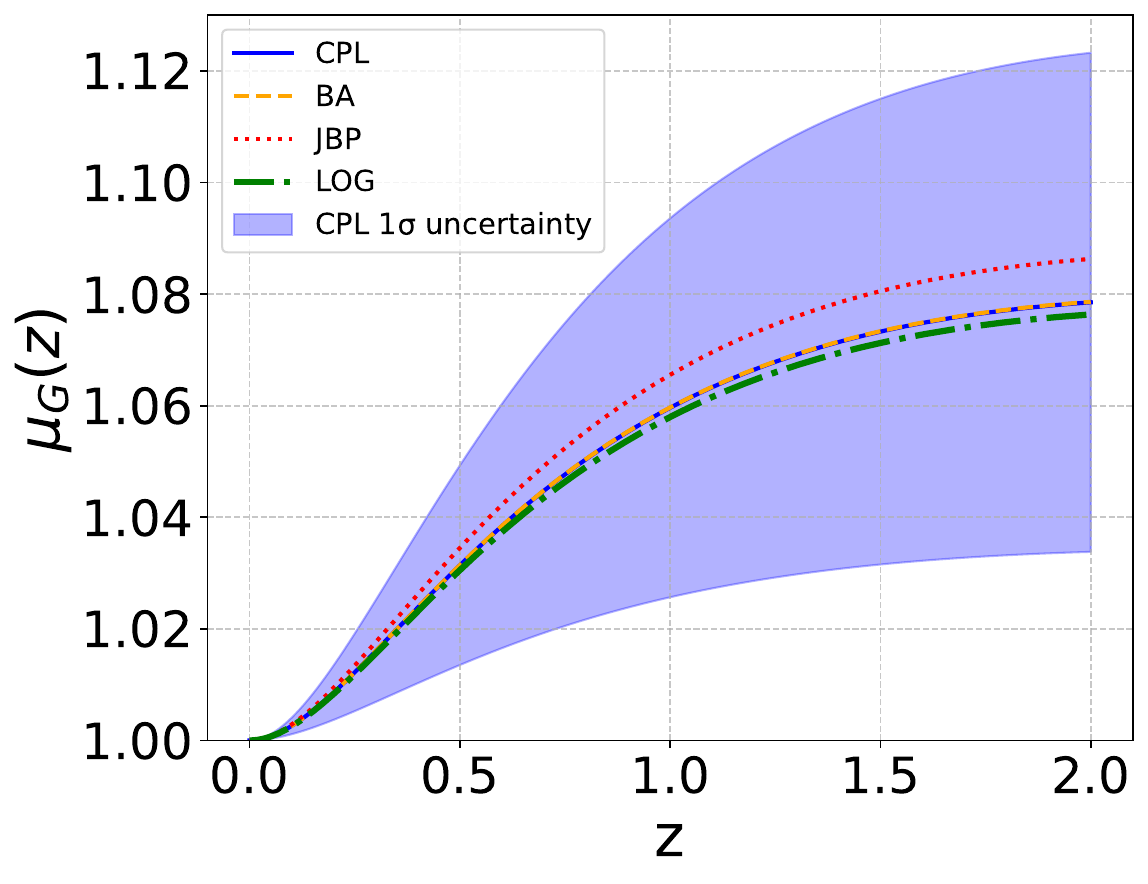}
 \caption{The quantity $\mu_G$ plotted according to Eq.(\ref{pantazisparam}) for each of the four models of dark energy selected with respect to the redshift $z$ with with the best-fit values for $g_a$ for each model taken from Table \ref{tab:fs8_constraint}.}
        \label{mu}
    \end{subfigure}
     \begin{subfigure}{\linewidth}
        \centering
        \includegraphics[width=1\linewidth]{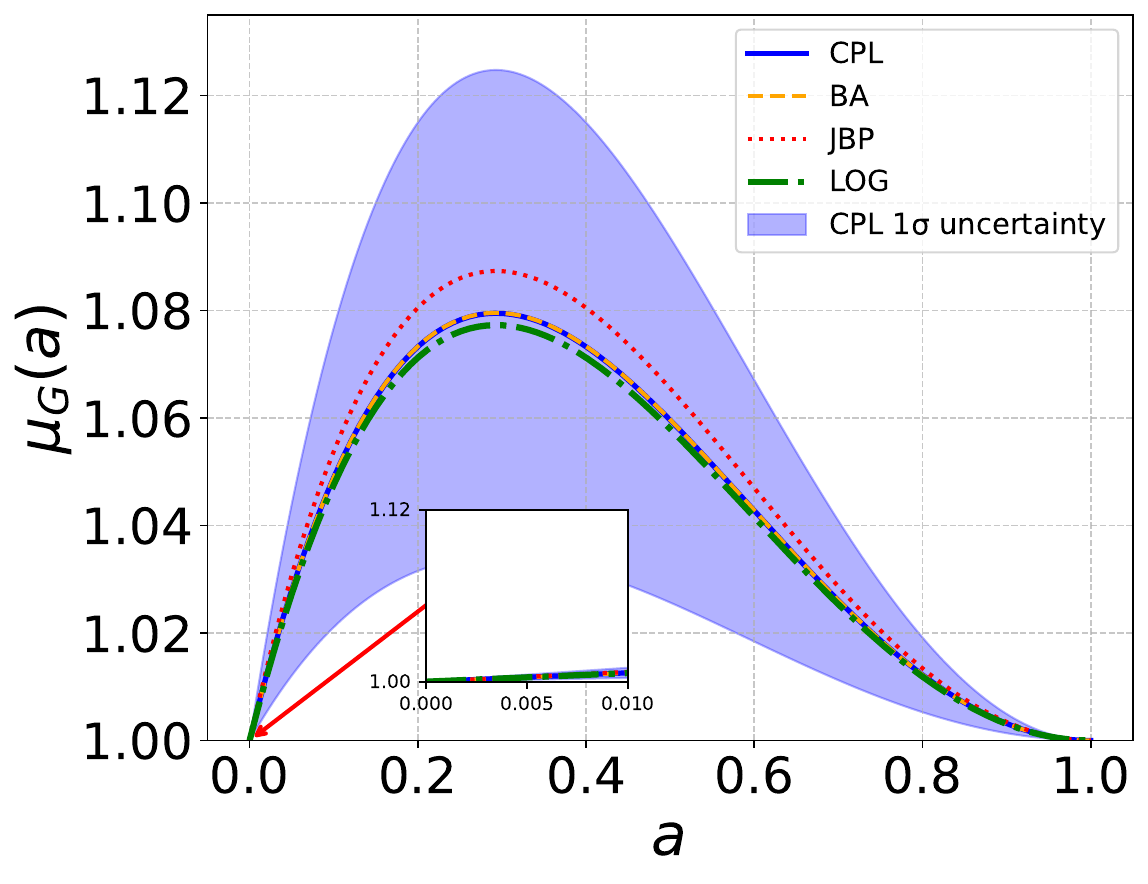}
        \caption{The quantity $\mu_G$ plotted according to Eq.(\ref{pantazisparam}) for each of the four models of dark energy selected with respect to the scale factor $a$ .}
        \label{framed}
        \end{subfigure}
         \caption{Evolution of the effective gravitational constant for the four dark energy parametrizations, shown as functions of (a) redshift $z$ and (b) scale factor $a$. Shaded regions represent $1\sigma$ confidence bands of CPL. The transient enhancement of gravity at intermediate times is a robust prediction across all parametrizations, driven by the positive best-fit values $g_a > 0$. The dataset combination used here is: Pantheon+, CMB, DESI DR2, RSD, and BBN constraint.}
   \label{Fig:2}
\end{figure}

\begin{figure}
\begin{subfigure}{\linewidth}
        \centering
        \includegraphics[width=1\linewidth]{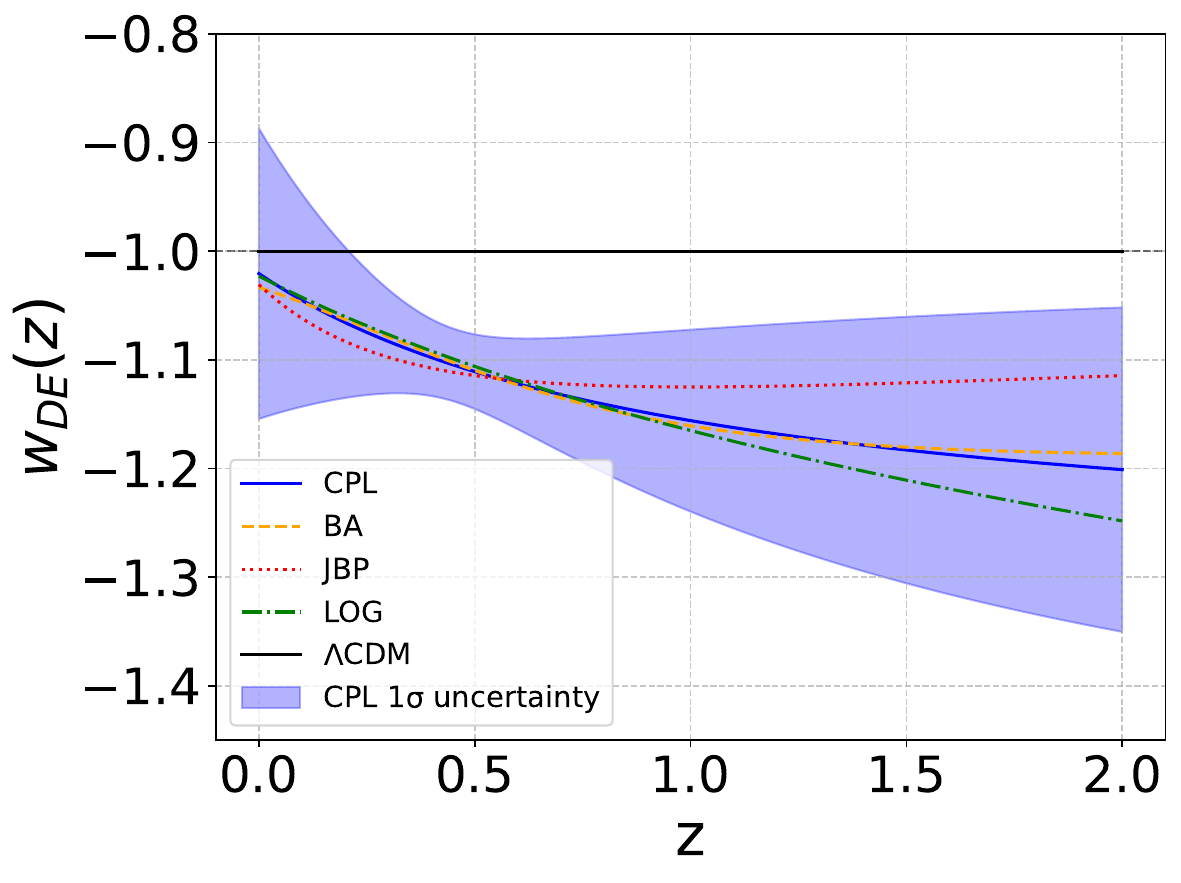}
        \caption{Modified gravity: different dark energy parametrizations $w(z)$ plotted according to their best fit values of $w_0$ and $w_a$ from the observational data (Table \ref{tab:fs8_constraint}). Every parametrization favors phantom crossing in a future time.}
 \label{4wz}
    \end{subfigure}
\begin{subfigure}{\linewidth}
        \centering
        \includegraphics[width=1\linewidth]{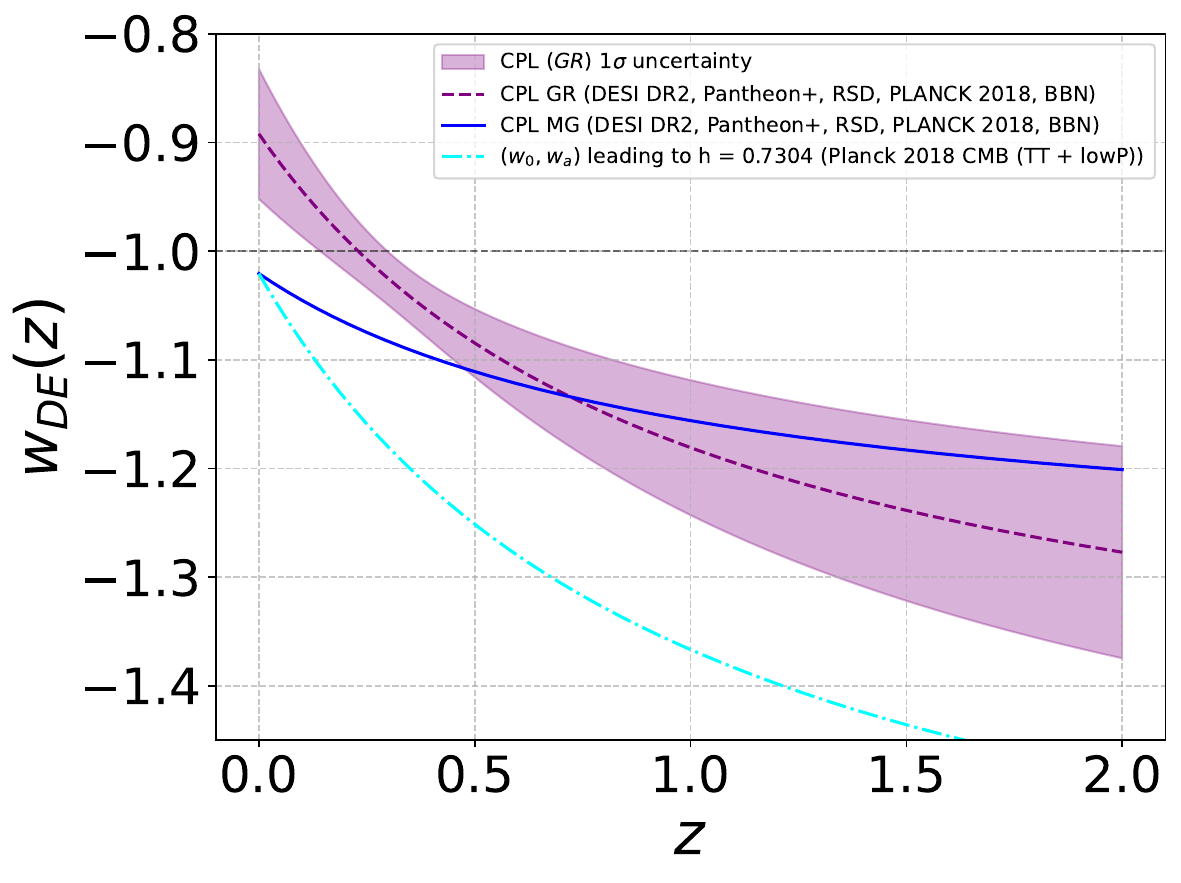}
        \caption{ We contrast three CPL parametrizations for $w(z)$: dynamical dark energy ($g_a=0$), modified gravity, and the $H_0$-tension alleviating case (shown dash-dotted, for $H_0=73.04$ km/s/Mpc) where $w_0$ is the CPL best-fit value and $w_a$ follows $w_a \simeq -5.978 - 6.051w_0 - 0.854w_0^2$ \citep{Alestas:2020mvb}.}
        \label{cpl_wz}
        \end{subfigure}
      \caption{The equation of state $w(z)$ plotted for all four models selected in panel (a) and in three cases of the CPL parametrization in panel (b). The dataset combination used here is: Pantheon+, CMB, DESI DR2, RSD, and BBN constraint.}
      \label{fig:3}
\end{figure}

\subsubsection{\label{subsec:muG_results}Evolution of the Effective Gravitational Constant}

As a result of the data analysis and the determination of the best-fit values of $w_0$, $w_a$, and $g_a$ for each dynamical dark energy parametrization, Figures~\ref{Fig:2} and~\ref{fig:3} were produced. 
Figure~\ref{Fig:2} shows the model-independent evolution of the effective gravitational constant, expressed as $\mu_G(z) = G_{\rm eff}(z)/G_N$ from Eq.(\ref{pantazisparam}), for the four parametrizations, plotted as a function of redshift $z$ (panel a) and scale factor $a = 1/(1+z)$ (panel b). 
For the CPL parametrization with $g_a = 0.318\pm0.18$ (see Table \ref{tab:fs8_constraint}), the effective gravitational constant reaches a maximum of $\mu_G(z=2) \approx 1.08$, corresponding to a $8\%$ enhancement relative to Newton’s constant.
As $z \to \infty$,  the parametrization approaches $\mu_G \to 1$, ensuring consistency with Big Bang Nucleosynthesis constraints $G_{\rm eff}/G_N = 1.09 \pm 0.2$ at $z \sim 10^9$. As can be seen in the small frame of Fig.\ref{framed} the derivative $d\mu_G/da \approx 0$ as $\mu_G$ almost asymptotically approaches 1.

\subsubsection{\label{subsec:wDE_evolution}Evolution of the Dark Energy Equation of State}

Figure~\ref{4wz} compares the evolution of the dark energy equation-of-state parameter $w(z)$ for the four parametrizations (CPL, BA, Logarithmic, JBP) using the best-fit values from Table~\ref{tab:fs8_constraint}. The shaded regions represent $1\sigma$ confidence bands for the CPL parametrization.
All four parametrizations exhibit phantom crossing in future time (negative redshift):
\begin{itemize}
    \item \textbf{CPL (solid blue):} Crosses the phantom divide at $z_{\rm cross} \approx -0.071$, with $w(z=2) \approx -1.2$ and $w(z=0) \approx -1.02$.
    
    \item \textbf{BA (dashed orange):} BA never crosses the phantom divide line, reaching $w(z=2) \approx -1.19$ before returning to $w(z=0) \approx -1.03$.
    
    \item \textbf{Logarithmic (dot-dashed green):} For the Logaritmhic $z_{\rm cross} \approx -0.11$, achieving $w(z=2) \approx -1.25$ exhibiting the strongest phantom behavior at intermediate redshifts and $w(z=0) \approx -1.02$.
    
    \item \textbf{JBP (dotted red):} Exhibits the mildest phantom behavior at intermediate redshifts, with $w(z=2) \approx -1.11$ and $w(z=0)\approx -1.03$, while the crossing happens at $z_{\rm cross} \approx -0.07$.
\end{itemize}

Figure \ref{cpl_wz} shows a direct comparison between the CPL parametrizations constrained under GR, within the MG framework defined by Eq.(\ref{pantazisparam}), and an alleviating $(w_0, w_a)$ curve addressing the Hubble tension \cite{Alestas:2020mvb}. We derive the $w(z)$ curve that relieves the $H_0$ tension (cyan dash-dotted line) using the geometric degeneracy method of Ref.~\cite{Alestas:2020mvb}. Imposing the equality condition 
$d_A(\bar{\omega}_m, \bar{\omega}_r, \bar{\omega}_b, h_{\Lambda\mathrm{CDM}}, w=-1)
=
d_A(\bar{\omega}_m, \bar{\omega}_r, \bar{\omega}_b, h_{\mathrm{local}}, w_0, w_a)$,
with $h_{\Lambda\mathrm{CDM}} = 0.6736$ (Planck) and $h_\mathrm{local}= 0.7304$ ~\cite{Riess:2021jrx}, and solving numerically for the CPL parametrization $w(z) = w_0 + w_a \frac{z}{1+z}$, we obtain the second-order degeneracy relation $w_a \simeq -5.978 - 6.051 w_0 - 0.854 w_0^2$. The fixed Planck parameter combinations are $\bar{\omega}_m = 0.1430$, $\bar{\omega}_b = 0.02237$, and $\bar{\omega}_r = 4.64 \times 10^{-5}$ \cite{Planck:2018vyg}. The resulting degeneracy curve is $w(z) = -1.02 - 0.69 \left( \frac{z}{1+z} \right)$.

 In the GR case ($g_a = 0$), represented by the purple curve and its $1\sigma$ uncertainty band, the model is constrained using the combined DESI DR2, Pantheon+, CMB, and RSD datasets and BBN constraint, yielding best-fit parameters $w_0 \simeq -0.89 \pm 0.06$ and $w_a \simeq -0.58 \pm 0.26$ (see Table \ref{tab:constraint}). The best-fit reconstruction of the dark energy equation of state yields $w_{\mathrm{DE}}(0) \simeq -0.89$ at redshift $z=0$, decreasing to $w_{\mathrm{DE}}(z=2) \simeq -1.28$ in the past, which represents the standard CPL case under GR that favors a phantom crossing.
The MG case, shown by the dark-blue curve, follows a similar evolution to the CPL (GR) case but exhibits a less steep trajectory, thereby predicting a future-time phantom crossing.

\begin{figure}[htbp]
    \centering
    \includegraphics[width=\columnwidth]{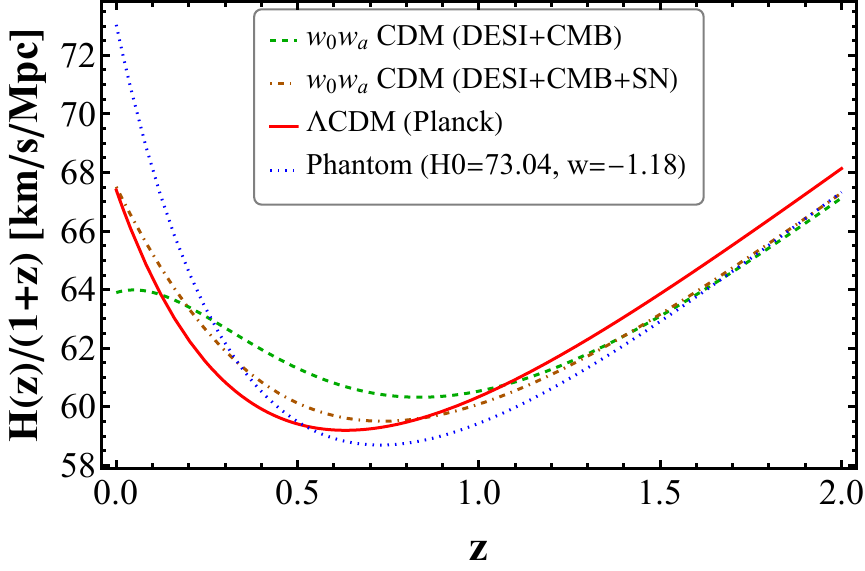} 
    \caption{The evolution of the comoving Hubble rate, $H(z)/(1+z)$, for different cosmological models. The solid red line represents the best-fit Planck $\Lambda$CDM model. The dashed green and dot-dashed brown curves show the best-fit $w_0w_a$CDM models from the DESI+CMB and DESI+CMB+SNe joint analyses, respectively \cite{DESI:2025zgx}. At redshift $z = 0$, the value of this quantity is the Hubble constant, $H_0$. The DESI-preferred dynamical dark energy models predict a lower $H_0$ than Planck $\Lambda$CDM, thus exacerbating the Hubble tension with local measurements like SH0ES. For comparison, the dotted blue line shows a phantom model with $H_0=73.04$ km s$^{-1}$ Mpc$^{-1}$ that aligns with the SH0ES measurement.}
    \label{fig:hubble_tension_worsens}
\end{figure}

\begin{figure}[htbp]
    \centering
    \includegraphics[width=\columnwidth]{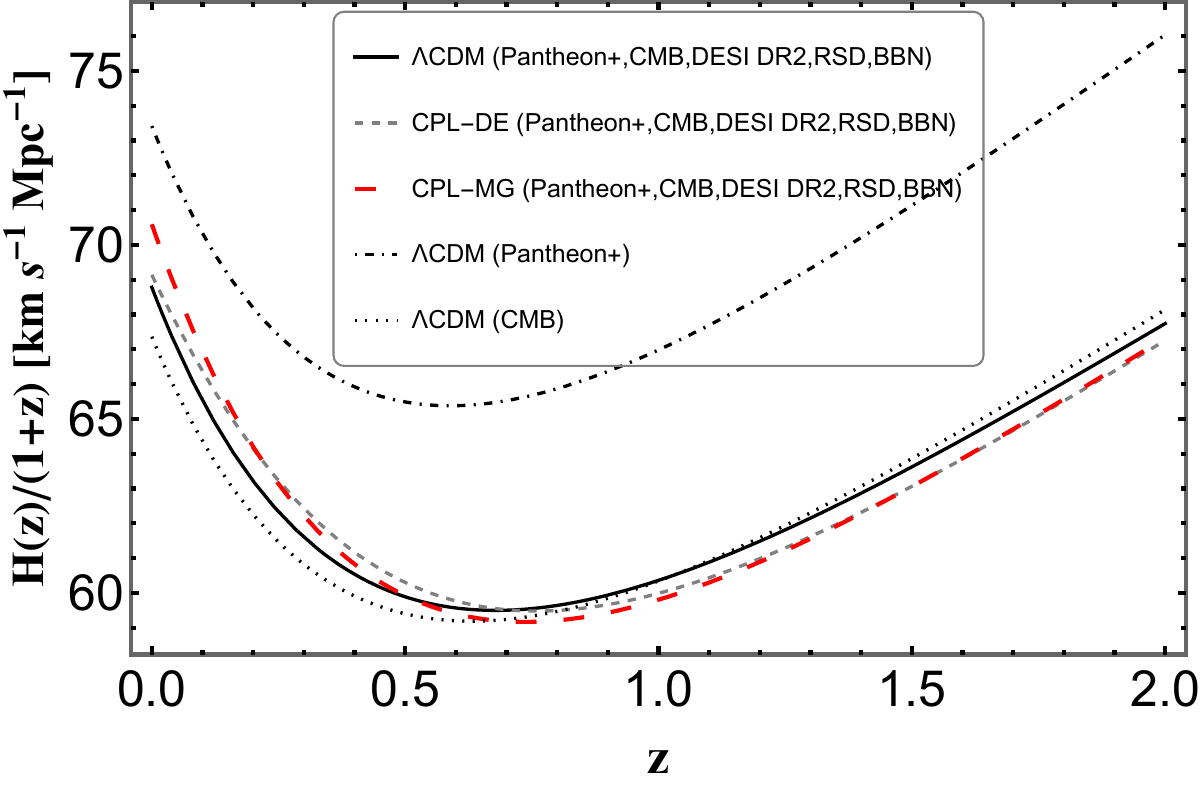} 
    \caption{Evolution of the comoving Hubble rate, $H(z)/(1+z)$, for different cosmological models. 
    The  dashed (large dashing) line represents the scalar-tensor CPL (CPL-MG) model, computed for the best-fit values obtained from the combined datasets (Pantheon+, CMB, DESI DR2, RSD, BBN). 
    The  dashed line shows the dynamical dark energy CPL (CPL-DE) model using the same data combination, while the solid line corresponds to the $\Lambda$CDM model for the full dataset (Pantheon+, CMB, DESI DR2, RSD, BBN). 
    For comparison, the dot-dashed  line illustrates the $\Lambda$CDM prediction constrained only by the Pantheon+ sample, and the dotted  line represents $\Lambda$CDM constrained by CMB data alone.  }
    \label{fig:hubble_tension}
\end{figure}

We assess model performance using the Akaike Information Criterion (AIC) and the Bayesian Information Criterion (BIC), which balance goodness-of-fit against model complexity \cite{Kass_Raftery_1995,Burnham_Anderson_2002} (see also \cite{Shi:2012ma}):
\[
\mathrm{AIC}=\chi^2_{\min}+2k,\qquad
\mathrm{BIC}=\chi^2_{\min}+k\ln N,
\]
where $k$ is the number of free parameters and $N=1738$ is the total number of data points. As shown in Table~\ref{tab:aic_bic_results}, the Log model ($g_a\neq0$) achieves the  minimum, ${\rm AIC}_{\min}=1585.53$, which corresponds to $\Delta{\rm AIC} = {\rm AIC}_i - {\rm AIC}_{\Lambda{\rm CDM}}=-12.62$ relative to the base $\Lambda$CDM model. Since a difference of $\Delta{\rm AIC} < -10$ indicates that the base model has essentially no empirical support compared to the extended model, the AIC strongly favors our modified-gravity parameterizations. In contrast, the BIC penalizes additional parameters more heavily, selecting the economical $\Lambda$CDM (${\rm BIC}_{\min}=1625.45$); our extended models yield $\Delta{\rm BIC}\in[3.76,4.90]$, representing only ``positive'' evidence against them ($2<\Delta{\rm BIC}\leq6$) and not reaching strong exclusion ($\Delta{\rm BIC}>6$). Overall, $\Lambda$CDM remains the simplest model, but the strong AIC preference shows that our dynamical MG parameterizations capture real features in the data that $\Lambda$CDM misses, making them robust predictive alternatives.

\section{\label{sec:tensions}Implications for Cosmological Tensions}
    
The proposed solutions to the Hubble constant ($H_0$) discrepancy can generally be divided into two main categories, depending on whether they preserve the precisely measured value of the angular acoustic scale $\theta_A$ (Eq.(\ref{thetaA})). The first category includes models that reduce the sound horizon $r_s(z_*)$ in order to compensate for the shorter angular diameter distance implied by a higher $H_0$.  The second category involves late-time modifications to the expansion history after recombination, ensuring that the angular diameter distance to recombination remains unchanged. 
In contrast, resolving the $S_8$ tension requires suppressing the growth of cosmic structure. 

It is important to clarify our assumptions. In Eq.(\ref{pantazisparam}), the parameter $\mu_G$ describes a scale-independent modification of the effective gravitational coupling that governs the growth of matter perturbations and, in the absence of a screening mechanism, can also affect the intrinsic luminosity $L$ of Type Ia supernovae. 
This parametrization implies that gravity remains effectively indistinguishable from general relativity both at the present epoch and at the time of recombination, while allowing deviations at intermediate redshifts. 
To address the ongoing $H_0$ and $S_8 \equiv \sigma_8 \sqrt{\Omega_{m0}/0.3}$ tensions, we explore the impact of allowing the gravitational strength to deviate from GR while simultaneously permitting deformations in the Hubble expansion rate $H(z)$.

\subsubsection{\label{subsubsec:hubble_tension}The Hubble Tension}

\begin{table}[htbp]
\centering
\caption{Comparison of Hubble constant determinations from different methods and models (including the BBN constraint).}
\label{tab:H0_comparison}
\begin{ruledtabular}
\begin{tabular}{lc}
Model & $H_0$ [km s$^{-1}$ Mpc$^{-1}$] \\
\hline
\multicolumn{2}{c}{RSD+DESI DR2+BBN+CMB+Pantheon+} \\
\hline
$\Lambda$CDM & $68.8\pm 0.3$\\
CPL (GR) & $69.1 \pm 0.6$\\
CPL (MG) & $70.6 \pm 1.4$ \\
BA  (MG) & $70.6 \pm 1.6$ \\
Logarithmic (MG) & $70.6 \pm 1.4$ \\
JBP (MG) & $70.8 \pm 1.0$ \\
\end{tabular}
\end{ruledtabular}
\end{table}

The $\Lambda$CDM value inferred from the CMB remains in strong disagreement with the SH0ES determination 
($H_0 = 73.04 \pm 1.04$), exhibiting a tension of approximately $4.9\sigma$. 
In comparison, the CPL model—according to Table~VII of \citet{DESI:2025zgx}—yields 
$H_0 = 63.7^{+1.7}_{-2.1}$ from the Planck + DESI combination, corresponding to a 
$\sim 4.3\sigma$ discrepancy relative to SH0ES (see Fig.\ref{fig:hubble_tension_worsens}). 
Similarly, the DESI BAO+CMB+Pantheon+ analysis infers a value of 
$H_0 = 67.54 \pm 0.59$, which is about $4.6\sigma$ lower than SH0ES.

In our analysis, from Table~\ref{tab:H0_comparison}, it is evident that the standard $\Lambda$CDM model, when all datasets are combined (RSD+DESI DR2+BBN constraint+CMB+Pantheon+), yields a value of 
$H_0 =( 68.8 \pm 0.3)\,\mathrm{km\,s^{-1}\,Mpc^{-1}}$, which is approximately 
$4.4\sigma$ lower than the local Pantheon+ best-fit result of 
$H_0 = (73.4 \pm 1.0)\,\mathrm{km\,s^{-1}\,Mpc^{-1}}$ derived from our analysis. 
The CPL (GR) model  yields 
$H_0 = (69.1 \pm 0.6)\,\mathrm{km\,s^{-1}\,Mpc^{-1}}$, which corresponds to a $3.7\sigma$ discrepancy—representing only a modest improvement over the $\Lambda$CDM prediction when all datasets are combined\footnote{Please note that this result is a feature of the specific dataset combination used in the present analysis (RSD+DESI DR2+BBN+CMB+Pantheon+), particularly when the Pantheon+ sample is included. Furthermore, our findings are fully consistent with the trend reported in the DESI DR2 analysis \citep{DESI:2025zgx}, which also observes a lower $H_0$ value in the CPL model compared to $\Lambda$CDM when using RSD+DESI DR2+CMB data. This comparison is shown in Table~\ref{tab:H0_comparison2}.}.  Additionally, we note that when the Pantheon+ dataset is included in the compilation, the predicted $H_0$ value tends to be slightly higher, with a correspondingly larger $1\sigma$ uncertainty compared to the $\Lambda$CDM case.

In contrast, our CPL-Modified Gravity (MG) models predict a Hubble constant of 
$H_0 \approx 70.6 \pm 1.4\,\mathrm{km\,s^{-1}\,Mpc^{-1}}$, representing a 
$\sim 2.2\sigma$ upward shift relative to the Planck $\Lambda$CDM estimate 
($H_0 = 67.36 \pm 0.54$) and a $\sim 1.4\sigma$ downward shift with respect to the SH0ES measurement 
($H_0 = 73.04 \pm 1.04$). 
This result is in excellent agreement with the TRGB distance-ladder determination 
($H_0 = 69.8 \pm 1.9$)~\citep{Freedman:2021ahq}, differing by only $0.3\sigma$. In addition,  MG models, such as CPL (MG), BA (MG), and JBP (MG), produce higher central values of $H_0$ 
(around $70.6$–$70.8\,\mathrm{km\,s^{-1}\,Mpc^{-1}}$) with larger $1\sigma$ uncertainties ($\sim1.0$–$1.6$). 
This broader and higher range shifts these models closer to the local measurement, thereby alleviating the Hubble tension. 
 In our analysis, we compare models using the dataset combination RSD+DESI DR2+BBN constraint+CMB+Pantheon+.  Although the Hubble tension is commonly defined within the context of the $\Lambda$CDM model (see Fig.~\ref{fig:hubble_tension}), our conclusions are drawn strictly within this specific dataset compilation (RSD+DESI DR2+BBN constraint+CMB+Pantheon+). This choice is motivated by quasi-degeneracies in the determination of the coupling parameter $g_a$, which prevent a meaningful reduction of the dataset. 
Under this common dataset compilation, we consistently compare all Modified Gravity (MG) models with the CPL (GR) dynamical dark energy parametrization\footnote{For completeness, additional dark energy parametrizations (JBP, BA, and Logarithmic) are presented in the Appendix \ref{AppA}.} and with the $\Lambda$CDM model.  For the combined dataset (RSD+DESI DR2+BBN constraint+CMB+Pantheon+), the standard $\Lambda$CDM model gives 
$H_0 = 68.8 \pm 0.3\,\mathrm{km\,s^{-1}\,Mpc^{-1}}$, about $4.4\sigma$ lower than the local Pantheon+ result of $73.4 \pm 1.0\,\mathrm{km\,s^{-1}\,Mpc^{-1}}$. The CPL (GR) model yields $H_0 = 69.1 \pm 0.6\,\mathrm{km\,s^{-1}\,Mpc^{-1}}$, corresponding to a $3.7\sigma$ discrepancy. We find that MG models systematically prefer higher values of $H_0$, indicating their potential to alleviate the Hubble tension. In particular, the CPL (MG) model predicts
$$H_0 \approx 70.6 \pm 1.4\,\mathrm{km\,s^{-1}\,Mpc^{-1}},$$
corresponding to an upward shift of approximately $2.2\sigma$ relative to the Planck $\Lambda$CDM estimate ($H_0 = 67.36 \pm 0.54$)\cite{Planck:2018vyg} and a downward shift of approximately $1.4\sigma$ compared to the SH0ES measurement  ($H_0 = 73.04 \pm 1.04$)\cite{Riess:2021jrx}. 
Although the Hubble tension is not completely resolved, it is alleviated, i.e., in the considered class of modified gravity models exhibiting a $\mu_G(z)$-like behavior (with dark energy parametrizations CPL, BA, logarithmic, and JBP) relative to general relativity with the same parametrizations, as well as $\Lambda$CDM, under the combined dataset comprising RSD, DESI DR2, BBN constraint,  CMB, and Pantheon+.

The physical origin of this alleviation arises from the data’s preference for a cosmological phase in which an enhanced effective gravitational strength coexists with a phantom-like dark energy component. 
 In order to simultaneously account for the observed structure formation and supernova luminosity data, modified gravity (MG) models require the duration of the phantom phase to be slightly longer than in general relativity with a purely dynamical dark energy component. 
This prolonged phantom epoch naturally results in a higher inferred value of $H_0$ (see also \citep{Lee:2022cyh}).

\subsubsection{\label{subsubsec:S8_tension}The $S_8$ Tension}

The $S_8$ parameter, defined as 
$S_8 \equiv \sigma_8(\Omega_{m0}/0.3)^{0.5}$,
and it quantifies the amplitude of matter fluctuations on $8\,h^{-1}$ Mpc scales and is particularly sensitive to modifications of gravity.

Table~\ref{tab:S8_comparison} presents a comparison of values from various cosmological models. When all datasets are combined (CMB, DESI DR2, BBN, Pantheon+, RSD), the dynamical dark energy model (CPL) yields \( S_8 = 0.777 \pm 0.043 \), whereas the standard \(\Lambda\)CDM model results in \( S_8 = 0.798 \pm 0.043 \). 
 In particular, the MG scenario prefers $\sigma_8 \approx 0.78 \pm 0.03$ with $\Omega_{\rm m0} \simeq 0.29 \pm 0.01$, compared to the $\Lambda$CDM values $\sigma_8 \approx 0.80 \pm 0.03$ and $\Omega_{\rm m0} \simeq 0.297 \pm 0.004$, and the CPL (GR) case with $\sigma_8 \approx 0.78 \pm 0.03$ and $\Omega_{\rm m0} \simeq 0.298 \pm 0.006$. 
Consequently, the modified gravity (MG) models successfully reproduce the observed $f\sigma_8(z)$ measurements and consistently predict a lower best-fit value of $S_8 \simeq 0.76 \pm 0.05$.

\begin{table}[htbp]
\centering
\caption{Comparison of $S_8$ determinations from CMB and large-scale structure (LSS) measurements. 
The $S_8$ uncertainty is computed using the standard error propagation formula $\sigma_{S_8}^2 = 
\sigma_{\Omega_{\mathrm{m0}}}^2
\left( \frac{\partial S_8}{\partial \Omega_{\mathrm{m0}}} \right)^2
+ \sigma_{\sigma_8}^2
\left( \frac{\partial S_8}{\partial \sigma_8} \right)^2
+ 2\,\sigma_{\Omega_{\mathrm{m0}}\sigma_8}
\left( \frac{\partial S_8}{\partial \Omega_{\mathrm{m0}}} \right)
\left( \frac{\partial S_8}{\partial \sigma_8} \right)$~\cite{Bevington2003}.
}
\label{tab:S8_comparison}
\begin{ruledtabular}
\begin{tabular}{lc}
Model & $S_8$ \\

\hline
\multicolumn{2}{c}{RSD+DESI DR2+BBN+CMB+Pantheon+} \\
\hline
$\Lambda$CDM  & $0.798 \pm 0.043$ \\
CPL DE  & $0.777 \pm 0.043$ \\
 CPL (MG) & $0.763 \pm 0.049$ \\
BA (MG) & $0.763 \pm 0.051$ \\
Logarithmic (MG) & $0.765 \pm 0.048$ \\
JBP (MG) & $0.759 \pm 0.045$ \\
\end{tabular}
\end{ruledtabular}
\vspace{6pt}
\end{table}

\section{A recipe to reconstruct the scalar tensor Lagrangian}\label{sec:theory}

Note that our data analysis and physical interpretation thus far have been rather generic, relying on the chosen parametrization of $\mu_G$ (Eq.(\ref{pantazisparam})) and the specific parametrizations of dynamical dark energy (CPL, BA, Log, JBP). 
In essence, we have examined how allowing the gravitational strength to deviate from general relativity affects the deformation of the Hubble expansion history. 
We now proceed to investigate whether scalar–tensor gravity, with the Lagrangian density given in Eq.(\ref{eq:lagrangian}), can reproduce the reconstructed evolutions of both $H(z)$ and $\mu_G(z)$.  

By varying the action corresponding to the Lagrangian in Eq.(\ref{eq:lagrangian}) with respect to the metric $g^{\mu\nu}$, we obtain the field equations~\citep{Esposito-Farese:2000pbo}:
\begin{widetext}
\begin{equation}\label{fieldequations}
    F(\Phi)G_{\mu\nu} = T_{\mu\nu}^{\text{matter}} 
    + \nabla_\mu \nabla_\nu F(\Phi) 
    - g_{\mu\nu} \nabla^\sigma\nabla_\sigma F(\Phi) 
    - \frac{1}{2} g_{\mu\nu} \partial_\mu \Phi \partial^\mu \Phi+\partial_\mu\Phi\partial_\nu\Phi-U(\Phi)g_{\mu\nu}.
\end{equation}
\end{widetext}

Adopting a flat Friedmann--Robertson--Walker (FRW) metric given by
\begin{equation}\label{FRLWmetric}
    ds^2 = -dt^2 + a^2(t)\left[dr^2 + r^2\left(d\theta^2 + \sin^2\theta \, d\phi^2\right)\right],
\end{equation}
substituting Eq.\eqref{FRLWmetric} into Eq.\eqref{fieldequations} yields the following coupled system of equations~\cite{Esposito-Farese:2000pbo}:
\begin{subequations}\label{eq:friedmann}
\begin{align}
3F(\Phi)H^2 &= \rho + \frac{1}{2}\dot{\Phi}^{2} - 3H\dot{F}(\Phi) + U(\Phi), \label{eq:friedmann1} \\[1ex]
-2F(\Phi)\dot{H} &= \rho + p + \dot{\Phi}^{2} + \ddot{F}(\Phi) - H\dot{F}(\Phi). \label{eq:friedmann2}
\end{align}
\end{subequations}
Assuming a perfect fluid matter content ($p=0$), we convert the Friedmann equations (Eqs.\eqref{eq:friedmann1} and \eqref{eq:friedmann2}) to redshift space with the rescaled potential $U \rightarrow U \cdot H_0^2$ to obtain (we denote as $'\equiv d/dz$):
\begin{widetext}
\begin{equation}\label{fried12}
F'' + \left[ \frac{q'}{2q} - \frac{4}{1+z} \right] F' + \left[ \frac{6}{(1+z)^2} - \frac{2}{(1+z)} \frac{q'}{2q} \right] F = \frac{2U}{(1+z)^2 q} +3 \frac{1+z}{q} \Omega_{\rm m0},
\end{equation}
\begin{equation}\label{fried22}
\Phi'^2 = - \frac{6F'}{1+z} + \frac{6F}{(1+z)^2} - \frac{2U}{(1+z)^2 q} -6 \frac{1+z}{q} \Omega_{\rm m0}. 
\end{equation}
\end{widetext}

Note that we have made the following identification in Eqs.(\ref{fried12}--\ref{fried22}):

\begin{equation}
\label{Friedman}
\begin{split}
q(z) \equiv \frac{H^2(z)}{H_0^2} & = \Omega_{\mathrm{r}} (1+z)^4 + \Omega_{\mathrm{m0}} (1+z)^3 \\
& \quad + \Omega_{\mathrm{DE}} \exp\left(\int_0^z \frac{3\left[1+w_{\mathrm{DE}}(z')\right]}{1+z'} \, dz'\right)
\end{split}
\end{equation}
 and we are employing various phenomenological parametrizations for $w_{\mathrm{DE}}(z)$ (Eqs.(\ref{cpl}--\ref{jbp}) correspond to the CPL, BA, Logarithmic, and JBP parametrizations, respectively).

In contrast to General Relativity, where the expansion history $H(z)$ uniquely determines the scalar potential $U(z)$, scalar–tensor theories require the additional specification of the coupling function $F(z)$. This necessitates the collection of appropriate observational datasets that can collectively constrain the corresponding quantities, enabling the reconstruction of the fundamental functions $(H(z), F(z), U(z))$ needed to build the theory. By combining the reconstruction frameworks presented in~\citep{Boisseau:2000pr,Perivolaropoulos:2005yv}, we ultimately reconstruct the scalar potential \( U(\Phi) \) and the non-minimal coupling function \( F(\Phi) \). 
After some manipulation can rewrite the above Eqs.(\ref{fried12}-\ref{fried22})
\begin{widetext}
    \begin{equation}
    U(z)=
\frac{q}{2}\,(1+z)^2\,F''(z)
+ \biggl(\frac{(1+z)^2\,q'}{4} \;-\; 2\,q\,(1+z)\biggr)\,F'(z)
+ \biggl(3\,q \;-\; \frac{(1+z)\,q'}{2}\biggr)\,F(z)
- \frac{3}{2}\,(1+z)^3\,\Omega_{0m},\label{EqU}
\end{equation}
\begin{equation}
    \Phi'^2(z)=-\,F''(z)
- \biggl(\frac{2}{1+z} + \frac{q'}{2\,q}\biggr)\,F'(z)
+ \frac{q'}{q\,(1+z)}\,F(z)
- \frac{3\,(1+z)}{q}\,\Omega_{0m}.\label{EqPhi}
\end{equation}
\end{widetext}

Here, we introduce an additional assumption based on the approximation that on sub-horizon scales within the quasi-static regime—where spatial derivatives dominate over temporal ones—the scalar perturbation equations simplify, rendering $G_{\mathrm{eff}}$ effectively scale-independent (see also \citep{Boisseau:2000pr,Amendola:2007rr,Tsujikawa:2007gd}). We constrain the effective Newton constant $G_{\mathrm{eff}}$ using both growth and Type~Ia supernova (SN~Ia) data (in unscreened environments). For the data analysis, we employed a scale-independent phenomenological parametrization, as given by Eq.~(\ref{pantazisparam}) \citep{Nesseris:2017vor} (see Section~\ref{sec:datanalysisprocedure}). This parametrization is suitable for a massless scalar–tensor theory of gravity reconstructed through the effective gravitational constant, which is defined as
\begin{equation}\label{gefftheory}
  G_{\mathrm{eff}}
  = \frac{1}{8 \pi F}
    \frac{2F + 4\left(\tfrac{dF}{d\Phi}\right)^2}
         {2F + 3\left(\tfrac{dF}{d\Phi}\right)^2}.
\end{equation}
Here, $G_{\mathrm{eff}}$ represents the effective gravitational coupling between two test masses in the presence of a massless dilaton \citep{Boisseau:2000pr,Tsujikawa:2007gd} (see also Appendix~\ref{appC}).

From the definition in Eq.\eqref{eq:geff_param} and  Eq.\eqref{gefftheory}  we can derive the following expression:
\begin{equation}\label{eqnumerical}
    \Phi'^2=\frac{F'^2(4-3\mu_G F)}{2F(\mu_G F-1)}.
\end{equation}
We observe that Eq.(\ref{eqnumerical}) becomes singular when the condition $\mu_G(z') = 1/F(z')$ is satisfied at some redshift $z'$.   We now demand only that $\Phi'^2 \geq 0$ and $\mu_G(z) \neq 1/F(z)$.  Under these conditions, we obtain the following result
\begin{equation}\label{inequalitygen}
   \frac{1}{F(z)} <\mu_{\rm G}(z)\leq \frac{4}{3 F(z)}.
\end{equation}
We maintain the initial condition $F(z=0)=1$, but we set by hand the present-day value to $\mu_{\mathrm{G},0} = 1.00002$. This choice is made to be consistent with local experimental constraints \cite{Boisseau:2000pr,Bertotti:2003rm,Negrelli:2020yps}, while also ensuring that Eq.\eqref{inequalitygen} is respected given the initial condition.


Plugging Eq.\eqref{EqPhi} in Eq.\eqref{eqnumerical} we can solve the second order differential equation to find $F(z)$:
\begin{widetext}
\begin{equation}\label{d.EqF}
-\,F''(z)
- \biggl(\frac{2}{1+z} + \frac{q'}{2\,q}\biggr)\,F'(z)
+ \frac{q'}{q\,(1+z)}\,F(z)
- \frac{3\,(1+z)}{q}\,\Omega_{0m}-\frac{F'(z)^2\left[4-3\mu_G(z) F(z)\right]}{2F(z)\left[\mu_G(z) F(z)-1\right]}=0
\end{equation}
\end{widetext}
Once the coupling function $F(z)$ has been determined, we substitute $F(z)$ together with the best-fit function $q(z)$ into Eq.(\ref{EqPhi}) and numerically integrate it to recover $U(z)$ via Eq.(\ref{EqU}). With the initial condition $\Phi(0)=0$, we invert to $z(\Phi)$ numerically. Note that we can write $U(z)=U(z(\Phi))$ and $F(z)=F(z(\Phi))$ given that $\Phi(z)$ is bijective \footnote{In our analysis, the functional form of $\mu_G(z)$ is a physically motivated assumption, and the reconstruction of $\Phi(z)$ is performed within this parametrized framework. The monotonic—and therefore bijective—behavior of $\Phi(z)$ results from the best-fit numerical reconstruction. Within the range of interest, the inversion to obtain single-valued functions $F(\Phi)$ and $U(\Phi)$ is valid only where $\Phi(z)$ remains monotonic;  if, for example, $\Phi(z)$ were oscillatory, then a global single-valued inversion would not be achievable, and the procedure would have to be suitably restricted or appropriately modified.} in the range of interest.

\begin{figure}
    \centering
    \includegraphics[width=1\linewidth]{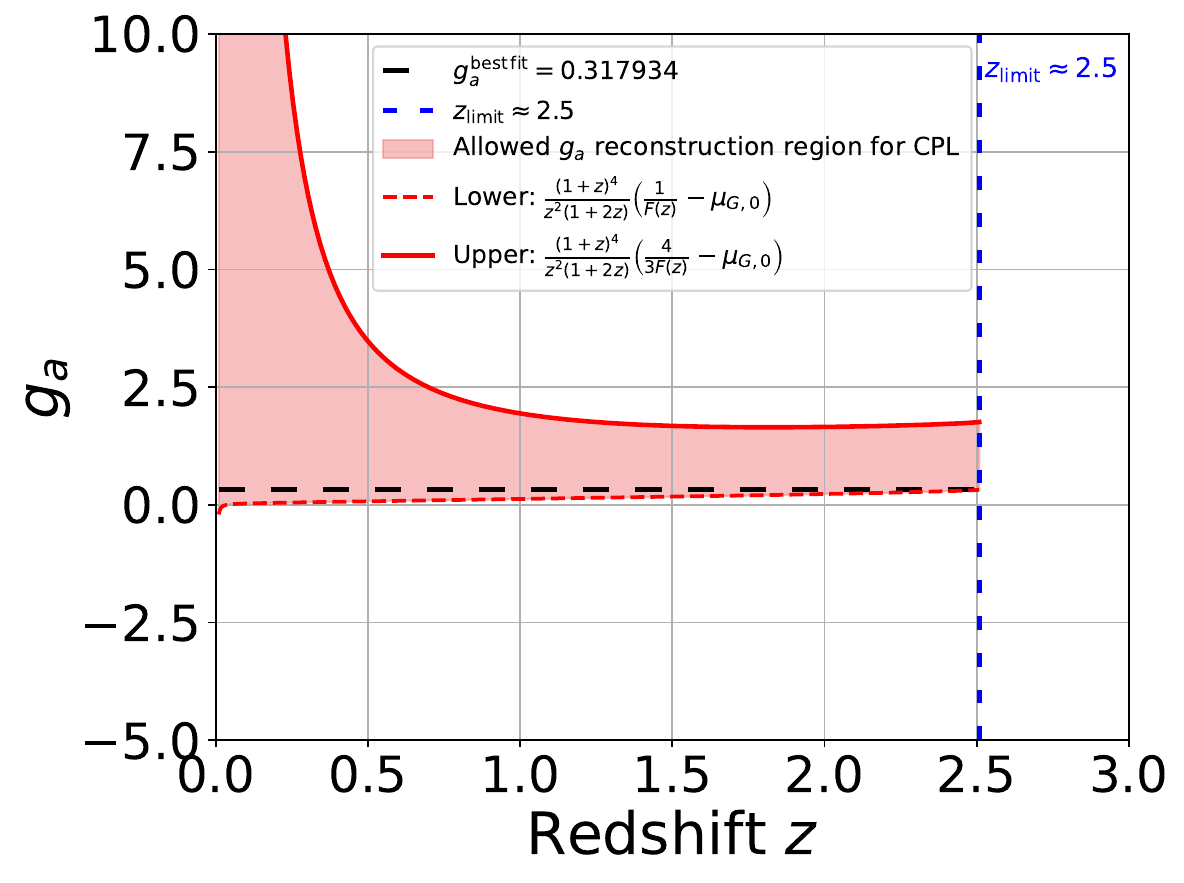}
    \caption{The dashed line indicates the constant best-fit value $g_a$ for the CPL parametrization. The shaded region represents the portion of parameter space (for the CPL parametrization) defined by Eq.(\ref{inequalityga}) where a consistent reconstruction of the scalar-tensor theory under consideration is possible. Note that the reconstruction becomes invalid beyond $z = z_{\rm limit}$, since at this redshift $\mu_{\rm G}(z_{\rm limit}) = 1/F(z_{\rm limit})$. The dataset combination used here is: Pantheon+, CMB, DESI DR2, RSD, and BBN constraint. }
    \label{fig:boundary}
\end{figure}

As can be seen in Figs.(\ref{Fig:2},\ref{fig:3},\ref{fig:boundary},\ref{fig:Fz_Uz_plots},\ref{fig:Fphi_Uphi_plots}),  using the phenomenological, scale--independent parametrization of Eq.(\ref{pantazisparam}) together with the quasi--static subhorizon expression for the effective gravitational coupling, Eq.(\ref{gefftheory}), we successfully reconstructed the functions $F(\Phi)$, $\Phi(z)$, and $U(\Phi)$ within the redshift range $0 \lesssim z \lesssim 2$. Since the reconstruction is reliable only up to $z \sim 2$, and not as far as $z = 1100$, we neglect radiation terms in all the above equations of scalar--tensor gravity. This reconstruction provides an effective, data--driven mapping between the observed expansion and growth histories and the underlying scalar--tensor degrees of freedom. It should, however, be interpreted as a local and phenomenological description of the theory: the parametrization assumes a massless (or very light scalar field), neglects any scale dependence of $G_{\mathrm{eff}}$, and relies on the quasi--static approximation valid only for linear, subhorizon modes. Consequently, the reconstructed $F$ and $U$ represent an effective late--time evolution  consistent with current large--scale structure and background data, but they do not uniquely specify the fundamental form of the underlying scalar--tensor Lagrangian or guarantee validity beyond the fitted redshift range.

By using Eq.(\ref{pantazisparam}) for $n=2$ and defining the shorthand $  x(z) \equiv (1+z)^4/(1+2z)z^2$ for the coefficient appearing in the inequality of Eq.\eqref{inequalitygen} after manipulation we get the following
\begin{equation}\label{inequalityga}
    x(z)\left(\frac{1}{F}-\mu_{\rm G,0}\right)< g_a\leq x(z)\left(\frac{4}{3 F}-\mu_{\rm G,0}\right)
\end{equation}
Using the best-fit $g_a$ parameter from Table~\ref{tab:fs8_constraint} for the CPL parametrization, we find that the allowed region ends at a limiting redshift of approximately $z_{\rm limit} \approx 2.5$, as illustrated in Figure~\ref{fig:boundary}.
  At this point, the best-fit line $g_a = \text{const.}$ of the inequality intersects the lower boundary, corresponding to the condition $\mu_G(z_{\rm limit}) F(z_{\rm limit}) = 1$. Beyond this redshift, the differential equation~\eqref{eqnumerical} develops a singularity, causing the reconstruction procedure to break down. Furthermore, note that $U(z)$ is related to $F''(z)$ via Eq.~\eqref{EqU}. As a result, the reconstruction of $U(z)$  begins to break down even before this redshift is reached. Consequently, the results as one approaches $\mu_G F\to 1$, where errors become extremely large, should not be considered physically meaningful.  Comparable limiting redshifts are obtained for the other dark energy parametrizations (BA, Logarithmic, and JBP).

It is important to note that this breakdown does not necessarily invalidate the scalar--tensor class of theories themselves. Throughout the analysis, several simplifying assumptions have been made---for example, the use of a scale--independent parametrization for $\mu_G(z)$ in Eq.\eqref{pantazisparam}, the assumed consistency between the phenomenological, scale--independent parametrization of Eq.\eqref{pantazisparam} and the quasi--static subhorizon expression for the effective gravitational coupling in Eq.\eqref{gefftheory}, as well as the adoption of the fitting relation in Eq.\eqref{eq:luminosity_scaling}, which links the luminosity ratio to $\mu_G$ in the absence of a screening mechanism. These assumptions may, however, introduce additional limitations to the interpretation of the results.

Therefore, further investigation is required before drawing definitive conclusions, such as ruling out this scalar-tensor framework as a viable approach to alleviating existing cosmological tensions. Nevertheless, in scenarios where the underlying assumptions and approximations hold reasonably well, it would be worthwhile to extend the analysis to reconstructions of more general or alternative classes of modified gravity theories.

\subsection{\label{subsec:results_redshift}Reconstructed Functions in Redshift Space}

\begin{figure*}[htbp]
    \centering
    \begin{subfigure}{\columnwidth}
        \centering
        \includegraphics[width=\columnwidth]{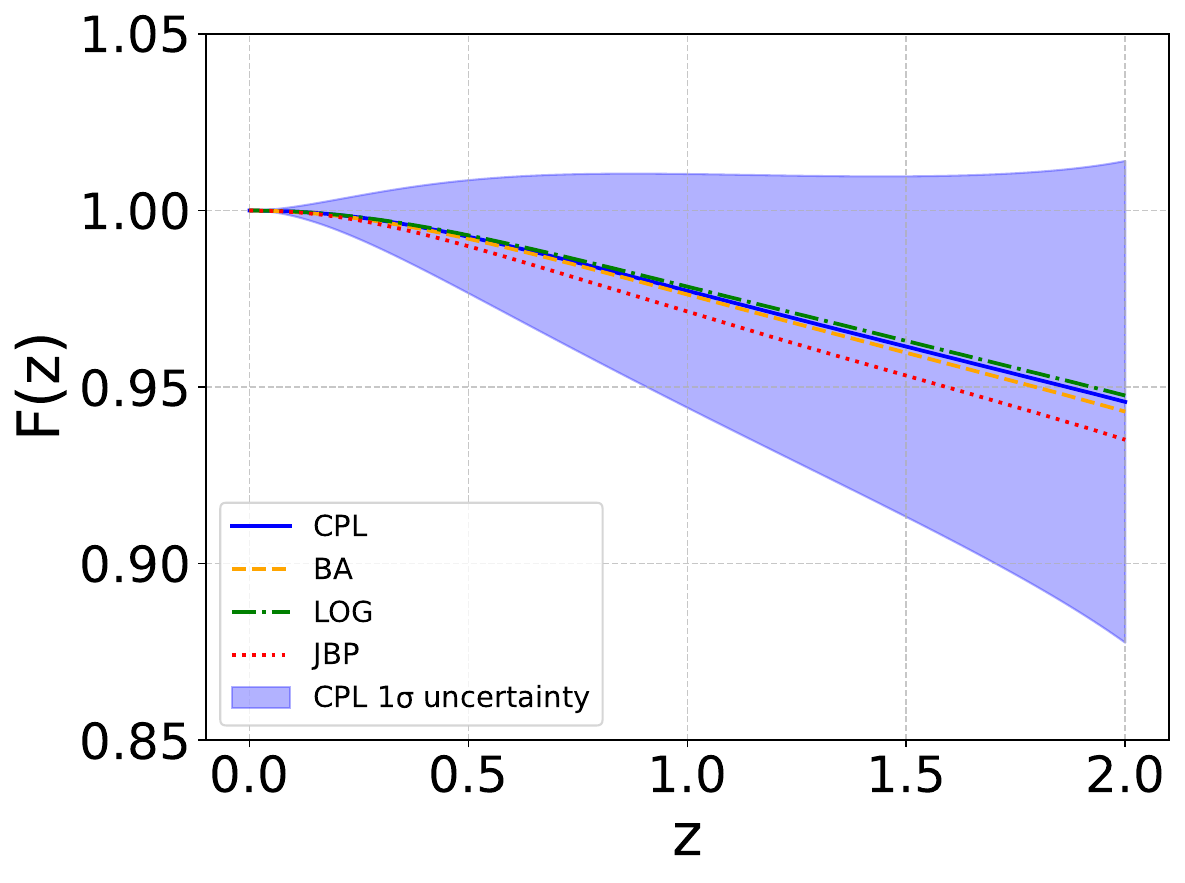}
        \caption{The reconstructed coupling function $F(z)$ showing deviations from the General Relativity value $F = 1$ at intermediate redshifts, with convergence toward unity at both $z = 0$ (enforced by boundary conditions). All four parametrizations predict $F(z) > 0$ throughout the valid range, with 3--5\% variations indicating weak modified gravity effects.}
        \label{fig:Fz_plot}
    \end{subfigure}
    \hfill
    \begin{subfigure}{\columnwidth}
        \centering
        \includegraphics[width=\columnwidth]{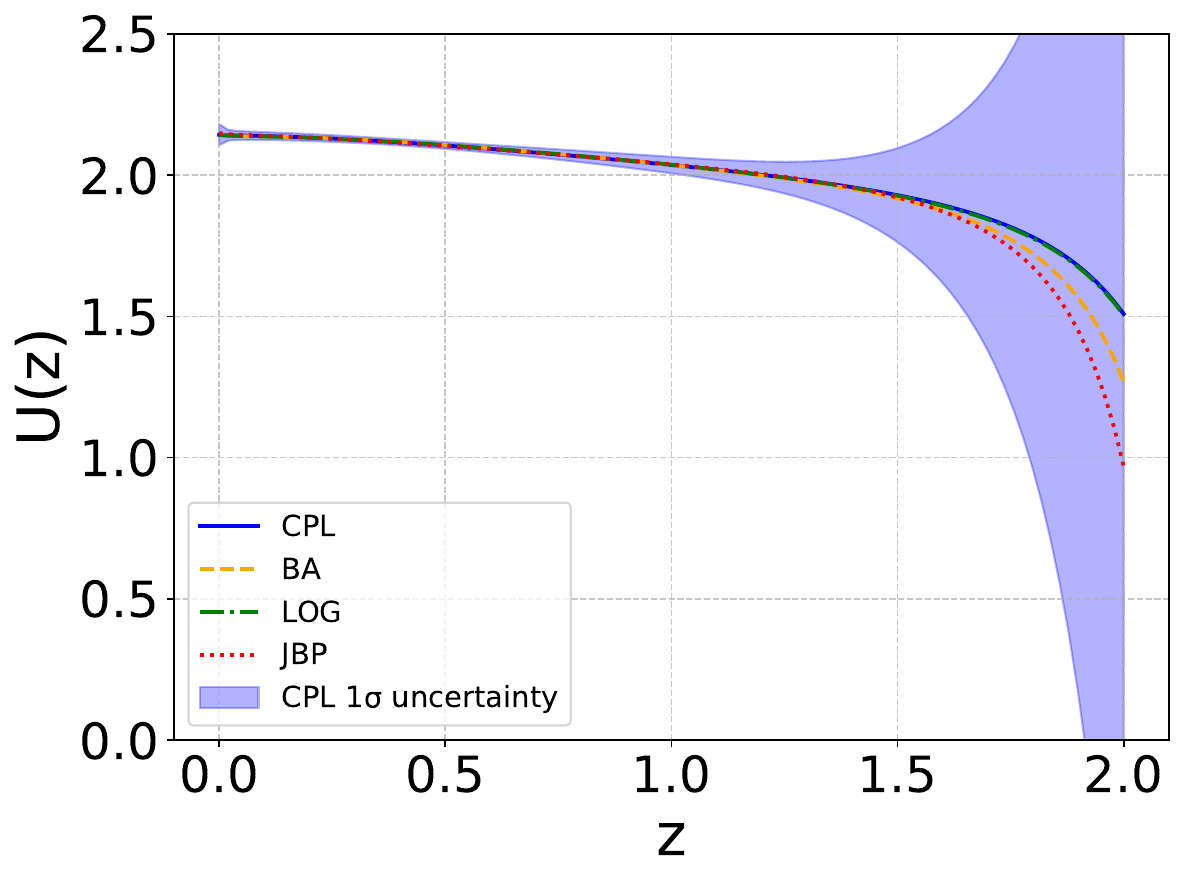}
        \caption{The reconstructed potential $U(z)$ in units of $H_0^2$, exhibiting monotonic decrease with increasing redshift.}
        \label{fig:Uz_plot}
    \end{subfigure}
    
    \vspace{1em}
    
    \begin{subfigure}{\columnwidth}
        \centering
        \includegraphics[width=\columnwidth]{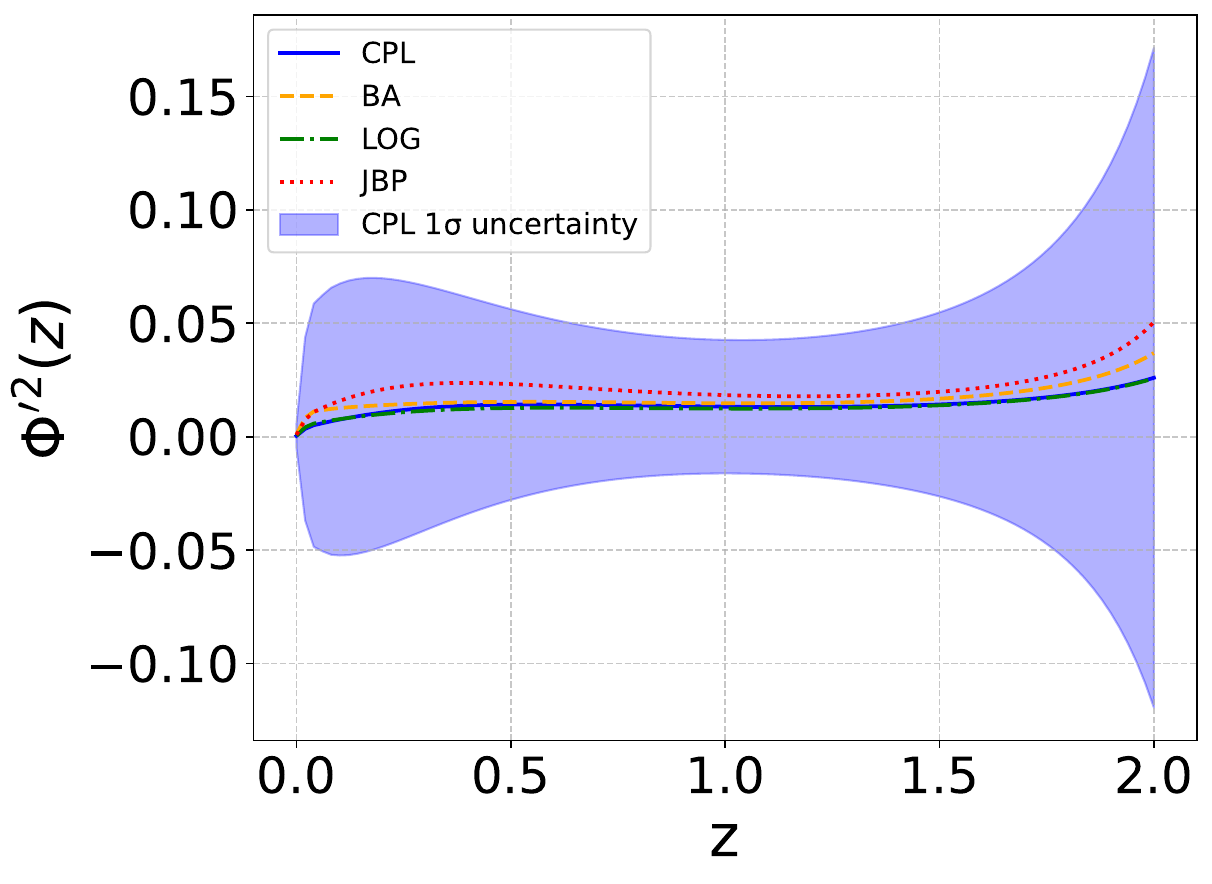}
        \caption{The kinetic term $\Phi'^2(z)$ demonstrating positivity throughout the reconstruction range $0 \leq z \lesssim 2$, confirming the physical viability of all four parametrizations. The kinetic energy increases toward higher redshift, indicating accelerated scalar field evolution in the past.}
        \label{fig:Phi2_plot}
    \end{subfigure}
    
    \caption{The reconstructed scalar-tensor functions as functions of redshift for the four dark energy parametrizations (CPL: solid blue, BA: dashed orange, Logarithmic: dot-dashed green, JBP: dotted red). Shaded regions represent $1\sigma$ confidence bands of CPL parametrization. All reconstructions satisfy the viability constraints $F(z) > 0$ and $\Phi'^2(z) \geq 0$ throughout the valid redshift range. The dataset combination used here is: Pantheon+, CMB, DESI DR2, RSD, and BBN constraint. }
    \label{fig:Fz_Uz_plots}
\end{figure*}

The fundamental scalar–tensor functions reconstructed from the derived $\mu_{G}(z)$ and $q(z)$ are the redshift-dependent quantities $\{F(z), U(z), \Phi'^{2}(z), \Phi(z)\}$, obtained for the four dark energy parametrizations: CPL, BA, Logarithmic, and JBP. 
Figure~\ref{fig:Fz_Uz_plots} shows the reconstructed coupling function $F(z)$ (panel a), scalar potential $U(z)$ (panel b), and kinetic term $\Phi'^2(z)$ (panel c) as functions of redshift. 
The shaded regions indicate the $1\sigma$ uncertainties for the CPL case, while all panels include results for the four parametrizations for direct comparison.

Panel~(a) of Figure~\ref{fig:Fz_Uz_plots} shows the reconstructed coupling function $F(z)$ in the range $0 \leq z \lesssim 2$, quantifying deviations from General Relativity via the effective gravitational coupling. By construction, all models satisfy $F(0) = 1$, ensuring consistency with local tests. At higher redshifts, $F(z)$ systematically departs from unity with similar qualitative trends across parametrizations. For example, in the CPL model (solid blue curve), $F(z)$ decreases from $F(0) = 1$ to $F(z \sim 1) \approx 0.97$--$0.975$, implying a slightly stronger gravity in the past.

Panel~(b) of Figure~\ref{fig:Fz_Uz_plots} displays the rescaled scalar potential $U(z)$. At $z = 0$, all parametrizations yield $U(0) \simeq 2.10$--$2.15$. With increasing redshift, $U(z)$ declines monotonically, reaching $U(z \sim 2) \simeq 1$--$1.5$ for four models. This behaviour indicates that, as we approach the present epoch, dark energy progressively dominates over matter.

\begin{figure*}[htbp]
    \centering
    \begin{subfigure}{\columnwidth}
        \centering
        \includegraphics[width=\columnwidth]{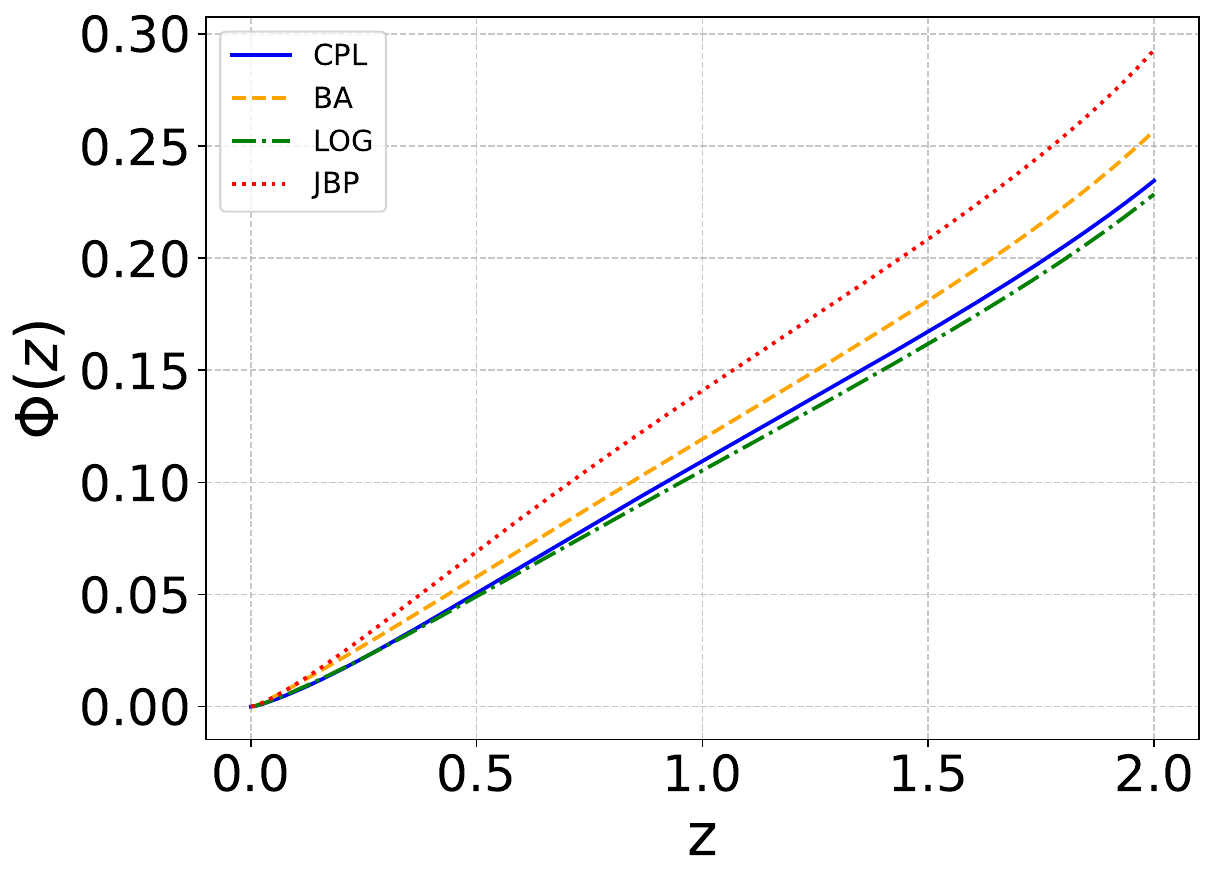}
        \caption{The scalar field $\Phi(z)$ obtained by integrating the kinetic term from Figure~\ref{fig:Phi2_plot}. The monotonic increase from $\Phi(0) = 0$ to $\Phi(z \sim 2) \sim 0.25$--$0.3$ confirms invertibility, enabling the transformation to field-dependent functions $F(\Phi)$ and $U(\Phi)$.}
        \label{fig:Phiz_plot}
    \end{subfigure}
    \hfill
    \begin{subfigure}{\columnwidth}
 \centering
        \includegraphics[width=\columnwidth]{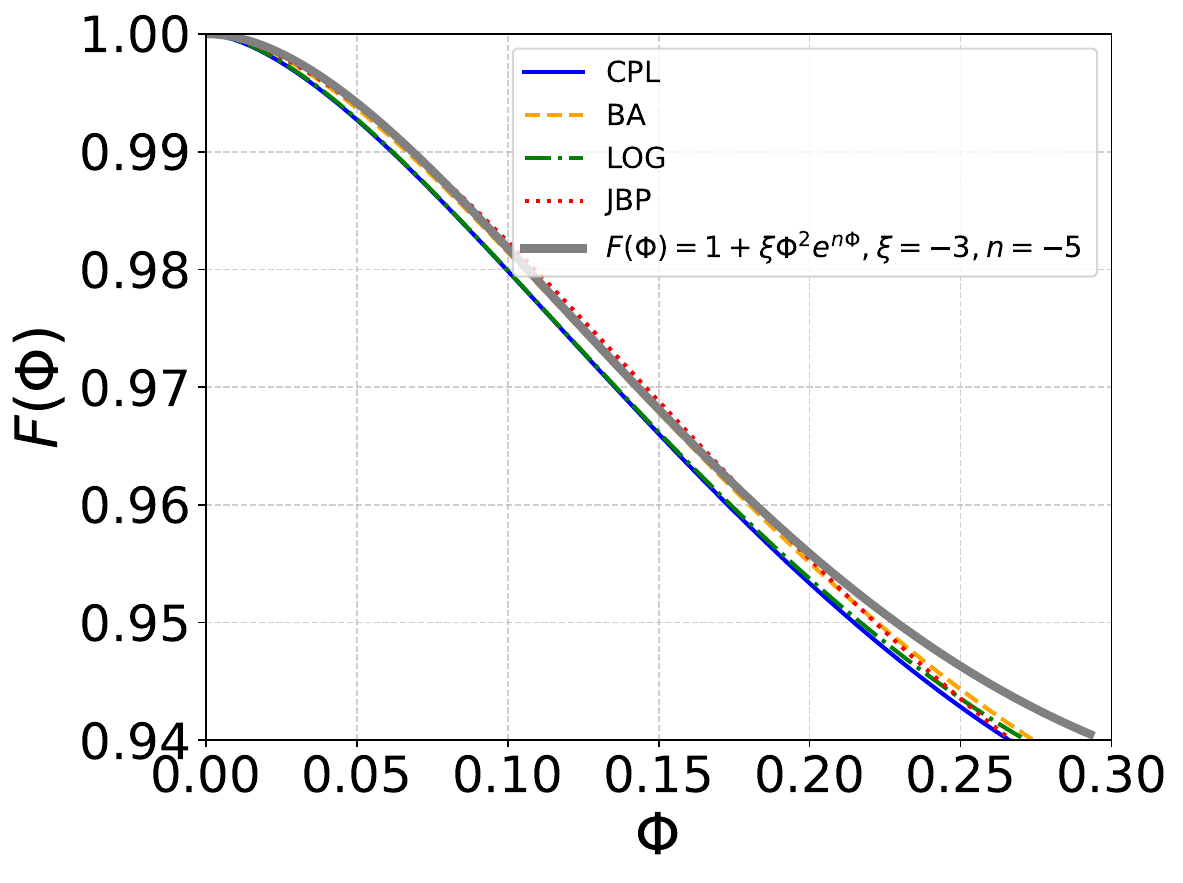}
        \caption{The coupling function $F(\Phi)$ obtained by inverting $\Phi(z)$ and substituting into $F(z)$, for the four different dark energy parameterizations. With the thick gray line the fitting functional form is shown.}
        \label{fig:Fphi_plot}
    \end{subfigure}
    
    \vspace{1em}
    
    \begin{subfigure}{\columnwidth}
        \centering
    \includegraphics[width=\columnwidth]{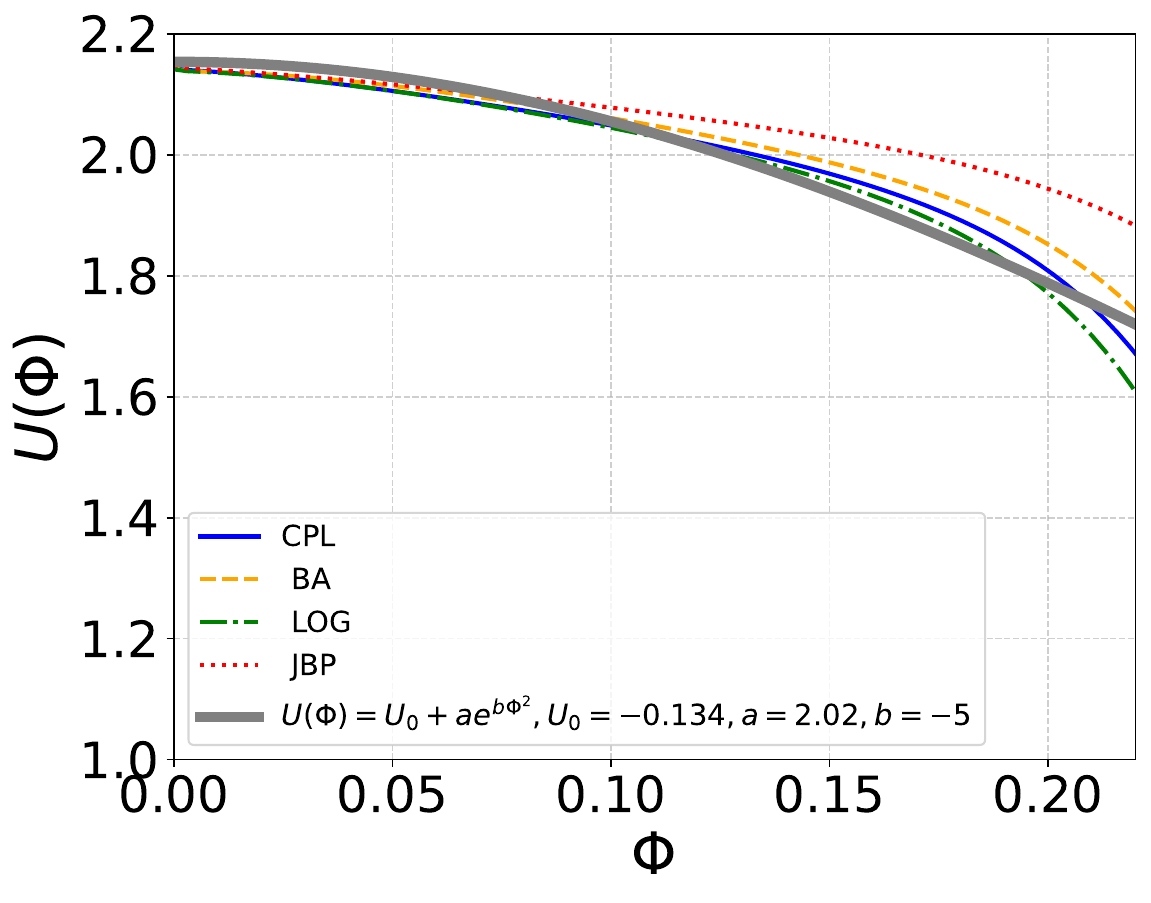}
        \caption{The potential $U(\Phi)$ obtained by inverting $\Phi(z)$ and substituting into $U(z)$ for the four different dark energy parameterizations. With the thick gray line the fitting functional form is shown.}
        \label{fig:Uphi_plot}
    \end{subfigure}
    
    \caption{The scalar field evolution and fundamental Lagrangian functions. Panel (a) shows $\Phi(z)$, confirming monotonic evolution enabling inversion. Panels (b) and (c) present the reconstructed $F(\Phi)$ and $U(\Phi)$ that define the scalar-tensor theory in the Jordan frame. The dataset combination used here is: Pantheon+, CMB, DESI DR2, RSD, and BBN constraint.}
    \label{fig:Fphi_Uphi_plots}
\end{figure*}

Panel~(c) presents the kinetic term $\Phi'^2(z)$ for all models. As shown, $\Phi'^2(z) \geq 0$ throughout the range $0 \leq z \lesssim 2$, confirming physical consistency. The function $\Phi'(z)$ remains close to zero up to $z \approx 1.75$, indicating minimal scalar field evolution at the present epoch. Beyond this point, it increases for all parametrizations—most notably for the JBP model (red dotted line), which reaches $\sim 0.05$ at $z = 2$.
Figure~\ref{fig:Phiz_plot} shows the reconstructed scalar field $\Phi(z)$, obtained by integrating the reconstructed $\Phi'^2(z)$:
\begin{equation}
    \Phi(z) = \pm \int_0^z \sqrt{\Phi'^2(z')}\,dz',
    \label{eq:Phi_integration}
\end{equation}
with $\Phi(0) = 0$. We adopt the positive branch for plotting. The field increases monotonically with redshift, reaching $\Phi(z \sim 2) \simeq 0.01$--$0.05$.

\subsection{\label{subsec:lagrangian_functions}Reconstructed Lagrangian and Fitting Functions}
\begin{figure}
    \centering
    \includegraphics[width=1\linewidth]{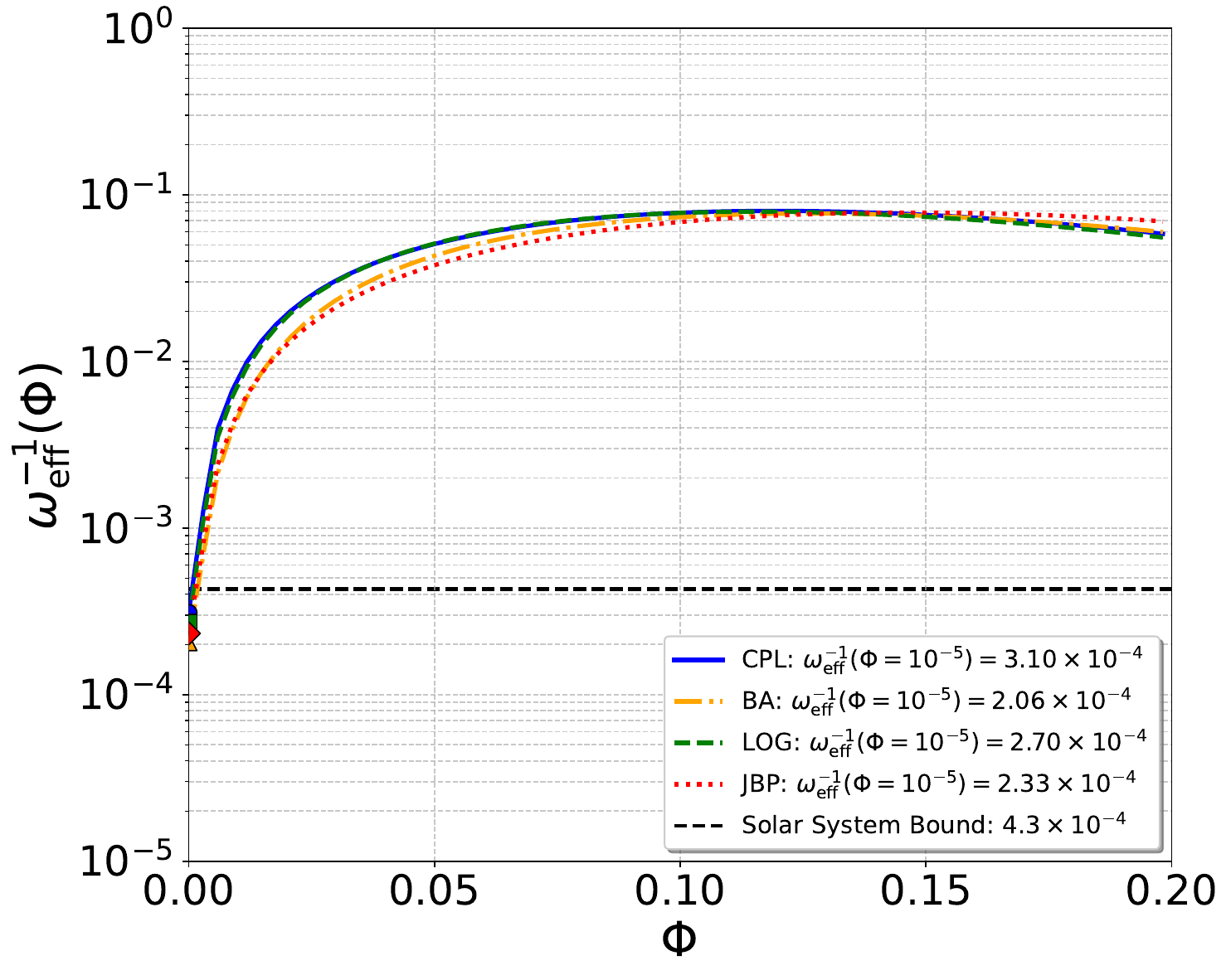}
    \caption{The evolution of $\omega_{\rm eff}^{-1}$ as a function of $\Phi$, for the dynamical dark energy models (MG), demonstrates consistency with the solar system constraint, namely $\omega_{\rm eff}^{-1}(0.00001) < 4.3 \times 10^{-4}$~\citep{Boisseau:2000pr,Amendola:2015ksp}. The dataset combination used here is: Pantheon+, CMB, DESI DR2, RSD, and BBN constraint.}
    \label{fig:solarsystemconstraints}
\end{figure}

While the redshift-dependent functions $\{F(z), U(z)\}$ in Section~\ref{subsec:results_redshift} show how observational data can constrain them, it is the field-dependent functions $F(\Phi)$ and $U(\Phi)$ that enter the Lagrangian density in Eq.\eqref{eq:lagrangian}.

Panels (b) and (c) of Figure~\ref{fig:Fphi_Uphi_plots} display the reconstructed coupling function $F(\Phi)$ and scalar potential $U(\Phi)$, obtained by numerically inverting the monotonic function $\Phi(z)$ from panel (a) and substituting $z(\Phi)$ into $F(z)$ and $U(z)$ from Figure~\ref{fig:Fz_Uz_plots} via 
\begin{align}
F(\Phi) &= F(z(\Phi)), \label{eq:F_Phi_construction}\\
U(\Phi) &= U(z(\Phi)). \label{eq:U_Phi_construction}
\end{align}

Figure \ref{fig:Fphi_plot} shows the reconstructed coupling function $F(\Phi)$, which determines how the effective strength of gravity varies with the scalar field configuration. At the present-day field value $\Phi = 0$, all four parametrizations satisfy $F(0) = 1$ by construction. As the field increases (moving backward in cosmic time), $F(\Phi)$ decreases approximately linearly, reaching $F(\Phi_{\rm limit}) \sim 0.94$ depending on the dark energy parametrization. The modest magnitude of the coupling variations---$\Delta F \sim 6\%$ over the accessible field range---is a direct consequence of Solar System constraints. Note that given the parametrization of $\mu_G$ the $dF/d\Phi|_{\Phi=0}=0$, leading to  Brans-Dicke\footnote{It is  convenient to recast a general scalar--tensor theory into a Brans--Dicke (BD)--like form in order to interpret its physical content in terms of a varying gravitational coupling and to compare with observational constraints.
Defining a new field $\psi \equiv F(\Phi)$ and using the chain rule
$\partial_{\mu}\psi = (dF/d\Phi)\,\partial_{\mu}\Phi$, the kinetic term becomes
$-\tfrac{1}{2}(\partial\Phi)^2
= -\tfrac{1}{2[dF(\Phi)/d\Phi]^{2}}(\partial\psi)^2$.
Writing this in BD form of the kinetic term of the Lagrangian in Eq.(\ref{eq:lagrangian}) i.e.
$-\tfrac{\omega_{\rm eff}(\Phi)}{2\psi}(\partial\psi)^2$,
one identifies the effective Brans--Dicke parameter as
$\omega_{\rm eff}(\Phi)=
\frac{F(\Phi)}{[dF(\Phi)/d\Phi]^{2}}$.
This correspondence helps to clarify how the theory deviates from General Relativity and the resulting observational consequences. As shown in Fig.\ref{fig:solarsystemconstraints}, the upper bounds from solar system measurements indicate that $(\gamma - 1) < 2.3 \times 10^{-5}$~\citep{Hoyle:2004cw}, which correspond to $\omega_{\rm eff}^{-1}(0) < 4.3 \times 10^{-4}$~\citep{Boisseau:2000pr,Amendola:2015ksp}.}  parameter $\omega_{\rm eff}^{-1}\to 0$ (see also Fig.\ref{fig:solarsystemconstraints}).

 A convenient fitting formula  for the reconstructed $F(\Phi)$ is given by the following function:
\begin{equation}
    F(\Phi) = 1 +\xi\Phi^2e^{n\Phi},
\end{equation}
which is shown in Fig.\ref{fig:Fphi_plot} (thick gray line), plotted for $\xi=-3$ and $n=-5$.  An interesting feature of this function is that $F(\Phi)$ approaches unity both at $\Phi = 0$ and asymptotically as $\Phi \to \infty$, and that its first derivative with respect to $\Phi$ vanishes at $\Phi = 0$ and also asymptotically as $\Phi \to \infty$.

In addition, the fitting function for the scalar potential \(U(\Phi)\), shown as the gray solid line, is expressed as
\begin{equation}
   U(\Phi) = U_0+ a e^{b\Phi^2}.
\end{equation}
Figure~\ref{fig:Uphi_plot} displays the reconstructed scalar potentials \(U(\Phi)\) corresponding to each dark energy parametrization, with the fitting functional form plotted for $U_0=-0.134$, $a=2.02$ and $b=-5$.
 Note that, in order to perform the fitting procedure, we assume specific functional forms for $F(\Phi)$ and $U(\Phi)$, and study them systematically. For instance, in the case of $F(\Phi)$, we construct the following quantity:
\begin{equation*}
    \chi^2 = \sum_i \frac{\left[F(\Phi(z_i)) - F^{\mathrm{model}}(\Phi(z_i))\right]^2}{\sigma_i^2},
\end{equation*}
where $F(\Phi(z_i))\equiv F(z_i)$ represents the reconstructed values of the function in the range $0 \lesssim z \lesssim 2$, and $\sigma_i$ denotes the corresponding $1\sigma$ uncertainty of each $F(z_i)$. We then define the total chi-squared as
\begin{equation*}
    \chi^2_{\mathrm{tot}} = \chi^2_{\mathrm{CPL}} + \chi^2_{\mathrm{BA}} + \chi^2_{\mathrm{Log}} + \chi^2_{\mathrm{JBP}},
\end{equation*}
which is subsequently minimized. Although, in the case of $U(\Phi)$, we repeat the fitting procedure, we instead use a uniform error, since the very large $1\sigma$ uncertainties calculated at higher redshifts (around $z=2$) tend to obscure the fit.


\section{Conclusion}\label{sec:conclusion}

In this work, we have established a systematic reconstruction framework for Modified Gravity (MG) theories, demonstrating that they can provide a viable and theoretically consistent explanation for the mild deviations from the $\Lambda$CDM paradigm suggested by recent cosmological observations. Our reconstruction framework suggests that the class of viable MG theories---those that can, at least approximately, be described by the scale-independent parametrization of $G_{\mathrm{eff}}$ in Eq.~(\ref{pantazisparam})---is capable of reproducing cosmic evolution by invoking an enhanced gravitational strength over the redshift range $z \sim 0 - z_{\rm CMB}$. This holds provided that the effective dark energy component exhibits a prolonged phantom-like behavior and possibly undergoes a phantom crossing, consistent with tentative indications from current observational data. 
The alleviation of cosmological tensions originates from a phase in which enhanced gravitational strength coexists with phantom-like dark energy. These effects act in opposition: the stronger gravity promotes structure growth, while the phantom dark energy suppresses it.
To reconcile the observed structure formation and supernova luminosities, MG models require the epoch of phantom behavior to persist slightly longer than in General Relativity with a purely dynamical dark energy component. 
This extended phantom phase naturally leads to a higher inferred value of $H_0$. 
Nonetheless, we stress that our models do not fully resolve the Hubble tension, but only alleviate it, i.e., when considering the dataset combination (RSD+DESI DR2+BBN+CMB+Pantheon+), the MG models provide a better fit than CPL and $\Lambda$CDM and yield a slightly higher value of $H_0 \simeq 70.6\pm1.4\,\mathrm{km\,s^{-1}\,Mpc^{-1}}$ compared to dynamical dark energy and $\Lambda$CDM. Although a complete resolution is not achieved within this framework, the tension is reduced for modified gravity models characterized by a $\mu_G(z)$-like behavior.

Specifically, by analyzing the DESI~DR2 dataset, the Pantheon+ Type~Ia supernova compilation, Cosmic Microwave Background (CMB) measurements, growth-rate (RSD) data, together with the Big Bang Nucleosynthesis constraint, we numerically reconstruct observationally motivated parametrizations of the effective gravitational coupling $G_{\rm eff}$ through Eq.~(\ref{pantazisparam}), as well as the dynamical dark energy equation of state $w(z)$, adopting the CPL, BA, Log, and JBP parametrizations.
A key finding of our analysis is the identification of a  quasi-degeneracy between the matter fluctuation amplitude $\sigma_8$ and the modified gravity parameter $g_a$.   
We propose a potential method to break this quasi-degeneracy by incorporating an independent constraint on the effective gravitational constant $G_{\mathrm{eff}}$ via its impact on Type Ia supernova luminosity. By reanalyzing the Pantheon+ dataset with the luminosity scaling relation $L/L_0 \simeq (G_{\mathrm{eff}}/G_N)^{1.46}$, we introduced a direct constraint on $\mu_G$ independent of growth-rate data, which successfully resolves the parameter degeneracy.

From Table~\ref{tab:H0_comparison}, the standard $\Lambda$CDM model with the combined datasets (RSD+DESI DR2+BBN+CMB+Pantheon+) yields $H_0 = 68.8 \pm 0.3\,\mathrm{km\,s^{-1}\,Mpc^{-1}}$, about $4.4\sigma$ below the local Pantheon+ result ($H_0 = 73.4 \pm 1.0$). The CPL (GR) model gives $H_0 = 69.1 \pm 0.6$, a modest $3.7\sigma$ improvement.   
In contrast, CPL-MG models predict $H_0 \simeq 70.6 \pm 1.4$, $\sim 2.2\sigma$ above the Planck $\Lambda$CDM value and $\sim 1.4\sigma$ below SH0ES, in excellent agreement with TRGB ($H_0 = 69.8 \pm 1.9$). Other MG parametrizations (BA, JBP) similarly yield $H_0 \sim 70.6$–$70.8$ with larger uncertainties ($\sim1.0$–$1.6$), shifting the inferred $H_0$ closer to local measurements and alleviating the Hubble tension (Fig.~\ref{fig:hubble_tension}).

The $S_8$ parameter, defined as 
$S_8 \equiv \sigma_8(\Omega_{m0}/0.3)^{0.5}$,
and it quantifies the amplitude of matter fluctuations on $8\,h^{-1}$ Mpc scales and is particularly sensitive to modifications of gravity. 
Table~\ref{tab:S8_comparison} presents a comparison of values from various cosmological models. When all datasets are combined (CMB, DESI DR2, Pantheon+, RSD, BBN), the dynamical dark energy model (CPL) yields \( S_8 = 0.777 \pm 0.043 \), whereas the standard \(\Lambda\)CDM model results in \( S_8 = 0.798 \pm 0.043 \). 
The additional degree of freedom, arising from allowing the gravitational strength to deviate over the redshift range $z \sim 0 - z_{\rm CMB}$ and from the extended phantom-like phase, effectively compensates for the lower fluctuation amplitude, $\sigma_8 \approx 0.78$, compared to the $\Lambda$CDM value of $\sigma_8 \approx 0.80$. As a result, the MG models successfully reproduce the observed $f\sigma_8(z)$ data and consistently yield a lower best-fit value of $S_8 \simeq 0.76 \pm 0.05$.

By revisiting the reconstruction using a specific class of scalar-tensor gravities defined by Eq.(\ref{eq:lagrangian}) in the Jordan frame, as discussed in previous works \citep{Boisseau:2000pr,Esposito-Farese:2000pbo,Perivolaropoulos:2005yv,Nesseris:2017vor,Kazantzidis:2019nuh}, we further reconstruct the scalar field $\Phi(z)$, the scalar potential $U(z)$, and the coupling function $F(z)$ in redshift space. As shown in Figs. \ref{Fig:2},\ref{fig:3},\ref{fig:boundary}--\ref{fig:Fphi_Uphi_plots}, using the phenomenological, scale-independent parametrization of Eq.(\ref{pantazisparam}) together with the quasi-static subhorizon expression for the effective gravitational coupling, Eq.(\ref{gefftheory}), we successfully reconstructed $F(\Phi)$, $\Phi(z)$, and $U(\Phi)$ within the redshift range $0 \lesssim z \lesssim 2$. This reconstruction provides a data-driven mapping between observed expansion and growth histories and the underlying scalar-tensor degrees of freedom. Functional forms that provide good fits to the reconstructed coupling function and scalar potential are given by  
\begin{align}
    F(\Phi) &=1 +\xi\Phi^2e^{n\Phi}, \nonumber \\
   U(\Phi) &= U_0+ a e^{b\Phi^2},
\end{align}
which together yield accurate representations of the numerically reconstructed functions within the range $0 \lesssim z \lesssim 2$.

It is important to emphasize that these limitations do not definitively invalidate the scalar–tensor class of theories defined in Eq.(\ref{eq:lagrangian}). 
Throughout our analysis, several simplifying assumptions were adopted. 
For instance, the scale-independent parametrization of $\mu_G(z)$ in Eq.(\ref{pantazisparam}) is set equal to Eq.(\ref{gefftheory}), which corresponds to a massless (or very light) scalar field, neglects any scale dependence of $G_{\mathrm{eff}}$, and relies on the quasi-static approximation, valid only for linear subhorizon modes. 
Additionally, the luminosity scaling relation in Eq.(\ref{eq:luminosity_scaling}) is assumed in the absence of potential screening mechanisms.

Accordingly, the reconstruction of the fundamental functions $F(\Phi)$ and $U(\Phi)$ should be interpreted as a local and phenomenological description of the theory over the redshift range $0 \lesssim z \lesssim 2$. The reconstructed functions represent an effective late-time evolution consistent with current large-scale structure and background data, but they do not uniquely determine the fundamental form of the underlying scalar-tensor Lagrangian, nor do they guarantee validity beyond the fitted redshift range.

Further investigation is therefore required before drawing definitive conclusions, such as ruling out the scalar-tensor framework defined by Eq.(\ref{eq:lagrangian}) as a viable model for explaining observational data and existing cosmological tensions. Nevertheless, in scenarios where these assumptions hold reasonably well, extending the reconstruction analysis to other classes of modified gravity theories represents a promising direction for future work. Future work should explore alternative parameterizations of $\mu_G(z)$\footnote{Specifically, a systematic study of additional parameterizations of $G_{\rm eff}$ would be valuable in assessing the extent to which our results may be biased by the particular functional form of the effective gravitational constant assumed, as well as in exploring other classes of modified gravity theories.}, revisit our underlying assumptions to assess their limitations, and investigate other modified gravity scenarios to test the robustness and generality of these findings.

\section*{Data Availability Statement}

The figure-generating Mathematica (v13) and Python (3.12.12.) code will be publicly accessible in the GitHub repository and licensed under MIT in \citep{Mathematicafile}.

\begin{acknowledgments}
We are grateful to Savvas Nesseris for sharing his code and for the subsequent helpful discussions. We are also grateful to Theodoros Katsoulas for fruitful conversations regarding this work, and to Eoin Ó Colgáin for his valuable feedback on the manuscript. We also thank Mikel Artola for carefully reading our manuscript, reproducing our results, and for his helpful feedback. This research was supported by COST Action CA21136 - Addressing observational tensions in cosmology with systematics and fundamental physics (CosmoVerse), supported by COST (European Cooperation in Science and Technology). Furthermore, Dimitrios Efstratiou was supported by a scholarship from the University of Ioannina under the project ``Enhancement and Support of the Operational, Research, and Educational Activities of the University of Ioannina'' (Project Code: 82985/144059/$\beta$6.$\varepsilon$).
\end{acknowledgments}

\appendix

\section{$\dot G_{\text{eff}}/G_{\text{eff}}$ in CPL Cosmology}

\begin{figure}
    \centering
    \includegraphics[width=1\linewidth]{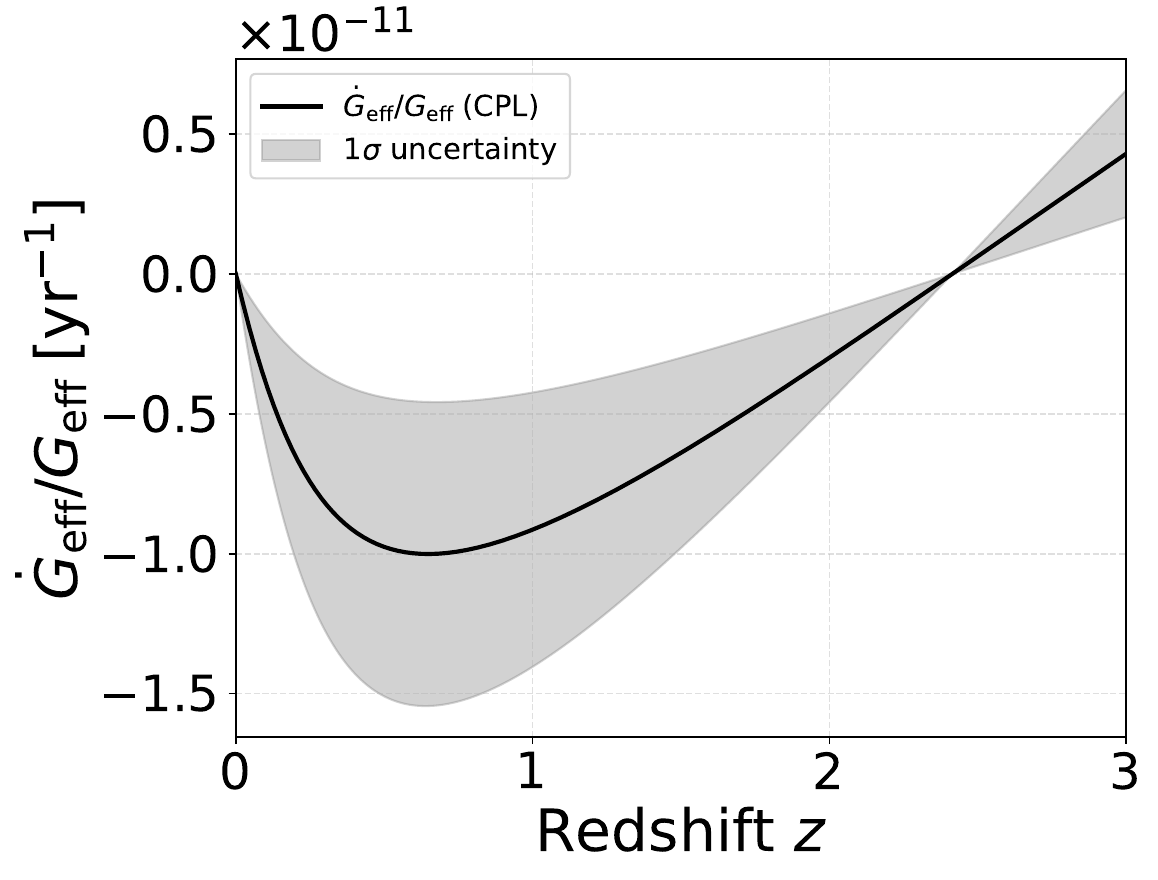}
    \caption{The quantity $\dot G_\text{eff}/G_\text{eff}$ plotted for the CPL parametrization with respect to the redshift $z$, along with its $1\sigma$ uncertainty. The dataset combination used here is: Pantheon+, CMB, DESI DR2, RSD, and BBN constraint.}
    \label{dotGeff_Geff_CPL}
\end{figure}

\begin{table*}
\caption{Comparison between observational constraints on the variation of the gravitational constant \citep{Alestas:2021nmi} and the CPL model predictions. The second column lists the characteristic timescales of each probe (as in the observational compilation), and the third column shows the corresponding redshift $z_{\rm CPL}$ obtained from the CPL cosmology. The fourth column gives the maximum observational limits on $|\dot G_{\rm eff}/G_{\rm eff}|$, while the fifth shows the CPL prediction at that epoch.}
\label{tab:comparison_methods}
\setlength{\tabcolsep}{5pt}
\renewcommand{\arraystretch}{1.15}
\begin{tabular}{lccccc} 
\hline\hline
Method & Time Scale [yr] & $z_{\rm CPL}$ & $|\dot G_{\rm eff}/G_{\rm eff}|_{\rm obs}^{\rm max}$ [yr$^{-1}$] & $|\dot G_{\rm eff}/G_{\rm eff}|_{\rm CPL}$ [yr$^{-1}$] & References \\
\hline
Lunar ranging                    & $24$                & $1.73\times10^{-9}$ & $7.1\times10^{-14}$   & $7.92\times10^{-20}$  & \cite{Hofmann:2018myc} \\
Solar system                     & $50$                & $3.61\times10^{-9}$ & $4.6\times10^{-14}$   & $1.65\times10^{-19}$  & \cite{Pitjeva:2013xxa,Pitjeva:2021hnc} \\
Pulsar timing                    & $1.5$               & $1.08\times10^{-10}$ & $3.2\times10^{-13}$   & $4.95\times10^{-21}$  & \cite{Bussieres:2025aml} \\
Strong lensing (local)           & $0.6$               & $4.33\times10^{-11}$ & $10^{-2}$             & $1.98\times10^{-21}$  & \cite{Giani:2020fpz} \\
Orbits of binary pulsar          & $22$                & $1.59\times10^{-9}$ & $1.0\times10^{-12}$   & $7.26\times10^{-20}$  & \cite{Zhu:2018etc} \\
Ephemeris of Mercury             & $7$                 & $5.05\times10^{-10}$ & $4\times10^{-14}$     & $2.31\times10^{-20}$  & \cite{Genova:2018} \\
Exoplanetary motion              & $4$                 & $2.89\times10^{-10}$ & $10^{-6}$             & $1.32\times10^{-20}$  & \cite{Masuda:2016ggi} \\
Viking lander ranging            & $6$                 & $4.33\times10^{-10}$ & $4\times10^{-12}$     & $1.98\times10^{-20}$  & \cite{PhysRevLett.51.1609} \\
Pulsating white dwarfs           & $0$                 & $0$               & $2.0\times10^{-12}$   & $0$                   & \cite{Corsico:2013ida} \\
Hubble diagram (SNe~Ia)          & $1.00\times10^{8}$ & $0.007$           & $1.0\times10^{-11}$   & $3.28\times10^{-13}$  & \cite{Gaztanaga:2001fh} \\
Gravitochemical heating          & $1.00\times10^{8}$ & $0.007$           & $4.0\times10^{-12}$   & $3.28\times10^{-13}$  & \cite{Jofre:2006ug} \\
Gravitational waves              & $1.30\times10^{8}$ & $0.009$           & $1.0\times10^{-9}$    & $4.26\times10^{-13}$  & \cite{Sun:2023bvy} \\
Helioseismology                  & $4.00\times10^{9}$ & $0.365$           & $1.25\times10^{-13}$  & $8.91\times10^{-12}$  & \cite{Bonanno:2017dcx} \\
Paleontology                     & $4.00\times10^{9}$ & $0.365$           & $2.0\times10^{-11}$   & $8.91\times10^{-12}$  & \cite{Uzan:2002vq} \\
Globular clusters                & $1.00\times10^{10}$ & $1.771$           & $3.5\times10^{-11}$   & $4.60\times10^{-12}$  & \cite{DeglInnocenti:1995hbi} \\
Binary pulsar masses             & $1.00\times10^{10}$ & $1.771$           & $4.8\times10^{-12}$   & $4.60\times10^{-12}$  & \cite{Thorsett:1996fr} \\
Strong lensing (cosmic)          & $1.00\times10^{10}$ & $1.771$           & $3.0\times10^{-1}$    & $4.60\times10^{-12}$  & \cite{Giani:2020fpz} \\
\hline\hline
\end{tabular}
\end{table*}
In Figure \ref{dotGeff_Geff_CPL}, we show the variation of the gravitational constant $\dot{G}_{\rm eff}/G_{\rm eff}$ in a CPL background universe, together with its $1\sigma$ uncertainty band computed via standard error propagation. Note that for the reconstructed quantities, $q \equiv q_{\rm CPL}(z, \Omega_{\rm m0}, w_0, w_a, h)$ is defined in Eq.(\ref{Friedman}), while $\mu_G$ is defined in Eq.(\ref{pantazisparam}). The corresponding expression reads:

\begin{align}
\frac{\dot{G}_{\rm eff}}{G_{\rm eff}} &= -3.23\,h\,\text{yr}^{-1} \times10^{-18}\,(3.156\times10^{7})\,\times \notag \\
&\qquad\qquad\qquad \qquad   (1+z)\,q_{\rm CPL}^{1/2} \frac{\mathrm{d}\mu_G(z,2,g_a^{\rm CPL})/\mathrm{d}z}{\mu_G(z,2,g_a^{\rm CPL})},
\label{eq:dotGoverG}
\end{align}
and it is evaluated using the best-fit parameter values of the specific model reported in Table~\ref{tab:fs8_constraint}.

On Eq.\eqref{eq:dotGoverG} we also have:
\begin{equation}
    \frac{\mathrm{d}\mu_G(z,2,g_a^{\rm CPL})}{\mathrm{d}z}=-\frac{2\,g_a\,z\,(z-3)(z-1)}{(1 + z)^5}.
\end{equation}
We propagate parameter uncertainties to the inferred time-variation of the effective Newton constant, denoted by $\dot G_{\rm eff}/G_{\rm eff}$, using the standard linear (first-order) error propagation. Let $\mathbf p=(\Omega_m,w_0,w_a,g_a,h)$ be the vector of model parameters. To first order the variance of $\dot G_{\rm eff}/G_{\rm eff}$ is
\begin{equation}
\sigma^2_{\dot G/G}
=
\sum_{i,j}
\frac{\partial}{\partial p_i}\!\left(\frac{\dot G_{\rm eff}}{G_{\rm eff}}\right)
\, c_{ij} \,
\frac{\partial}{\partial p_j}\!\left(\frac{\dot G_{\rm eff}}{G_{\rm eff}}\right),
\end{equation}
where the partial derivative components are evaluated at the best-fit parameter vector $\mathbf p_{\rm best}$. 
For the error propagation procedure, we used the $5\times5$ covariance matrix $[c_{ij}]$ as
\begin{widetext}
\[
[c_{ij}] = \begin{bmatrix}
0.00015596 & 0.0015654 & -0.0044129 & -0.0020125 & -0.00017596 \\
0.0015654 & 0.017891 & -0.053435 & -0.02132 & -0.0017547 \\
-0.0044129 & -0.053435 & 0.17022 & 0.059871 & 0.0047732 \\
-0.0020125 & -0.02132 & 0.059871 & 0.032793 & 0.0023041 \\
-0.00017596 & -0.0017547 & 0.0047732 & 0.0023041 & 0.00020265
\end{bmatrix}
\]
\end{widetext}
that embodies the parameters of interest $\{\Omega_{m0},w_0,w_a,g_a,h\}$.

Table~\ref{tab:comparison_methods} summarizes observational constraints on $\dot{G}_{\rm eff}/G_{\rm eff}$ and compares them with predictions from the CPL parametrization. It includes various astrophysical and cosmological probes, their characteristic timescales, and corresponding CPL redshifts. Observational limits on $|\dot{G}_{\rm eff}/G_{\rm eff}|$ range from precise solar system and lunar laser measurements ($\sim 10^{-14}\,\mathrm{yr}^{-1}$) to cosmological bounds from strong lensing ($\sim 10^{-1}\,\mathrm{yr}^{-1}$). In contrast, CPL predictions are consistently smaller by several orders of magnitude, indicating a mild evolution of $G_{\rm eff}$ and broad agreement with current observations across diverse epochs.

\section{Perturbative Derivation of $G_{\rm eff}$ in Scalar–Tensor Theories}\label{appC}

It has been shown~\citep{Boisseau:2000pr,Amendola:2007rr,Tsujikawa:2007gd} that the effective gravitational constant between two test masses is given by
\begin{equation}\label{geff}
    G_{\rm eff}
    = \frac{1}{8\pi F}
      \frac{2F + 4\bigl(\tfrac{dF}{d\Phi}\bigr)^2}
           {2F + 3\bigl(\tfrac{dF}{d\Phi}\bigr)^2}.
\end{equation}

To derive this relation in a simplified manner, we consider scalar perturbations of the metric (for $a=1$) of the form
\begin{equation}\label{metric}
    g_{00} = -1 - 2\Psi_g, \qquad 
    g_{ij} = \delta_{ij}(1 + 2\Phi_g), \qquad 
    g_{0i} = 0,
\end{equation}
and perturbations of the scalar field
\begin{equation}
    \Phi = \Phi_0 + \delta \Phi(\mathbf{x}),
\end{equation}
where $\Phi_0 \approx \text{const.}$ and $U(\Phi_0)$ is negligible at scales dominated by gravity. 

Note that, the variation of the action (with Lagrangian density as defined in Eq.(\ref{eq:lagrangian})) with respect to the scalar field $\Phi$ produces \citep{Esposito-Farese:2000pbo}
\begin{equation}\label{eqfield}
    -\frac{dU(\Phi)}{d\Phi} 
    + \nabla^{\mu}\nabla_{\mu}\Phi 
    + \frac{R}{2}\frac{dF}{d\Phi} = 0.
\end{equation}

While one could formally begin with an FLRW background by assuming 
$\Phi = \Phi_0(t) + \delta\Phi(\mathbf{x}, t)$ and taking the sub-horizon limit 
($H^2 \ll k^2 / a^2$) together with the quasi-static approximation 
($\dot{\Phi}_g, \ddot{\Phi}_g \ll k^2 \Phi_g / a$ and 
$\dot{\Psi}_g, \ddot{\Psi}_g \ll k^2 \Psi_g / a$), 
we instead adopt a Minkowski background to considerably simplify the calculations.

Then, the functions $F(\Phi)$ and $U(\Phi)$ can be expanded to first order in the perturbation:
\begin{equation}
    F(\Phi) = F(\Phi_0) + \frac{dF}{d\Phi}\Big|_{\Phi_0} \delta\Phi, 
    \qquad 
    U(\Phi) = U(\Phi_0) + \frac{dU}{d\Phi}\Big|_{\Phi_0} \delta\Phi.
\end{equation}
From the unperturbed field equation~(\ref{eqfield}), it follows that $dU/d\Phi|_{\Phi_0} = 0$.

 Given the metric Eq.(\ref{metric}), we obtain:
 \begin{widetext}
      \begin{equation}
    R_{00} = \nabla^2\Psi_g, \quad 
    R_{ij} = -\delta_{ij}\nabla^2\Phi_g-\partial_i\partial_j(\Phi_g+\Psi_g), \quad 
    R = -2\nabla^2(2\Phi_g+\Psi_g),
\end{equation}
 \end{widetext}

and consequently, the Einstein tensor components are
\begin{equation}
    G_{00} = -2\nabla^2\Phi_g, \quad 
    G_{ij} = (\delta_{ij}\nabla^2-\partial_i\partial_j)(\Phi_g+\Psi_g).
\end{equation}
From Eq.\eqref{fieldequations}, we deduce that anisotropic stress leads to:
\begin{equation}\label{anisotropicstress}
    \Psi_g+\Phi_g=-\frac{F_0'}{F_0}\delta\Phi.
\end{equation}
Note also that, we denote the derivative $\frac{dF}{d\Phi}\big|_{\Phi_0} \equiv F'_0$ for convenience.
  First, for a massless scalar field 
($d^2U/d\Phi^2|_{\Phi_0} = 0$), the perturbed field equation reads
\begin{equation}\label{eq0}
    F_0'\nabla^2\Psi_g=-\left(1+\frac{2F_0'^2}{F_0}\right)\nabla^2(\delta\Phi).
\end{equation}
Furthermore, the trace of the perturbed Eq.(\ref{fieldequations}), 
\begin{equation}
    2F_0\nabla^2( 2\Phi_g+\Psi_g)=-\rho-3F_0'\nabla^2(\delta\Phi),
\end{equation}
combined with Eq.(\ref{eq0}), gives (for $F_0'\neq0$)
\begin{equation}\label{eq2}
    \nabla^2(\delta\Phi)
    = -\frac{F_0'}
           {2F_0 + 3F_0'^2}
      \rho.
\end{equation}
Finally,  the $00$-component of Eq.(\ref{fieldequations}) yields
\begin{equation}\label{00componenteinstein}
    -2 F_0\nabla^2\Phi_g=\rho+F_0'\nabla^2(\delta \Phi)
\end{equation}

Finally, substituting Eq.(\ref{eq2}) into Eq.(\ref{00componenteinstein}) yields (and using Eq.(\ref{anisotropicstress}))
\begin{equation}\label{poissonpsimg}
    \nabla^2\Psi_g
    = \frac{1}{2F_0}
      \frac{
        2F_0 + 4\bigl(F_0')^2
      }{
        2F_0 + 3\bigl(F_0')^2
      }\rho.
\end{equation}
Comparing the above Eq.(\ref{poissonpsimg}) with the Poisson equation, which corresponds to the Newtonian limit of General Relativity (where anisotropic stress vanishes i.e. $\Psi_g = -\Phi_g$),
\begin{equation}
    \nabla^2 \Psi_g = \frac{1}{2}\rho, \quad (G_N = 1/8\pi),
\end{equation}
we can identify the effective gravitational constant $G_{\rm eff}$ in scalar--tensor theories, as defined in Eq.~(\ref{geff}).


\section{Data analysis for $w_{\rm DE}$ parametrizations}\label{AppA}

\begin{itemize}
    \item \textbf{CPL (Chevallier--Polarski--Linder)}~\cite{Chevallier:2000qy,Linder:2002et}:
    \begin{equation}
        w(z) = w_0 + w_a \frac{z}{1+z}
    \end{equation}
    
    \item \textbf{BA (Barboza--Alcaniz)}~\cite{Barboza:2008rh,Barboza:2011gd}:
    \begin{equation}
        w(z) = w_0 + w_a \frac{z(1+z)}{1+z^2}
    \end{equation}
    
    \item \textbf{Logarithmic}~\cite{Feng:2011zzo}:
    \begin{equation}
        w(z) = w_0 + w_a \ln(1 + z)
    \end{equation}
    
    \item \textbf{JBP (Jassal--Bagla--Padmanabhan)}~\cite{Jassal:2004ej}:
    \begin{equation}
        w(z) = w_0 + w_a \frac{z}{(1+z)^2}
    \end{equation}
\end{itemize}
\begin{figure}
    \centering
    \includegraphics[width=1\linewidth]{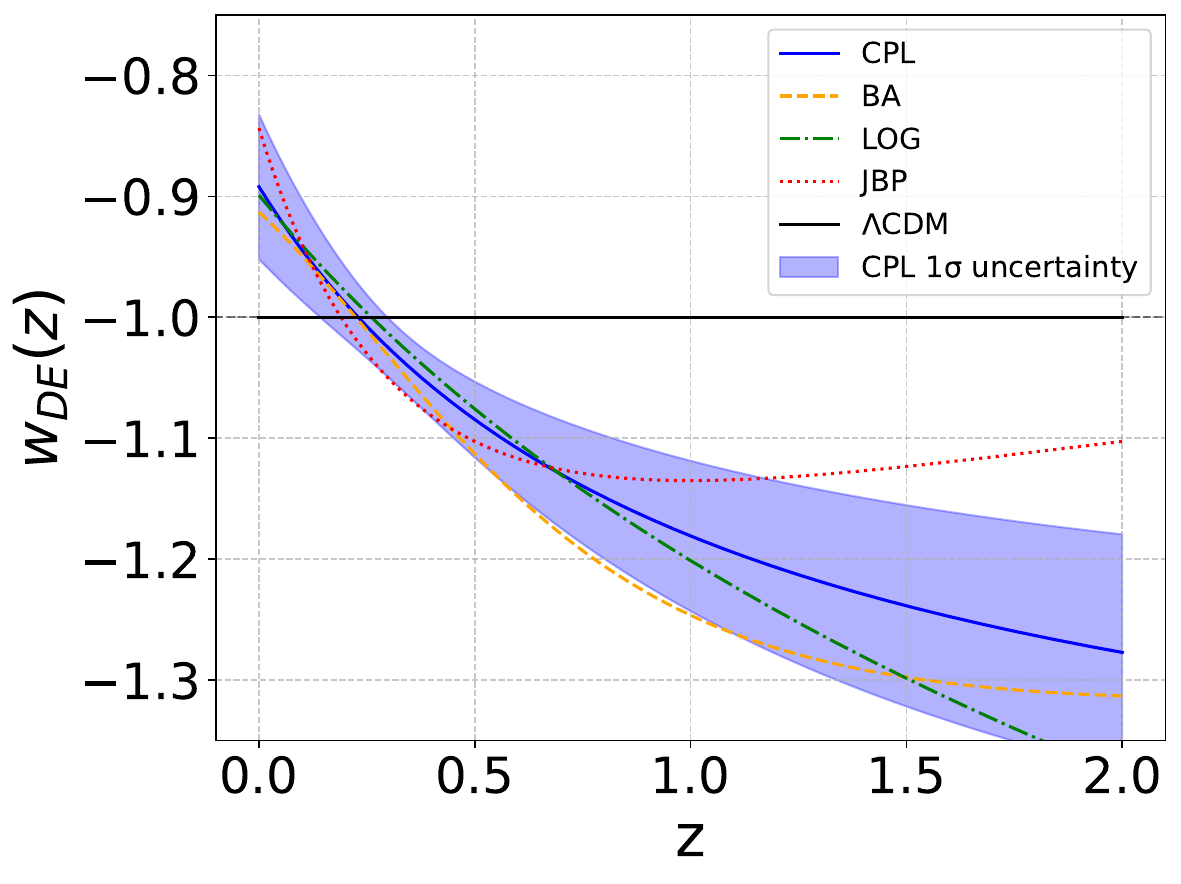}
    \caption{Different dark energy parametrizations $w(z)$ plotted in the case where $g_a=0$ according to their best fit values of $w_0$ and $w_a$ from the observational data (Table \ref{tab:ga equal 0}). Every parametrization favors phantom crossing. The dataset combination used here is: Pantheon+, CMB, DESI DR2, RSD, and BBN constraint.}
    \label{4models_wz_ga_0}
\end{figure}

For the $\Lambda$CDM model, the best-fit parameters were determined to be: 
$M = -19.398 \pm 0.009$, 
$\Omega_{\mathrm{m0}} = 0.297 \pm 0.004$, 
$\Omega_b h^2 = 0.0227 \pm 0.0036$, 
$\sigma_8 = 0.802 \pm 0.026$, 
and $h = 0.688 \pm 0.003$, 
with a minimum chi-squared value of $\chi^2_{\mathrm{min}} = 1588$. 

\begin{table}[htbp]
\centering
\caption{Comparison of Hubble constant determinations from different methods and models.}
\label{tab:H0_comparison2}
\begin{ruledtabular}
\begin{tabular}{lc}
Method/Model & $H_0$ [km s$^{-1}$ Mpc$^{-1}$] \\
\hline
\multicolumn{2}{c}{\textit{Local Measurements}} \\
SH0ES (Cepheids + SNe) \cite{Riess:2021jrx} & $73.04 \pm 1.04$ \\
\hline
\multicolumn{2}{c}{\textit{CMB (GR)}} \\
Planck 2018 $\Lambda$CDM \cite{Planck:2018vyg} & $67.36 \pm 0.54$ \\
\hline
\multicolumn{2}{c}{\textit{Selected datasets }} \\
$\Lambda$CDM (Pantheon+) & $73.4\pm 1$\\
$\Lambda$CDM (CMB)& $67.36\pm 0.7$\\
$\Lambda$CDM (RSD+DESI DR2+CMB) & $68.5\pm 0.3$\\
CPL(GR) (RSD+DESI DR2+CMB) & $64.9\pm 0.4$\\
\end{tabular}
\end{ruledtabular}
\end{table}
\begin{table*}[!ht]
\centering
\caption{\label{tab:ga equal 0}Observational constraints (Pantheon+, CMB, DESI DR2, RSD, BBN constraint) on the parameters for $g_a=0$. The minimum chi-squared $\chi^2_{\rm min}$ for each model is shown in the last row. Errors are $1\sigma$ uncertainties from the covariance matrices.}
\begin{ruledtabular}
\begin{tabular}{ccccc}
 & CPL & BA & Log & JBP \\
\hline
$M$ & $-19.379 \pm 0.014$ & $-19.368 \pm 0.014$ & $-19.381 \pm 0.014$ & $-19.381 \pm 0.014$ \\
$\Omega_{\rm m0}$ & $0.298 \pm 0.006$ & $0.295 \pm 0.006$ & $0.299 \pm 0.006$ & $0.298 \pm 0.007$ \\
$\Omega_b h^2$ & $0.02247 \pm 0.00013$ & $0.02246 \pm 0.00013$ & $0.02245 \pm 0.00014$ & $0.02251 \pm 0.00014$ \\
$w_0$ & $-0.892 \pm 0.060$ & $-0.913 \pm 0.056$ & $-0.899 \pm 0.060$ & $-0.843 \pm 0.123$ \\
$w_a$ & $-0.578 \pm 0.224$ & $-0.334 \pm 0.128$ & $-0.436 \pm 0.186$ & $-1.166 \pm 0.731$ \\
$\sigma_8$ & $0.780 \pm 0.026$ & $0.785 \pm 0.026$ & $0.805 \pm 0.026$ & $0.800 \pm 0.026$ \\
$h$ & $0.691 \pm 0.006$ & $0.696 \pm 0.006$ & $0.690 \pm 0.006$ & $0.690 \pm 0.007$ \\
\hline
$\chi^2_{\rm min}$ & $1578.44$ & $1579.21$ & $1576.89$ & $1580.65$ \\
\end{tabular}
\end{ruledtabular}
\end{table*}
We constrain cosmological parameters using multiple observational probes, considering the case without modified gravity ($g_a = 0$). This enables comparison with established results in the literature (see Table \ref{tab:H0_comparison2}). Our methodology combines current measurements from baryon acoustic oscillations (BAO), cosmic microwave background (CMB), Type Ia supernovae (SNe Ia), and redshift-space distortions (RSD). The Hubble function \( H(z) \) is constructed by adopting parametric forms for the dark energy equation of state \( w_{\mathrm{DE}}(z) \), considering the following parametrizations:

A notable difference in $\chi^2_{\mathrm{min}}$ is observed between the dynamical dark energy scenarios and the $\Lambda$CDM case.

Figure \ref{4models_wz_ga_0} illustrates $w(z)$ for our four dynamical dark energy parametrizations in the case mentioned. The best-fit values for each model used for the creation of this plot are shown in Table \ref{tab:ga equal 0}. For example, in the CPL (GR) scenario (see Fig.\ref{4models_wz_ga_0}), the phantom divide ($w=-1$) is crossed at a positive redshift of approximately $z \sim 0.3$ when using the combined dataset (RSD + DESI DR2 + BBN constraint + CMB + Pantheon+). Notably, if Pantheon+ is excluded from the dataset, the phantom crossing shifts to $z \simeq 0.49$.

\section{Generation of Correlated Posterior Samples and Triangle Plot Visualization}\label{appB}

Figure~\ref{fig:corner_plots_cpl} illustrates the joint and marginalized posterior distributions of the eight cosmological parameters for the CPL (MG) parametrization. 
Under the Gaussian  approximation and in the asymptotic limit of large samples, 
the covariance matrix of the parameter estimates is given by  $\boldsymbol{\Sigma} = \mathbf{F}^{-1}$, where $\boldsymbol{F}$ denotes the Fisher information matrix, where $F_{ij} = -\langle \partial_i \partial_j \ln \mathcal{L} \rangle$ quantifies the local curvature of the log-likelihood surface around the maximum. 
Within this approximation, $\boldsymbol{\mu}$ coincides with the vector of best-fit parameter values.
Note that the approximated covariance is

\begin{widetext}
\[
[\Sigma_{ij}] = \begin{bmatrix}
0.00108 & -0.000394 & 1.22\times 10^{-6} & -0.00375 & 0.00982 & 0.00541 & -0.000227 & 0.000457 \\
-0.000394 & 0.000156 & -5.62\times 10^{-7} & 0.00157 & -0.00441 & -0.00201 & 7.92\times 10^{-5} & -0.000176 \\
1.22\times 10^{-6} & -5.62\times 10^{-7} & 1.93\times 10^{-8} & -5.83\times 10^{-6} & 2.25\times 10^{-5} & 5.84\times 10^{-6} & -6.80\times 10^{-8} & 5.73\times 10^{-7} \\
-0.00375 & 0.00157 & -5.83\times 10^{-6} & 0.0179 & -0.0534 & -0.0213 & 0.000901 & -0.00175 \\
0.00982 & -0.00441 & 2.25\times 10^{-5} & -0.0534 & 0.170 & 0.0599 & -0.00241 & 0.00477 \\
0.00541 & -0.00201 & 5.84\times 10^{-6} & -0.0213 & 0.0599 & 0.0328 & -0.00148 & 0.00230 \\
-0.000227 & 7.92\times 10^{-5} & -6.80\times 10^{-8} & 0.000901 & -0.00241 & -0.00148 & 0.000692 & -9.48\times 10^{-5} \\
0.000457 & -0.000176 & 5.73\times 10^{-7} & -0.00175 & 0.00477 & 0.00230 & -9.48\times 10^{-5} & 0.000203
\end{bmatrix}
\]
\end{widetext}

The contours are obtained under the assumption of a multivariate Gaussian likelihood of the form
\begin{equation}
P(\boldsymbol{\theta} \,|\, \mathrm{data}) \propto 
\exp\!\left[-\tfrac{1}{2}(\boldsymbol{\theta} - \boldsymbol{\mu})^{\mathrm{T}} 
\boldsymbol{\Sigma}^{-1} (\boldsymbol{\theta} - \boldsymbol{\mu})\right],
\end{equation}
with uniform (non-informative) priors.

The corner plot is generated by drawing a large ensemble of random realizations 
$\boldsymbol{\theta}^{(k)} \sim \mathcal{N}(\boldsymbol{\mu}, \boldsymbol{\Sigma})$ 
using the \texttt{NumPy} multivariate normal sampler. 

Each realization represents a statistically consistent set of cosmological parameters, and the resulting point cloud reconstructs the Gaussian posterior in parameter space. 
The \texttt{GetDist} library \cite{Lewis:2019xzd} is then used to compute the one- and two-dimensional marginalized distributions, producing the filled $1\sigma$ and $2\sigma$ confidence contours shown in Fig.\ref{fig:corner_plots_cpl}. 
In the two-dimensional projections, the elliptical shape and orientation of the contours directly reflect the parameter covariances encoded in $\boldsymbol{\Sigma}$, i.e.\ the degeneracy directions implied by the Fisher matrix. 
Thus, the figure provides a visual representation of both the parameter uncertainties and their mutual correlations under the Gaussian likelihood approximation.

\newpage
\bibliography{main}

\end{document}